\documentclass[english]{IEEEtran}
\usepackage[T3,T1]{fontenc}
\usepackage[latin9]{inputenc}
\usepackage{color}
\usepackage{babel}
\usepackage{array}
\usepackage{amsmath}
\usepackage{amsthm}
\usepackage{amssymb}
\usepackage{graphicx}
\usepackage[unicode=true,pdfusetitle,
 bookmarks=true,bookmarksnumbered=true,bookmarksopen=true,bookmarksopenlevel=3,
 breaklinks=false,pdfborder={0 0 1},backref=false,colorlinks=true]
 {hyperref}
\hypersetup{
 pdfborderstyle=,pdfborderstyle={},pdfborderstyle={},pdfborderstyle={},pdfborderstyle={},pdfborderstyle={},pdfborderstyle={},pdfborderstyle={},pdfborderstyle={},pdfborderstyle={},pdfborderstyle={},pdfborderstyle={},pdfborderstyle={},pdfborderstyle={},pdfborderstyle={},pdfborderstyle={},pdfborderstyle={},pdfborderstyle={},pdfborderstyle={},pdfborderstyle={},pdfpagelayout=OneColumn,pdfnewwindow=true,pdfstartview=XYZ,plainpages=false,linkcolor=blue,urlcolor=blue,citecolor=red,anchorcolor=blue,linkcolor=blue,urlcolor=blue,citecolor=red,anchorcolor=blue}

\makeatletter

\providecommand{\tabularnewline}{\\}

\theoremstyle{plain}
\newtheorem{thm}{\protect\theoremname}
\theoremstyle{remark}
\newtheorem{rem}{\protect\remarkname}
\theoremstyle{plain}
\newtheorem{lem}{\protect\lemmaname}
\theoremstyle{plain}
\newtheorem{cor}{\protect\corollaryname}
\theoremstyle{plain}
\newtheorem{prop}{\protect\propositionname}

\usepackage{babel}

\newcommand{\subsetsim}{\mathrel{%
  \ooalign{\raise0.2ex\hbox{$\subset$}\cr\hidewidth\raise-0.8ex\hbox{\scalebox{0.9}{$\sim$}}\hidewidth\cr}}}
\newcommand{\supsetsim}{\mathrel{%
  \ooalign{\raise0.2ex\hbox{$\supset$}\cr\hidewidth\raise-0.8ex\hbox{\scalebox{0.9}{$\sim$}}\hidewidth\cr}}}

\newcommand{\subsetapprox}{\mathrel{%
  \ooalign{\raise0.4ex\hbox{$\subset$}\cr\hidewidth\raise-0.8ex\hbox{\scalebox{0.9}{$\approx$}}\hidewidth\cr}}}
\allowdisplaybreaks[1]
\DeclareSymbolFont{tipa}{T3}{cmr}{m}{n}
\DeclareMathAccent{\invbreve}{\mathalpha}{tipa}{16}

\providecommand{\corollaryname}{Corollary}
\providecommand{\lemmaname}{Lemma}
\providecommand{\propositionname}{Proposition}
\providecommand{\remarkname}{Remark}
\providecommand{\theoremname}{Theorem}

\newcommand{\supp}{\mathrm{supp}}

\usepackage{lineno}

\def\DD{\mathsf{D}}

\newcommand{\calA}{\mathcal{A}}
\newcommand{\calB}{\mathcal{B}}
\newcommand{\calC}{\mathcal{C}}

\newcommand{\calV}{\mathcal{V}}

\makeatother

\providecommand{\corollaryname}{Corollary}
\providecommand{\lemmaname}{Lemma}
\providecommand{\propositionname}{Proposition}
\providecommand{\remarkname}{Remark}
\providecommand{\theoremname}{Theorem}

\begin{document}
\title{Graphs of Joint Types, Noninteractive Simulation, and Stronger Hypercontractivity }
\author{Lei Yu, Venkat Anantharam, and Jun Chen\thanks{L. Yu is with the School of Statistics and Data Science, LPMC, KLMDASR,
and LEBPS, Nankai University, Tianjin 300071, China (e-mail: leiyu@nankai.edu.cn).
V. Anantharam is with the Department of Electrical Engineering and
Computer Sciences, University of California, Berkeley, CA 94720, USA
(e-mail: ananth@berkeley.edu). J. Chen is with the Department of Electrical
and Computer Engineering, McMaster University, Hamilton, ON L8S 4K1,
Canada (e-mail: chenjun@mcmaster.ca). Research of the first two authors
was supported by the NSF grants CNS--1527846, CCF--1618145, CCF-1901004,
CIF-2007965, the NSF Science \& Technology Center grant CCF--0939370
(Science of Information), and the William and Flora Hewlett Foundation
supported Center for Long Term Cybersecurity at Berkeley. The first
author was also supported in part by the NSFC grant 62101286 and the
Fundamental Research Funds for the Central Universities of China (Nankai
University) under grant 054-63233073. This paper was presented in
part at the 2021 IEEE International Symposium on Information Theory
\cite{yu2021type}.}}
\maketitle
\begin{abstract}
In this paper, we study the type graph, namely, a bipartite graph
induced by a joint type. We investigate the maximum edge density of
induced bipartite subgraphs of this graph having a number of vertices
on each side on an exponential scale in the length $n$ of the type.
This can be seen as an isoperimetric problem. We provide asymptotically
sharp bounds for the exponent of the maximum edge density as the length
of the type goes to infinity. We also study the biclique rate region
of the type graph, which is defined as the set of $(R_{1},R_{2})$
such that there exists a biclique of the type graph which has respectively
$2^{nR_{1}}$ and $2^{nR_{2}}$ vertices on the two sides. We provide
asymptotically sharp bounds for the biclique rate region as well.
We then discuss the connections of these results to noninteractive
simulation and hypercontractivity inequalities. Furthermore, as an
application of our results, a new outer bound for the zero-error capacity
region of the binary adder channel is provided, which improves the
previously best known bound, due to Austrin, Kaski, Koivisto, and
Nederlof. Our proofs in this paper are based on the method of types
and linear algebra. 
\end{abstract}


\section{Introduction}

Let $\mathcal{X}$ and $\mathcal{Y}$ be two finite sets. Let $T_{X}$
be an $n$-type on $\mathcal{X}$, i.e., an empirical distribution
of sequences from $\mathcal{X}^{n}$. Let $\mathcal{T}_{T_{X}}^{(n)}$,
or $\mathcal{T}_{T_{X}}$ for short, be the $n$-type class with respect
to $T_{X}$, i.e., the set of sequences of length $n$ having the
type $T_{X}$. Similarly, let $T_{XY}$ be a joint $n$-type\footnote{We attribute the parameter $n$ to $T_{XY}$. }
on $\mathcal{X}\times\mathcal{Y}$ and $\mathcal{T}_{T_{XY}}^{(n)}$,
or $\mathcal{T}_{T_{XY}}$ for short, the joint $n$-type class with
respect to $T_{XY}$. Note that $\mathcal{T}_{T_{XY}}\subseteq\mathcal{T}_{T_{X}}\times\mathcal{T}_{T_{Y}}$,
where $T_{X},T_{Y}$ are the marginal types corresponding to the joint
type $T_{XY}$. In this paper, we consider the undirected bipartite
graph $G_{T_{XY}}$ whose vertex set is $\mathcal{T}_{T_{X}}\cup\mathcal{T}_{T_{Y}}$
and whose edge set can be identified with $\mathcal{T}_{T_{XY}}$,
defined as follows. Consider $\mathbf{x}\in\mathcal{T}_{T_{X}}$ and
$\mathbf{y}\in\mathcal{T}_{T_{Y}}$ as vertices of $G_{T_{XY}}$.
Two vertices $\mathbf{x},\mathbf{y}$ are joined by an edge if and
only if $(\mathbf{x},\mathbf{y})\in\mathcal{T}_{T_{XY}}$. The graph
$G_{T_{XY}}$ is termed the graph of $T_{XY}$ or, more succinctly,
a type graph \cite{nazari2009new}. For brevity, when there is no
ambiguity, we use the abbreviated notation $G$ for $G_{T_{XY}}$.

For subsets $\calA\subseteq\mathcal{T}_{T_{X}},\calB\subseteq\mathcal{T}_{T_{Y}}$,
we obtain an induced subgraph $G[\calA,\calB]$ of $G$, whose vertex
set is the union of $\calA$ and $\calB$, and where $\mathbf{x},\mathbf{y}$
are joined by an edge if and only if they are joined by an edge in
$G$. For the induced subgraph $G[\calA,\calB]$, the (edge) density
$\rho(G[\calA,\calB])$ is defined as 
\[
\rho(G[\calA,\calB]):=\frac{\#\textrm{ of edges in }G[\calA,\calB]}{|\calA||\calB|}.
\]
Thus we have $\rho(G[\calA,\calB])=\frac{|(\calA\times\calB)\cap\mathcal{T}_{T_{XY}}|}{|\calA||\calB|}$.
Since\footnote{Throughout this paper, we 
write $a_{n}\doteq b_{n}$ to denote $a_{n}=b_{n}2^{o(n)}$.}{} $|\mathcal{T}_{T_{X}^{(n)}}|\doteq2^{nH_{T^{(n)}}(X)}$ \cite{Csiszar},
it follows that $\rho(G)=\frac{|\mathcal{T}_{T_{XY}^{(n)}}|}{|\mathcal{T}_{T_{X}^{(n)}}||\mathcal{T}_{T_{Y}^{(n)}}|}\doteq2^{-nI_{T^{(n)}}(X;Y)}$
for any sequence of joint types $\{T_{XY}^{(n)}\}$, where $I_{T^{(n)}}(X;Y)$
denotes the mutual information of the pair $(X,Y)$ having the joint
distribution $T_{XY}^{(n)}$, taken to the base $2$. Moreover, if
we only fix $T_{X},T_{Y}$, $\calA$, and $\calB$, then $T_{XY}\in\calC_{n}(T_{X},T_{Y})\mapsto\rho(G_{T_{XY}}[\calA,\calB])$
forms a probability mass function, i.e., 
\begin{align*}
\rho(G_{T_{XY}}[\calA,\calB]) & \ge0,\\
\sum_{T_{XY}\in\calC_{n}(T_{X},T_{Y})}\rho(G_{T_{XY}}[\calA,\calB]) & =1,
\end{align*}
where $\calC_{n}(T_{X},T_{Y})$ denotes the set of joint types $T_{XY}$
with marginals $T_{X},T_{Y}$. We term this distribution a type distribution,
which roughly speaking can be considered as a generalization from
binary alphabets to arbitrary finite alphabets of the classic distance
distribution in coding theory; please refer to \cite{macwilliams1977theory}
for the distance distribution of a single code, and \cite{yu2019on}
for the distance distribution between two codes.

Given $1\le M_{1}\le|\mathcal{T}_{T_{X}}|,1\le M_{2}\le|\mathcal{T}_{T_{Y}}|$,
define the maximal density of subgraphs with size $(M_{1},M_{2})$
as 
\begin{align}
\Gamma_{n}(M_{1},M_{2}) & :=\max_{\substack{\calA\subseteq\mathcal{T}_{T_{X}},\calB\subseteq\mathcal{T}_{T_{Y}}:\\
|\calA|=M_{1},|\calB|=M_{2}
}
}\rho(G[\calA,\calB]).\label{eq:Gamma_overline}
\end{align}

Recall that $T_{X|Y}$ and $T_{Y|X}$ denote the conditional types
corresponding to the joint type $T_{XY}$. For a sequence $\mathbf{x}\in\mathcal{T}_{T_{X}}$,
let 
\[
\mathcal{T}_{T_{Y|X}}(\mathbf{x}):=\{\mathbf{y}\in\mathcal{Y}^{n}:(\mathbf{x},\mathbf{y})\in\mathcal{T}_{T_{XY}}\}
\]
denote the corresponding conditional type class. Since $N_{1}:=|\mathcal{T}_{T_{Y|X}}(\mathbf{x})|$
is independent of $\mathbf{x}\in\mathcal{T}_{T_{X}}$, the degrees
of the vertices $\mathbf{x}\in\mathcal{T}_{T_{X}}$ are all equal
to the constant $N_{1}$. Similarly, the degrees of the vertices $\mathbf{y}\in\mathcal{T}_{T_{Y}}$
are all equal to the constant $N_{2}:=|\mathcal{T}_{T_{X|Y}}(\mathbf{y})|$.
Hence we have 
\begin{align*}
 & |\calB|\rho(G[\calA,\calB])+|\calB^{c}|\rho(G[\calA,\calB^{c}])\\
 & =\frac{|(\calA\times\mathcal{T}_{T_{Y}})\cap\mathcal{T}_{T_{XY}}|}{|\calA|}\\
 & =\frac{\sum_{\mathbf{x}\in\calA}|\mathcal{T}_{T_{Y|X}}(\mathbf{x})|}{|\calA|}=N_{1},
\end{align*}
where $\calB^{c}:=\mathcal{T}_{T_{Y}}\backslash\calB$. Thus, over
$\calA,\calB$ with fixed sizes, maximizing $\rho(G[\calA,\calB])$
is equivalent to minimizing $\rho(G[\calA,\calB^{c}])$ (or $\rho(G[\calA^{c},\calB])$).
In other words, determining the maximal density is in fact an edge-isoperimetric
problem which concerns minimizing the number of or weighted sum of
edges between a set of vertices and its complement. Furthermore, given
$\calA\subseteq\mathcal{T}_{T_{X}}$ and $M_{2}$, we see that 
\begin{align*}
 & \max_{\calB\subseteq\mathcal{T}_{T_{Y}}:|\calB|=M_{2}}\rho(G[\calA,\calB])\\
 & =\frac{1}{|\calA|M_{2}}\max_{\calB\subseteq\mathcal{T}_{T_{Y}}:|\calB|=M_{2}}\sum_{\mathbf{y}\in\calB}|\calA\cap\mathcal{T}_{T_{X|Y}}(\mathbf{y})|,
\end{align*}
and the maximum is attained by $\calB^{*}$ such that\footnote{This condition is closely related to a classic concept, the $\eta$-image
of a set, which was exploited in the context of the image size characterization
in \cite{Ahls76,Csi97}. } $|\calA\cap\mathcal{T}_{T_{X|Y}}(\mathbf{y})|\geq|\calA\cap\mathcal{T}_{T_{X|Y}}(\mathbf{y}')|$
for any $\mathbf{y}\in\calB^{*},\mathbf{y}'\notin\calB^{*}$. Hence,
$M_{2}\mapsto\max_{\calB\subseteq\mathcal{T}_{T_{Y}}:|\calB|=M_{2}}\rho(G[\calA,\calB])$
is nonincreasing, which implies that $\Gamma_{n}(M_{1},M_{2})$ is
nonincreasing in one parameter given the other parameter.

Let\footnote{We use the notation $[m:n]:=\{m,m+1,...,n\}$ and $[n]:=[1:n]$.}
\begin{align}
\mathcal{R}_{X}^{(n)} & :=\{\frac{1}{n}\log M_{1}:M_{1}\in[|\mathcal{T}_{T_{X}}|]\},
\:\label{eq:rates}\\
\mathcal{R}_{Y}^{(n)} & :=\{\frac{1}{n}\log M_{2}:M_{2}\in[|\mathcal{T}_{T_{Y}}|]\},
\end{align}
where the logarithm is taken to the base $2$. Given a joint $n$-type
$T_{XY}$, define the exponent of maximal density for a pair $(R_{1},R_{2})\in\mathcal{R}_{X}^{(n)}\times\mathcal{R}_{Y}^{(n)}$
as 
\begin{equation}
E_{n}(R_{1},R_{2}):=-\frac{1}{n}\log\Gamma_{n}(2^{nR_{1}},2^{nR_{2}}).\label{eq:-1}
\end{equation}
If the edge density of a subgraph in a bipartite graph $G$ is equal
to $1$, then this subgraph is called a biclique of $G$. Along these
lines, we define the biclique rate region of $T_{XY}$ as 
\begin{align}
\mathcal{R}_{n}(T_{XY}) & :=\{(R_{1},R_{2})\in\mathcal{R}_{X}^{(n)}\times\mathcal{R}_{Y}^{(n)}:\nonumber \\
 & \qquad\qquad\Gamma_{n}(2^{nR_{1}},2^{nR_{2}})=1\}.\label{eq:-16}
\end{align}
Observe that any $n$-type $T_{XY}$ can also be viewed as a $kn$-type
for $k\ge1$. With an abuse of notation, we continue to use $T_{XY}$
to denote the corresponding $kn$-type. With this in mind, for an
$n$-type $T_{XY}$ define the asymptotic exponent of maximal density
for a pair $(R_{1},R_{2})\in\mathcal{R}_{X}^{(n)}\times\mathcal{R}_{Y}^{(n)}$
as\footnote{The limit exists because $\log\Gamma_{kn}(2^{knR_{1}},2^{knR_{2}})$
is subadditive in $k$ for a given $n$-type $T_{XY}$. Further, given
$T_{XY}$, the limit does not depend on the value of $n$ that we
attribute to $T_{XY}$. } 
\begin{equation}
E(R_{1},R_{2}):=\lim_{k\to\infty}-\frac{1}{kn}\log\Gamma_{kn}(2^{knR_{1}},2^{knR_{2}}),\label{eq:-1-1}
\end{equation}
and the asymptotic biclique rate region as\footnote{
Using a product construction, we see that $k\mathcal{R}_{kn}(T_{XY})$
is ``superadditive'' in $k$, i.e., $k_{1}\mathcal{R}_{k_{1}n}(T_{XY})+k_{2}\mathcal{R}_{k_{2}n}(T_{XY})\subseteq(k_{1}+k_{2})\mathcal{R}_{(k_{1}+k_{2})n}(T_{XY})$.
Hence $\mathcal{R}(T_{XY})=\mathrm{closure}\lim_{k\to\infty}\mathcal{R}_{kn}(T_{XY})$
and, moreover, $\mathcal{R}(T_{XY})$ is only dependent on $T_{XY}$
and independent 
of the value of $n$ that we attribute to $T_{XY}$. } 
\begin{equation}
\mathcal{R}(T_{XY}):=\mathrm{closure}\bigcup_{k\ge1}\mathcal{R}_{kn}(T_{XY}).\label{eq:asympbiclique}
\end{equation}
The blocklengths considered here are taken as multiples of $n$, since
the limit and union above are taken for fixed $T_{XY}$, but $T_{XY}$
is not always an $m$-type for an arbitrary integer $m$.

In this paper we are interested in characterizing the limits $E(R_{1},R_{2})$
and $\mathcal{R}(T_{XY})$, and in bounding the corresponding convergence
rates.

\subsection{Motivations}

Our motivations for studying the type graph have the following three
aspects. 
\begin{enumerate}
\item The method of types is a classic and powerful tool in information
theory. In this method, the basic unit is the (joint) type or (joint)
type class. To the authors' knowledge, it is not well understood how
the sequence pairs are distributed in a joint type class. The maximal
density (as well as the biclique rate region) measures how concentrated
are the joint-type sequence pairs by counting the number of joint-type
sequence pairs in each ``local'' rectangular subset. Hence, our
study of the type graph deepens the understanding of the distribution
(or structure) of sequence pairs in a joint type class. The first
study on this topic can be traced back to Han and Kobayashi's work
\cite{han1985maximal}, and it was also investigated in \cite{krithivasan2007large,nazari2009new,nazari2010typicality}
recently. In all these works, either a typicality graph (an approximate
version of the type graph) or an approximate version of a biclique
of the type graph was considered. In contrast, we consider the exact
version of a biclique of the type graph, which results in a rate region
different from theirs. 
\item Observe that if one starts with a pair sequence $(\mathbf{x},\mathbf{y})$
in the joint type class $\mathcal{T}_{T_{XY}}$, then the type graph
can be constructed from the set of all pair sequences resulting from
permutations of this pair sequence. Thus, unlike other well-studied
large graphs, the type graph is deterministic rather than stochastic.
There are relatively few works focusing on deterministic large graphs.
Hence, as a purely combinatorial problem, studying the type graph
is of independent interest. 
\item The maximal and minimal density problems for type graphs are closely
related to noninteractive simulation problems (or noise stability
problems) and hypercontractivity inequalities. Hence, studying the
type graph could provide more insights to these related topics. 
\end{enumerate}

\subsection{Related Works}

\label{subsec:related} Han and Kobayashi \cite{han1985maximal} introduced
a concept similar to the asymptotic biclique rate region defined in
this paper. However, roughly speaking, their definition is an approximate
version of our definition, in the sense that in their definition,
for a distribution $P_{XY}$ (not necessarily a type), type classes
are replaced with typical sets with respect to $P_{XY}$, and the
constraint $\Gamma_{n}(2^{nR_{1}},2^{nR_{2}})=1$ is replaced with
$\Gamma_{n}(2^{nR_{1}^{(n)}},2^{nR_{2}^{(n)}})\rightarrow1$ as $n\to\infty$
for a sequence of types $T_{XY}^{(n)}$ converging to $P_{XY}$ and
a sequence of pairs $(R_{1}^{(n)},R_{2}^{(n)})$ converging to $(R_{1},R_{2})$.
This approximate version was also investigated in \cite{krithivasan2007large,nazari2009new,nazari2010typicality}.

In fact, the maximal and minimal density problems on a type graph
are equivalent to the noninteractive simulation problem in some sense.
Given a joint distribution $P_{XY}^{n}$, the noninteractive simulation
problem concerns  estimating the maximal and minimal joint probability
$P_{XY}^{n}(\calA\times\calB)$ when the marginal probabilities $P_{X}^{n}(\calA)$
and $P_{Y}^{n}(\calB)$ are given. The study of the noninteractive
simulation problem dates back to G\'acs and K\"orner's and Witsenhausen's
seminal papers \cite{gacs1973common,witsenhausen1975sequences}. Most
of the existing works on this topic focus on doubly symmetric binary
sources (DSBSes). For the DSBS, by utilizing the tensorization property
of maximal correlation, Witsenhausen proved sharp bounds on $P_{XY}^{n}(\calA\times\calB)$
for the case $P_{X}^{n}(\calA)=P_{Y}^{n}(\calB)=\frac{1}{2}$, where
the upper and lower bounds are respectively attained by symmetric
$(n-1)$-subcubes (e.g., $\calA=\calB=\{\mathbf{x}:x_{1}=1\}$) and
anti-symmetric $(n-1)$-subcubes (e.g., $\calA=-\calB=\{\mathbf{x}:x_{1}=1\}$).
Recently, by combining Fourier analysis with a coding-theoretic result,
the first author and Tan \cite{yu2019on} derived the sharp upper
bound for the case $P_{X}^{n}(\calA)=P_{Y}^{n}(\calB)=\frac{1}{4}$,
where the upper bound is attained by symmetric $(n-2)$-subcubes (e.g.,
$\calA=\calB=\{\mathbf{x}:x_{1}=x_{2}=1\}$). Kahn, Kalai, and Linial
\cite{kahn1988influence} first applied the single-function version
of (forward) hypercontractivity inequalities to obtain bounds for
the noninteractive simulation problem, by replacing nonnegative functions
in the hypercontractivity inequalities with Boolean functions. Mossel
and O'Donnell \cite{mossel2006non,O'Donnell14analysisof} applied
the two-function version of hypercontractivity inequalities to obtain
bounds in a similar way. Kamath and the second author \cite{kamath2016non}
improved the use of hypercontractivity inequalities in a slightly
different way, specifically by replacing nonnegative functions with
two-valued functions (not restricted to be $\{0,1\}$-valued). Furthermore,
as mentioned previously, Ordentlich, Polyanskiy, and Shayevitz \cite{ordentlich2020note}
studied the regime in which $P_{X}^{n}(\calA_{n}),P_{Y}^{n}(\calB_{n})$
vanish exponentially fast, and they solved the limiting cases $\rho\to0,1$.
The symmetric case $P_{X}^{n}(\calA_{n})=P_{Y}^{n}(\calB_{n})$ in
this exponential regime was solved by Kirshner and Samorodnitsky \cite{kirshner2021moment}.
Furthermore, the noninteractive simulation problem for Gaussian sources
was investigated in \cite{borell1985geometric,mossel2015robust},
and the ones with Markov chain noise models and multi-terminal versions
of noninteractive simulation problems have also been studied in the
literature; e.g., \cite{mossel2006non,mossel2005coin}. We refer readers
to the monograph \cite{yu2021common} for a comprehensive introduction
to this topic.

 Brascamp--Lieb (BL) inequalities constitute a class of inequalities
that generalize the families of H\"older inequalities.  Hypercontractivity
inequalities are special cases of BL inequalities. Hypercontractivity
inequalities were investigated in \cite{bonami1968ensembles,kiener1969uber,schreiber1969fermeture,bonami1970etude,beckner1975inequalities,gross1975logarithmic,ahlswede1976spreading,borell1982positivity,bakry1994hypercontractivite,mossel2013reverse}
among others. Information-theoretic characterizations of the BL (and
hypercontractivity) inequalities can be traced back to Ahlswede and
G\'acs's seminal work \cite{ahlswede1976spreading}, where a related
quantity, known as the hypercontractivity constant, was expressed
in terms of relative entropies. The information-theoretic characterization
for the forward BL inequalities on Euclidean spaces was given in \cite{carlen2009subadditivity};
this was independently discovered later \cite{nair2014equivalent}
in the case of finite alphabets. An information-theoretic characterization
of the reverse BL inequalities for finite alphabets was provided in
\cite{kamath2015reverse,beigi2016equivalent,liu2016brascamp}. By
using Fenchel duality, the extension of the characterization for
forward inequalities to arbitrary measurable spaces and the extension
of the characterization for reverse inequalities to Polish spaces
under certain compactness conditions were done in \cite{liu2018forward}.
These compactness conditions were removed in \cite{yu2021strong}
by using large deviations theory.  

\subsection{Main Contributions}

Out main contribution in this paper is the complete characterization
of the asymptotic biclique rate region for any joint type defined
on finite alphabets. We observe that, in general, the asymptotic biclique
rate region defined by us is a subset (in general, a strict subset)
of the approximate one defined by Han and Kobayashi \cite{han1985maximal}.
In fact, their definition for a distribution $P_{XY}$ is equal to
the asymptotic rate region of a sequence of $n$-types $\{T_{XY}^{(n)}\}$
approaching $P_{XY}$, which satisfy the condition $E_{n}(R_{1}^{(n)},R_{2}^{(n)})\to0$
as $n\to\infty$. Our proof for the characterization of biclique rate
region combines information-theoretic techniques and linear algebra;
similar techniques were also used in \cite{lovasz1979shannon,liu2022minoration}.

We also characterize the asymptotic exponent of maximal density, and
interpret it in terms of noninteractive simulation, for which the
marginal probabilities are exponentially small. Note that this regime
was first explicitly studied by Ordentlich, Polyanskiy, and Shayevitz
\cite{ordentlich2020note}, who solved limiting cases for DSBSes.
In fact, a complete characterization (involving time-sharing random
variables) of this problem exists in the literature, which is a direct
consequence of the existing information-theoretic characterization
of Brascamp--Lieb inequalities. Applying this result to zero-error
coding for the binary adder channel yields a new bound on the zero-error
capacity.

Finally, we relax Boolean functions in noninteractive simulation problems
to any nonnegative functions, but still restrict their suppports to
be exponentially small. We obtain stronger (forward and reverse) Brascamp--Lieb
and hypercontractivity inequalities, which, in asymptotic cases, reduce
to the common ones when the exponents of the sizes of the supports
are zero. (Note that these stronger inequalities can be also derived
from the existing information-theoretic characterization of the classic
Brascamp--Lieb inequalities.) Similar inequalities were previously
derived by Polyanskiy--Samorodnitsky \cite{polyanskiy2019improved}
and by Kirshner--Samorodnitsky \cite{kirshner2021moment} by different
methods.

\subsection{Notation}

\label{subsec:notation}

We write $:=$ and occasionally $=:$ for equality by definition.
Throughout this paper, for two sequences of reals, we use $a_{n}\doteq b_{n}$
to denote $a_{n}=b_{n}2^{o(n)}$. We use $\calC(Q_{X},Q_{Y})$ to
denote the set of couplings $Q_{XY}$ with marginals $Q_{X},Q_{Y}$.
Given $Q_{X|UW}$ and $Q_{Y|VW}$, we use $\calC(Q_{X|UW},Q_{Y|VW})$
to denote the set of conditional couplings $Q_{XY|UVW}$ with conditional
marginals $Q_{X|UW},Q_{Y|VW}$. Note that, given $Q_{X|UW}$ and $Q_{Y|VW}$,
for any $Q_{XY|UVW}\in\calC(Q_{X|UW},Q_{Y|VW})$ and any $Q_{UVW}$,
the joint law $Q_{XYUVW}=Q_{XY|UVW}Q_{UVW}$ is such that $X\leftrightarrow(U,W)\leftrightarrow V$
and $Y\leftrightarrow(V,W)\leftrightarrow U$, where the notation
$X\leftrightarrow Y\leftrightarrow Z$ for a triple of random variables
$(X,Y,Z)$ denotes that $X$ and $Z$ are conditionally independent
given $Y$. For a length-$n$ sequence $\mathbf{x}$, we use $T_{\mathbf{x}}$
to denote the type of $\mathbf{x}$. For an $m\times n$ matrix $\mathbf{B}=(b_{i,j})$
and two subsets $\mathcal{H}\subseteq[m],\mathcal{L}\subseteq[n]$,
we use $\mathbf{B}_{\mathcal{H},\mathcal{L}}$ to denote $(b_{i,j})_{i\in\mathcal{H},j\in\mathcal{L}}$,
i.e., the submatrix of $\mathbf{B}$ consisting of the elements with
indices in $\mathcal{H}\times\mathcal{L}$. For a length-$n$ vector
or sequence $\mathbf{x}$ and a subset $\mathcal{J}\subseteq[n]$,
$\mathbf{x}_{\mathcal{J}}:=(x_{j})_{j\in\mathcal{J}}$ is defined
similarly. For a distribution $P_{X}$, we use $P_{X}^{n}$ to denote
the $n$-fold product of $P_{X}$. We will also use notations $H_{Q}(X)$
or $H(Q_{X})$ to denote the entropy of $X\sim Q_{X}$. If the distribution
is denoted by $P_{X}$, we sometimes write the entropy as $H(X)$
for brevity. We use $\supp(P_{X})$ to denote the support of $P_{X}$.
The logarithm $\log$ is taken to the base $2$, and $\ln$ is taken
to the natural base. Note that, as is the case for many other information-theoretic
results, the results in this paper can be viewed as independent of
the choice of the base of the logarithm as long as exponentiation
is interpreted as being with respect to the same base.

For a joint distribution $P_{XY}$ and for functions $f:\mathcal{X}\to[0,\infty)$
and $g:\mathcal{Y}\to[0,\infty)$, define their {\em inner product}
\begin{align}
\langle f,g\rangle & :=\mathbb{E}[f(X)g(Y)]=\sum_{(x,y)\in\mathcal{X}\times\mathcal{Y}}P_{XY}(x,y)f(x)g(y).\label{eqn:inner_prod}
\end{align}
The {\em $L^{p}$-norm} of $f$ for $p\in[1,\infty)$ and the {\em
pseudo $L^{p}$-norm} of $f$ for $p\in(0,1)$ are defined as 
\begin{align}
\Vert f\Vert_{p} & :=\big(\mathbb{E}[f(X)^{p}]\big)^{1/p}=\bigg(\sum_{x\in\mathcal{X}}P_{X}(x)f(x)^{p}\bigg)^{1/p}.\label{eqn:Lp}
\end{align}

\section{Type Graphs }

In this section, we completely characterize the asymptotic exponent
of maximal density and the asymptotic biclique rate region.

\subsection{Exponents}

The asymptotic behavior of the exponent of maximal density is characterized
in the following theorem, whose proof is provided in Appendix \ref{sec:Proof-of-Theorem-exponent}.
For all nonnegative pairs $(R_{1},R_{2})$, define 
\begin{equation}
F^{*}(R_{1},R_{2}):=\max_{\substack{P_{XYW}:P_{XY}=T_{XY},\\
H(X|W)\leq R_{1},H(Y|W)\leq R_{2}
}
}H(X,Y|W),\label{eq:F}
\end{equation}
and 
\begin{equation}
E^{*}(R_{1},R_{2}):=R_{1}+R_{2}-F^{*}(R_{1},R_{2}).\label{eq:E*}
\end{equation}

\begin{thm}
\label{thm:exponentoftypegraph} Given a joint $n$-type $T_{XY}$
with $n\ge2(|\mathcal{X}||\mathcal{Y}|+2)|\mathcal{X}||\mathcal{Y}|$,
for $(R_{1},R_{2})\in\mathcal{R}_{X}^{(n)}\times\mathcal{R}_{Y}^{(n)}$,
we have 
\begin{equation}
E^{*}(R_{1},R_{2})\le E_{n}(R_{1},R_{2})\leq E^{*}(R_{1},R_{2})+\varepsilon_{n},\label{eq:-10}
\end{equation}
where $\varepsilon_{n}:=\frac{(|\mathcal{X}||\mathcal{Y}|+2)|\mathcal{X}||\mathcal{Y}|}{n}\log\frac{(n+1)n^{6}}{|\mathcal{X}|^{4}|\mathcal{Y}|^{4}}$.
As a consequence, for any $n\ge1$ and any joint $n$-type $T_{XY}$,
we have 
\begin{equation}
E(R_{1},R_{2})=E^{*}(R_{1},R_{2}).\label{eq:asympexp}
\end{equation}
Without loss of optimality, the alphabet size of $W$ in the definition
of $F^{*}(R_{1},R_{2})$ can be assumed to be no larger than $|\mathcal{X}||\mathcal{Y}|+2$. 
\end{thm}
\begin{rem}
$E^{*}(R_{1},R_{2})$ can be also expressed as 
\begin{align*}
E^{*}(R_{1},R_{2}) & =R_{1}+R_{2}-H_{T}(XY)+G^{*}(R_{1},R_{2}),
\end{align*}
with 
\begin{equation}
G^{*}(R_{1},R_{2}):=\min_{\substack{P_{XYW}:P_{XY}=T_{XY},\\
H(X|W)\leq R_{1},H(Y|W)\leq R_{2}
}
}I(X,Y;W)\label{eq:G}
\end{equation}
corresponding to the minimum common rate given marginal rates $(R_{1},R_{2})$
in the Gray--Wyner source coding network \cite[Theorem 14.3]{Gamal}. 
\end{rem}
\begin{rem}
The explicit expression of $E^{*}$ for the doubly symmetric binary
source was given in Section \ref{subsec:Example:-DSBS}. 
\end{rem}
\begin{rem}
A slightly weaker statement, $E_{n}(R_{1},R_{2})=E^{*}(R_{1},R_{2})+O(\frac{\log n}{n})$,
can be recovered from a more general result given in \cite[(6) and (7)]{liu2019smoothing}
via the noninteractive simulation interpretation of the maximal density
problem; see Section \ref{subsec:Sources}. 
\end{rem}
Before proving Theorem \ref{thm:exponentoftypegraph}, we first list
several properties of $F^{*}(R_{1},R_{2})$ in the following lemma.
The proof is provided in Appendix \ref{sec:Proof-of-Lemma-F}. 
\begin{lem}
\label{lem:F}For any joint $n$-type $T_{XY}$ and $R_{1},R_{2}\ge0$,
the following properties of $F^{*}(R_{1},R_{2})$ hold. \\
 1) Given $R_{1}$, $F^{*}(R_{1},R_{2})$ is nondecreasing in $R_{2}$
and, given $R_{2}$, $F^{*}(R_{1},R_{2})$ is nondecreasing in $R_{1}$.
\\
 2) $F^{*}(R_{1},R_{2})\le\min\{H_{T}(X,Y),R_{1}+R_{2},R_{1}+H_{T}(Y|X),R_{2}+H_{T}(X|Y)\}$.
Moreover, $F^{*}(H_{T}(X),H_{T}(Y))=H_{T}(X,Y)$. \\
 3) $F^{*}(0,R_{2})=\min\{R_{2},H_{T}(Y|X)\}$ and, similarly, $F^{*}(R_{1},0)=\min\{R_{1},H_{T}(X|Y)\}$.
\\
 4) $F^{*}(R_{1},R_{2})$ is concave in $(R_{1},R_{2})$ on $\{(R_{1},R_{2}):R_{1}\ge0,R_{2}\ge0\}$.
\\
 5) For $\delta_{1},\delta_{2}\ge0$, we have $0\le F^{*}(R_{1}+\delta_{1},R_{2}+\delta_{2})-F^{*}(R_{1},R_{2})\leq\delta_{1}+\delta_{2}$
for all $R_{1}\ge0$, $R_{2}\ge0$. 
\end{lem}
Theorem \ref{thm:exponentoftypegraph} is an edge-isoperimetric result
for the bipartite graph induced by a joint $n$-type $T_{XY}$. For
the case in which $\mathcal{X}=\mathcal{Y}$ and $T_{X}=T_{Y}$, the
bipartite graph of $T_{XY}$ can be replaced by a non-bipartite one.
Consider a directed graph\footnote{ When we extend the bipartite graph to a non-bipartite one, we assume
the graph to be directed, in order to ensure that the pairs of sequences
$(\mathbf{x},\mathbf{y})$ and the edges in the graph are mapped to
each other in a one-to-one way. } (allowing self-loops if $X=Y$ under $T_{XY}$) in which the vertices
consist of $\mathbf{x}\in\mathcal{T}_{T_{X}}$ and there is a directed
edge from\footnote{Without of loss generality, we consider the edges from $\mathbf{x}$
to $\mathbf{y}$, since we can obtain a graph with edges from $\mathbf{y}$
to $\mathbf{x}$ if we consider the type $T_{YX}$ (instead of $T_{XY}$). } $\mathbf{x}$ to $\mathbf{y}$ if and only if $(\mathbf{x},\mathbf{y})\in\mathcal{T}_{T_{XY}}$.
Hence, for this case, Theorem \ref{thm:exponentoftypegraph} can be
also considered as an edge-isoperimetric result for a directed graph
induced by $T_{XY}$. Specifically, for a subset $\calA\subseteq\mathcal{T}_{T_{X}}$,
let $G[\calA]$ be the induced subgraph of the directed graph of $T_{XY}$.
The (edge) density $\rho(G[\calA])$ is defined as 
\begin{align*}
\rho(G[\calA]) & :=\frac{\#\textrm{ of directed edges in }G[\calA]}{|\calA|^{2}}\\
 & =\frac{|(\calA\times\calA)\cap\mathcal{T}_{T_{XY}}|}{|\calA|^{2}}.
\end{align*}
Given $1\le M\le|\mathcal{T}_{T_{X}}|$, define the maximal density
of subgraphs with size $M$ as\footnote{We use the same notation as the one in (\ref{eq:Gamma_overline})
for the bipartite graph case, but here 
the edge density has only one parameter. The 
difference between these two maximal densities is that in (\ref{eq:Gamma_overline})
the maximization is taken over a pair of sets $(\calA,\calB)$, but
here only over one set (equivalently, under the restriction $\calA=\calB$).} 
\begin{align*}
\Gamma_{n}(M) & :=\max_{\calA\subseteq\mathcal{T}_{T_{X}}:|\calA|=M}\rho(G[\calA]).
\end{align*}
Given a joint $n$-type $T_{XY}$, for $R\in\mathcal{R}_{X}^{(n)}$
as defined in \eqref{eq:rates}, define the exponent of maximal density
as 
\begin{equation}
E_{n}(R):=-\frac{1}{n}\log\Gamma_{n}(2^{nR}).\label{eq:-1-3}
\end{equation}
For any subsets $\calA,\calB$ of $\mathcal{X}^{n}$, we have 
\[
|\calA||\calB|\rho(G[\calA,\calB])\le|\calA\cup\calB|^{2}\rho(G[\calA\cup\calB]).
\]
On the other hand, 
\[
\Gamma_{n}(M)\le\Gamma_{n}(M,M).
\]
Hence 
\[
\frac{1}{4}\Gamma_{n}(\frac{M}{2},\frac{M}{2})\le\Gamma_{n}(M)\le\Gamma_{n}(M,M).
\]
Combining the inequalities above with Theorem \ref{thm:exponentoftypegraph}
yields the following result. 
\begin{cor}
\label{thm:exponentoftypegraph-2} For any joint $n$-type $T_{XY}$,
and $R\in\mathcal{R}_{X}^{(n)}$, we have 
\begin{equation}
E_{n}(R)=E^{*}(R,R)+O(\frac{\log n}{n}),\label{eq:-10-5}
\end{equation}
where the asymptotic constant in the $O(\frac{\log n}{n})$ term on
the right hand side depends only on $|\mathcal{X}|$, and $E^{*}(R_{1},R_{2})$
is defined in Theorem \ref{thm:exponentoftypegraph}. 
\end{cor}
For the case of $\mathcal{X}=\mathcal{Y}$ and $T_{X}=T_{Y}$, the
bipartite graph of $T_{XY}$ can be also considered as an undirected
graph (allowing self-loops if $X=Y$ under $T_{XY}$) in which the
vertices consist of $\mathbf{x}\in\mathcal{T}_{T_{X}}$ and $(\mathbf{x},\mathbf{y})$
is an edge if and only if $(\mathbf{x},\mathbf{y})$ or $(\mathbf{y},\mathbf{x})\in\mathcal{T}_{T_{XY}}$.
By a similar argument to the above, Corollary \ref{thm:exponentoftypegraph-2}
still holds for this case, which hence can be considered as a generalization
of \cite[Theorem 1.6]{kirshner2021moment} from binary alphabets to
arbitrary finite alphabets.

\subsection{Biclique Rate Region}

The asymptotic behavior of the biclique rate region is characterized
in the following theorem, whose proof is provided in Appendix \ref{sec:Proof-of-Theorem-exponent-1}.
Define 
\begin{align}
\mathcal{R}^{*}(T_{XY}) & :=\bigcup_{\substack{0\le\alpha\le1,P_{XY},Q_{XY}:\\
\alpha P_{XY}+(1-\alpha)Q_{XY}=T_{XY}
}
}\{(R_{1},R_{2}):\nonumber \\
 & \qquad\qquad R_{1}\le\alpha H_{P}(X|Y),\nonumber \\
 & \qquad\qquad R_{2}\le(1-\alpha)H_{Q}(Y|X)\}.\label{eq:Ri-1}
\end{align}

\begin{thm}
\label{thm:bicliquerateregion} For any $n\ge8(|\mathcal{X}||\mathcal{Y}|)^{7/5}$
and any $T_{XY}$, 
\begin{align}
 & (\mathcal{R}^{*}(T_{XY})-[0,\varepsilon_{1,n}]\times[0,\varepsilon_{2,n}])\cap(\mathcal{R}_{X}^{(n)}\times\mathcal{R}_{Y}^{(n)})\nonumber \\
 & \subseteq\mathcal{R}_{n}(T_{XY})\nonumber \\
 & \subseteq\mathcal{R}^{*}(T_{XY})\cap(\mathcal{R}_{X}^{(n)}\times\mathcal{R}_{Y}^{(n)})\label{eq:bicliquerateregion}
\end{align}
where $\mathcal{R}_{n}(T_{XY})$ is defined in \eqref{eq:-16}, ``$-$''
is the Minkowski difference (i.e., for $\calA,\calB\subseteq\mathbb{R}^{m}$,
$\calA-\calB:=\bigcap_{b\in\calB}(\calA-b)$), $\varepsilon_{1,n}:=\frac{|\mathcal{X}||\mathcal{Y}|}{n}\log\frac{n^{4}(n+1)}{16|\mathcal{X}|}$,
and $\varepsilon_{2,n}:=\frac{|\mathcal{X}||\mathcal{Y}|}{n}\log\frac{n^{4}(n+1)}{16|\mathcal{Y}|^{2}}$.
In particular, 
\begin{equation}
\mathcal{R}(T_{XY})=\mathcal{R}^{*}(T_{XY}),\label{eq:-3}
\end{equation}
where $\mathcal{R}(T_{XY})$ is the asymptotic biclique rate region,
defined in \eqref{eq:asympbiclique}. 
\end{thm}
\begin{rem}
Theorem \ref{thm:bicliquerateregion} can be easily generalized to
the $k$-variables case with $k\ge3$. For this case, let $T_{X_{1},...,X_{k}}$
be a joint $n$-type. Then the graph $G$ induced by $T_{X_{1},...,X_{k}}$
is in fact a $k$-partite hypergraph. The (edge) density of the subgraph
of $G$ with vertex sets $(\calA_{1},...,\calA_{k})$ is defined as
\[
\rho(G[\calA_{1},...,\calA_{k}]):=\frac{|(\prod_{i=1}^{k}\calA_{i})\cap\mathcal{T}_{T_{X_{1},...,X_{k}}}|}{\prod_{i=1}^{k}|\calA_{i}|}.
\]
It is interesting to observe that $\rho(G)\doteq2^{-nI_{T^{(n)}}(X_{1};...;X_{k})}$
for a sequence of joint types $\{T_{X_{1},...,X_{k}}^{(n)}\}$, where
$I_{T^{(n)}}(X_{1};...;X_{k}):=\sum_{i=1}^{k}H_{T^{(n)}}(X_{i})-H_{T^{(n)}}(X_{1},...,X_{k})$.
Given a joint $n$-type $T_{X_{1},...,X_{k}}$, we define the $k$-clique
rate region as 
\begin{align*}
\mathcal{R}_{n}(T_{X_{1},...,X_{k}}) & :=\{(\frac{1}{n}\log|\calA_{1}|,...,\frac{1}{n}\log|\calA_{k}|):\\
 & \qquad\qquad\rho(G[\calA_{1},...,\calA_{k}])=1\}.
\end{align*}
Following similar steps to our proof of Theorem \ref{thm:bicliquerateregion},
for this case we have 
\begin{align*}
 & (\mathcal{R}^{*}(T_{X_{1},...,X_{k}})-[0,O(\frac{\log n}{n})]^{k})\cap(\prod_{i=1}^{k}\mathcal{R}_{X_{i}}^{(n)})\\
 & \subseteq\mathcal{R}_{n}(T_{X_{1},...,X_{k}})\\
 & \subseteq\mathcal{R}^{*}(T_{X_{1},...,X_{k}})\cap(\prod_{i=1}^{k}\mathcal{R}_{X_{i}}^{(n)}),
\end{align*}
where 
\begin{align*}
\mathcal{R}^{*}(T_{X_{1},...,X_{k}}) & :=\bigcup_{\substack{\alpha_{i}\ge0,P_{X_{1},...,X_{k}}^{(i)},i\in[k]:\sum_{i=1}^{k}\alpha_{i}=1\\
\sum_{i=1}^{k}\alpha_{i}P_{X_{1},...,X_{k}}^{(i)}=T_{X_{1},...,X_{k}}
}
}\\
 & \qquad\{(R_{1},...,R_{k}):R_{i}\le\alpha_{i}H_{P^{(i)}}(X_{i}|X^{\backslash i})\},
\end{align*}
with $X^{\backslash i}:=(X_{1},...,X_{i-1},X_{i+1},...,X_{k})$. 
\end{rem}
\begin{prop}
\label{prop:convex}Given $T_{XY}$, $\mathcal{R}^{*}(T_{XY})$ is
a closed convex set. 
\end{prop}
\begin{IEEEproof}
Using the continuity of $H_{P}(X|Y)$ in $P_{XY}$ it can be established
that $\mathcal{R}^{*}(T_{XY})$ is closed. Convexity follows by the
following argument. For any $(R_{1},R_{2}),(\hat{R}_{1},\hat{R}_{2})\in\mathcal{R}^{*}(T_{XY})$,
there exist $(\alpha,P_{XY},Q_{XY})$ and $(\hat{\alpha},\hat{P}_{XY},\hat{Q}_{XY})$
such that 
\begin{align*}
 & \alpha P_{XY}+(1-\alpha)Q_{XY}=T_{XY},\\
 & \hat{\alpha}\hat{P}_{XY}+(1-\hat{\alpha})\hat{Q}_{XY}=T_{XY},\\
R_{1} & \le\alpha H_{P}(X|Y),R_{2}\le(1-\alpha)H_{Q}(Y|X),\\
\hat{R}_{1} & \le\hat{\alpha}H_{\hat{P}}(X|Y),\hat{R}_{2}\le(1-\hat{\alpha})H_{\hat{Q}}(Y|X).
\end{align*}
Then for any $\lambda\in[0,1]$, 
\begin{align}
 & \lambda R_{1}+(1-\lambda)\hat{R}_{1}\nonumber \\
 & \le\lambda\alpha H_{P}(X|Y)+(1-\lambda)\hat{\alpha}H_{\hat{P}}(X|Y)\\
 & \le\beta H_{P^{(\theta)}}(X|Y),\label{eq:-50}
\end{align}
where $\beta=\lambda\alpha+(1-\lambda)\hat{\alpha}$, and $P_{XY}^{(\theta)}:=\theta P_{XY}+(1-\theta)\hat{P}_{XY}$
with $\theta=\frac{\lambda\alpha}{\lambda\alpha+(1-\lambda)\hat{\alpha}}$
if $\beta>0$; $P_{XY}^{(\theta)}$ is chosen as an arbitrary distribution
if $\beta=0$. Here (\ref{eq:-50}) follows since $H_{P}(X|Y)$ is
concave in $P_{XY}$. By symmetry, $\lambda R_{2}+(1-\lambda)\hat{R}_{2}\le(1-\beta)H_{Q^{(\hat{\theta})}}(Y|X)$,
where $Q_{XY}^{(\hat{\theta})}:=\hat{\theta}P_{XY}+(1-\hat{\theta})\hat{P}_{XY}$
with $\hat{\theta}=\frac{\lambda(1-\alpha)}{\lambda(1-\alpha)+(1-\lambda)(1-\hat{\alpha})}$
if $\beta<1$; $Q_{XY}^{(\hat{\theta})}$ is chosen as an arbitrary
distribution if $\beta=1$. Since $\beta P_{XY}^{(\theta)}+(1-\beta)Q_{XY}^{(\hat{\theta})}=T_{XY}$,
it follows that $\lambda(R_{1},R_{2})+(1-\lambda)(\hat{R}_{1},\hat{R}_{2})\in\mathcal{R}^{*}(T_{XY})$,
i.e., $\mathcal{R}^{*}(T_{XY})$ is convex. 
\end{IEEEproof}
Since $\mathcal{R}^{*}(T_{XY})$ is convex, an extreme case of $\mathcal{R}^{*}(T_{XY})$
is a triangle region. We next study when the asymptotic biclique rate
region is a triangle region. We obtain the following necessary and
sufficient condition. The proof is provided in Appendix \ref{sec:Proof-of-Proposition}. 
\begin{prop}
\label{prop:line} Let $T_{XY}$ be a joint $n$-type such that $H_{T}(X|Y),H_{T}(Y|X)>0$.
Then the asymptotic biclique rate region $\mathcal{R}(T_{XY})$ is
a triangle region, i.e., 
\begin{align*}
\mathcal{R}(T_{XY}) & =\mathcal{R}_{\bigtriangleup}(T_{XY})\\
 & :=\bigcup_{0\le\alpha\le1}\{(R_{1},R_{2}):R_{1}\le\alpha H_{T}(X|Y),\\
 & \qquad\qquad R_{2}\le(1-\alpha)H_{T}(Y|X)\},
\end{align*}
if and only if $T_{XY}$ satisfies that $T_{X|Y}(x|y)^{1/H_{T}(X|Y)}=T_{Y|X}(y|x)^{1/H_{T}(Y|X)}$
for all $x,y$. 
\end{prop}
The condition in Proposition \ref{prop:line} is satisfied by the
joint $n$-types $T_{XY}$ which have marginals $T_{X}=\mathrm{Unif}(\mathcal{X}),T_{Y}=\mathrm{Unif}(\mathcal{Y})$
and satisfy at least one of the following two conditions:\\
 1) $|\mathcal{X}|=|\mathcal{Y}|$;\\
 2) $X,Y$ are independent under the distribution $T_{XY}$.\\

\noindent \textbf{Example (DSBS):} A typical example that satisfies
these conditions is the DSBS, whose distribution is given in Table
\ref{tab:The-distribution-of}. Hence, the asymptotic biclique rate
region is the triangle region $\{(R_{1},R_{2}):R_{1}+R_{2}\le h(\frac{1-\rho}{2})\}$
if the joint $n$-type $T_{XY}$ is a DSBS with correlation coefficient
$\rho\in[0,1]$. Here, $h:t\mapsto-t\log t-(1-t)\log(1-t)$ denotes
the binary entropy function.


\begin{table}
\begin{centering}
\begin{tabular}{|p{1cm}|p{1cm}|p{1cm}|}
\hline 
\[
X\backslash Y
\]
 & 
\[
0
\]
 & 
\[
1
\]
\tabularnewline
\hline 
\[
0
\]
 & 
\[
\frac{1+\rho}{4}
\]
 & 
\[
\frac{1-\rho}{4}
\]
\tabularnewline
\hline 
\[
1
\]
 & 
\[
\frac{1-\rho}{4}
\]
 & 
\[
\frac{1+\rho}{4}
\]
\tabularnewline
\hline 
\end{tabular}
\par\end{centering}
\caption{\label{tab:The-distribution-of}The distribution of a DSBS with correlation
coefficient $\rho$.}
\end{table}

For a joint type $T_{XY}$ the Han and Kobayashi region (which is
of course defined for any joint distribution, not necessarily a type)
is given by \cite{han1985maximal} 
\begin{align}
\mathcal{R}^{**}(T_{XY}) & :=\bigcup_{\substack{P_{XYW}:P_{XY}=T_{XY},\\
X\leftrightarrow W\leftrightarrow Y
}
}\{(R_{1},R_{2}):\nonumber \\
 & \qquad\qquad R_{1}\le H(X|W),R_{2}\le H(Y|W)\}.\label{eq:HKregion}
\end{align}
By Theorem \ref{thm:exponentoftypegraph}, $\mathcal{R}^{**}(T_{XY})$
also coincides with the region $\{(R_{1},R_{2}):E^{*}(R_{1},R_{2})=0\}$.
This implies that $\mathcal{R}^{*}(T_{XY})\subseteq\mathcal{R}^{**}(T_{XY})$,
i.e., that for any joint type $T_{XY}$ the asymptotic biclique rate
region defined in this paper is a subset of Han and Kobayashi's approximate
version. This can also be seen by directly comparing the definition
of $\mathcal{R}^{*}(T_{XY})$ in \eqref{eq:Ri-1} to the definition
of $\mathcal{R}^{**}(T_{XY})$ in \eqref{eq:HKregion}. This can be
seen as follows. For a joint type $T_{XY}$, let $Q_{XY}^{(0)}$,
$Q_{XY}^{(1)}$, and $0\le\alpha\le1$ be such that $\alpha Q_{XY}^{(0)}+(1-\alpha)Q_{XY}^{(1)}=T_{XY}$,
and let $(R_{1},R_{2})$ be a rate pair such that $R_{1}\le\alpha H_{Q^{(0)}}(X|Y)$
and $R_{2}\le(1-\alpha)H_{Q^{(1)}}(Y|X)$. Without loss of generality,
we assume that $\mathcal{X}$ and $\mathcal{Y}$ are disjoint, i.e.,
$\mathcal{X}\cap\mathcal{Y}=\emptyset$, since otherwise, we can respectively
map them to another two sets satisfying this requirement by bijections.
Let $(X,Y,W)$ be a tuple of random variables such that $W$ takes
values in $\mathcal{X}\cup\mathcal{Y}$, and $W\in\mathcal{Y}$ with
probability $\alpha$ and $W\in\mathcal{X}$ with probability $1-\alpha$.
Moreover, under the condition $W\in\mathcal{Y}$, it holds that $W=Y$
and $(X,Y)\sim Q_{XY}^{(0)}$; under the condition $W\in\mathcal{X}$,
it holds that $W=X$ and $(X,Y)\sim Q_{XY}^{(1)}$. It can be checked
that $(X,Y)\sim T_{XY}$ and we have $X\leftrightarrow W\leftrightarrow Y$,
$R_{1}\le H(X|W)$, and $R_{2}\le H(Y|W)$. This inclusion can be
strict. For example, when the joint $n$-type $T_{XY}$ is a DSBS
with a positive crossover probability, the region $\mathcal{R}^{**}(T_{XY})$,
which is computed in \cite[Section 4]{han1985maximal}, strictly contains
the asymptotic biclique region $\mathcal{R}^{*}(T_{XY})$, which,
by Proposition \ref{prop:line}, is a triangle region. Another family
of examples where the asymptotic biclique region is strictly contained
in the region of Han and Kobayashi is when the joint $n$-type $T_{XY}$
is $\mathrm{Unif}(\mathcal{X}\times\mathcal{Y})$. Here $\mathcal{R}^{**}(T_{XY})$
equals the rectangle region $[0,H(X)]\times[0,H(Y)]$, while Proposition
\ref{prop:line} implies that $\mathcal{R}^{*}(T_{XY})$ is a triangle
region.

The difference between the exact and approximate versions of asymptotic
biclique rate regions is caused by the ``type overflow'' effect,
which was crystallized by the first author and Tan in \cite{yu2020exact}.
Let $(R_{1},R_{2})$ be a pair such that $E^{*}(R_{1},R_{2})=0$.
Let $(\calA,\calB)$ be an optimal pair of subsets attaining $E^{*}(R_{1},R_{2})$.
All the sequences in $\calA$ have type $T_{X}$, and all the sequences
in $\calB$ have type $T_{Y}$. However, in general, the joint types
of $(\mathbf{x},\mathbf{y})\in\calA\times\calB$ might ``overflow''
from the target joint type $T_{XY}$. The number of non-overflowed
sequence pairs (i.e., $|(\calA\times\calB)\cap\mathcal{T}_{T_{XY}}|$)
has exponent $R_{1}+R_{2}$, since $E^{*}(R_{1},R_{2})=0$. This means
that not too many sequence pairs have overflowed. However, if type
overflow is forbidden, then we must reduce the rates of $\calA$ and
$\calB$ to satisfy this requirement. This leads to the exact version
of the asymptotic biclique rate region being strictly smaller than
the approximate version. In other words, the exact asymptotic biclique
rate region is more sensitive to the type overflow effect than the
approximate version. A similar conclusion was previously drawn by
the first author and Tan in \cite{yu2020exact} for the common information
problem. Technically speaking, the type overflow effect corresponds
to the fact that optimization over couplings is involved in our expressions.
Intuitively, it is caused by the Markov chain constraints in the problem.
We believe that the type overflow effect usually accompanies problems
involving Markov chains.

\section{Noninteractive Simulation}

In this section, we connect the maximal density problem on type graphs
to the noninteractive simulation (or noise stability) problem. We
focus on two noninteractive simulation problems, one with sources
uniformly distributed over a joint $n$-type and the other with memoryless
sources.

\subsection{\label{subsec:Sources}Sources $\mathrm{Unif}(\mathcal{T}_{T_{XY}})$}

In this subsection, we assume $(\mathbf{X},\mathbf{Y})\sim P_{\mathbf{X},\mathbf{Y}}:=\mathrm{Unif}(\mathcal{T}_{T_{XY}})$.
Given two marginal probabilities $P_{\mathbf{X}}(\calA)$ and $P_{\mathbf{Y}}(\calB)$,
what are the possible maximal and minimal values of the joint probability
$P_{\mathbf{X},\mathbf{Y}}(\calA\times\calB)$? This problem is termed
the noninteractive binary simulation problem{} or the (two-set version
of) noise stability problem.

Define $\mathcal{E}_{X}^{(n)}:=\{\frac{1}{n}\log|\mathcal{T}_{T_{X}}|-R_{1}\::\:R_{1}\in\mathcal{R}_{X}^{(n)}\}$
and $\mathcal{E}_{Y}^{(n)}:=\{\frac{1}{n}\log|\mathcal{T}_{T_{Y}}|-R_{2}\::\:R_{2}\in\mathcal{R}_{Y}^{(n)}\}$,
where $\mathcal{R}_{X}^{(n)}$ and $\mathcal{R}_{Y}^{(n)}$ are defined
in \eqref{eq:rates}. Given a joint $n$-type $T_{XY}$, for $(E_{1},E_{2})\in\mathcal{E}_{X}^{(n)}\times\mathcal{E}_{Y}^{(n)}$,
define the exponents of the maximal and minimal noise stability as
\begin{align}
\underline{\Upsilon}_{n}(E_{1},E_{2}) & :=-\frac{1}{n}\log\max_{\substack{\calA\subseteq\mathcal{T}_{T_{X}},\calB\subseteq\mathcal{T}_{T_{Y}}:\\
P_{\mathbf{X}}(\calA)=2^{-nE_{1}},\\
P_{\mathbf{Y}}(\calB)=2^{-nE_{2}}
}
}P_{\mathbf{X},\mathbf{Y}}(\calA\times\calB),\label{eq:-72g}\\
\overline{\Upsilon}_{n}(E_{1},E_{2}) & :=-\frac{1}{n}\log\min_{\substack{\calA\subseteq\mathcal{T}_{T_{X}},\calB\subseteq\mathcal{T}_{T_{Y}}:\\
P_{\mathbf{X}}(\calA)=2^{-nE_{1}},\\
P_{\mathbf{Y}}(\calB)=2^{-nE_{2}}
}
}P_{\mathbf{X},\mathbf{Y}}(\calA\times\calB).\label{eq:-71g}
\end{align}
The noninteractive binary simulation problem is to determine these
two quantities. This problem originates from G\'acs and K\"orner's and
Witsenhausen's seminal works \cite{gacs1973common,witsenhausen1975sequences}
in the study of the G\'acs--K\"orner--Witsenhausen common information.
This topic has also attracted independent interest from the computer
science community, due to the connection with the analysis of Boolean
functions \cite{ODonnell14analysisof}. We refer readers to the related
works mentioned in Section \ref{subsec:related} or the monograph
\cite{yu2021common} for more information on this problem.

We determine the asymptotic behavior of $\underline{\Upsilon}_{n}$
in the following theorem. However, the asymptotic behavior of $\overline{\Upsilon}_{n}$
is currently unclear; see the discussion in Section \ref{sec:Concluding-Remarks}.
For $0\le s\le H(X),0\le t\le H(Y)$, define 
\begin{align*}
\underline{\Upsilon}^{*}(s,t) & :=\min_{\substack{P_{XYW}:P_{XY}=T_{XY},\\
I(X;W)\ge s,I(Y;W)\ge t
}
}I(XY;W).
\end{align*}

\begin{thm}
\label{thm:NS_Type}For any $T_{XY}$ and $(E_{1},E_{2})\in\mathcal{E}_{X}^{(n)}\times\mathcal{E}_{Y}^{(n)}$,
we have 
\begin{equation}
\underline{\Upsilon}_{n}(E_{1},E_{2})=\underline{\Upsilon}^{*}(E_{1},E_{2})+O(\frac{\log n}{n}),\label{eq:-10-4}
\end{equation}
where the asymptotic constant in the $O(\frac{\log n}{n})$ bound
depends only on $|\mathcal{X}|,|\mathcal{Y}|$. 
\end{thm}
In fact, this result can be recovered from a more general result given
in \cite[(6) and (7)]{liu2019smoothing}. The latter generalizes the
$\underline{\Upsilon}_{n}$ for the uniform distribution over a type
class to the infimum of $\underline{\Upsilon}_{n}$ over all distributions
not too far from a product distribution. 
\begin{IEEEproof}
Observe that 
\begin{align}
P_{\mathbf{X}}(\calA) & =|\calA|\big/|\mathcal{T}_{T_{X}}|,\label{eq:-2}\\
P_{\mathbf{Y}}(\calB) & =|\calB|\big/|\mathcal{T}_{T_{Y}}|,\\
P_{\mathbf{X},\mathbf{Y}}(\calA\times\calB) & =\rho(G[\calA,\calB])|\calA||\calB|\big/|\mathcal{T}_{T_{XY}}|.\label{eq:}
\end{align}
So, Theorem \ref{thm:NS_Type} is implied by Theorem \ref{thm:exponentoftypegraph}. 
\end{IEEEproof}

\subsection{\label{subsec:Noninteractive-Simulation-with-IID}Sources $P_{XY}^{n}$}

In this subsection, we consider the noninteractive simulation problem
with $(\mathbf{X},\mathbf{Y})\sim P_{XY}^{n}$, where $P_{XY}$ is
a joint distribution defined on $\mathcal{X}\times\mathcal{Y}$. We
still assume that $\mathcal{X},\mathcal{Y}$ are finite sets of cardinality
at least $2$ and that $P_{X}(x)>0$, $P_{Y}(y)>0$ for all $(x,y)\in\mathcal{X}\times\mathcal{Y}$,
where $P_{X},P_{Y}$ denote the marginal distributions of $P_{XY}$.
Ordentlich, Polyanskiy, and Shayevitz \cite{ordentlich2020note} focused
on binary symmetric distributions $P_{XY}$, and studied the exponent
of $P_{XY}^{n}(\calA\times\calB)$ given that $P_{X}^{n}(\calA),P_{Y}^{n}(\calB)$
vanish exponentially fast with exponents $E_{1},E_{2}$, respectively.
Let 
\[
E_{1,\max}:=-\log P_{X,\min},\quad E_{2,\max}:=-\log P_{Y,\min},
\]
where $P_{X,\min}:=\min_{x}P_{X}(x)$, $P_{Y,\min}:=\min_{y}P_{Y}(y)$.
In this subsection, we consider an arbitrary distribution $P_{X,Y}$
satisfying $P_{X}(x)>0$, $P_{Y}(y)>0$ for all $(x,y)\in\mathcal{X}\times\mathcal{Y}$
and, for $E_{1}\in[0,E_{1,\max}],E_{2}\in[0,E_{2,\max}]$, we aim
at characterizing\footnote{\label{fn:By-time-sharing-arguments,}By time-sharing arguments, given
$(E_{1},E_{2})$, $\{n\underline{\Theta}_{n}(E_{1},E_{2})\}_{n\ge1}$
is subadditive. Hence, by Fekete's Subadditive Lemma, the first limit
in (\ref{eq:-66}) exists 
and equals $\inf_{n\ge1}\underline{\Theta}_{n}(E_{1},E_{2})$. Similar
observations serve to define the second limit in (\ref{eq:-66}).} 
\begin{align}
\underline{\Theta}(E_{1},E_{2}) & :=\lim_{n\to\infty}\underline{\Theta}_{n}(E_{1},E_{2}),\label{eq:-66}\\
\overline{\Theta}(E_{1},E_{2}) & :=\lim_{n\to\infty}\overline{\Theta}_{n}(E_{1},E_{2}),
\end{align}
where the exponents of the maximal and minimal noise stability are
defined by 
\begin{align}
\underline{\Theta}_{n}(E_{1},E_{2}) & :=-\frac{1}{n}\log\max_{\substack{\calA\subseteq\mathcal{X}^{n},\calB\subseteq\mathcal{Y}^{n}:\\
P_{X}^{n}(\calA)\leq2^{-nE_{1}},\\
P_{Y}^{n}(\calB)\leq2^{-nE_{2}}
}
}P_{XY}^{n}(\calA\times\calB),\label{eq:-72}\\
\overline{\Theta}_{n}(E_{1},E_{2}) & :=-\frac{1}{n}\log\min_{\substack{\calA\subseteq\mathcal{X}^{n},\calB\subseteq\mathcal{Y}^{n}:\\
P_{X}^{n}(\calA)\geq2^{-nE_{1}},\\
P_{Y}^{n}(\calB)\geq2^{-nE_{2}}
}
}P_{XY}^{n}(\calA\times\calB).\label{eq:-71}
\end{align}

For $E_{1}\in[0,E_{1,\max}],E_{2}\in[0,E_{2,\max}]$, define 
\begin{align}
 & \underline{\Theta}^{*}(E_{1},E_{2})\label{eq:thetastar}\\
 & :=\min_{\substack{Q_{XYW}:D(Q_{X|W}\|P_{X}|Q_{W})\ge E_{1},\\
D(Q_{Y|W}\|P_{Y}|Q_{W})\ge E_{2}
}
}D(Q_{XY|W}\|P_{XY}|Q_{W})\\
 & =\min_{\substack{Q_{W},Q_{X|W},Q_{Y|W}:\\
D(Q_{X|W}\|P_{X}|Q_{W})\ge E_{1},\\
D(Q_{Y|W}\|P_{Y}|Q_{W})\ge E_{2}
}
}\DD(Q_{X|W},Q_{Y|W}\|P_{XY}|Q_{W}),\\
 & \overline{\Theta}^{*}(E_{1},E_{2})\label{eq:-6}\\
 & :=\max_{\substack{Q_{W},Q_{X|W},Q_{Y|W}:\\
D(Q_{X|W}\|P_{X}|Q_{W})\leq E_{1},\\
D(Q_{Y|W}\|P_{Y}|Q_{W})\leq E_{2}
}
}\DD(Q_{X|W},Q_{Y|W}\|P_{XY}|Q_{W}),
\end{align}
where 
\begin{align*}
 & \DD(Q_{X|W},Q_{Y|W}\|P_{XY}|Q_{W})\\
 & :=\min_{Q_{XY|W}\in\calC(Q_{X|W},Q_{Y|W})}D(Q_{XY|W}\|P_{XY}|Q_{W})
\end{align*}
and the notation $\calC(Q_{X|W},Q_{Y|W})$ is defined in Subsection
\ref{subsec:notation}. Without loss of optimality, the alphabet size
of $W$ in either \eqref{eq:thetastar} or \eqref{eq:-6} can be assumed
to be no larger than $3$. This is a consequence of the support lemma
in \cite{Gamal}.

The asymptotic exponents $\underline{\Theta}$ and $\overline{\Theta}$
are characterized in the following theorem. Note that the following
theorem is not new, since it is a direct consequence of the information-theoretic
characterization of Brascamp--Lieb inequalities given in \cite{ahlswede1976spreading,carlen2009subadditivity,nair2014equivalent,liu2018information}
for the forward part and \cite{kamath2015reverse,beigi2016equivalent,liu2016brascamp,liu2018information,yu2021strong}
for the reverse part. For more details see \cite[Section 10.3]{yu2021common}. 
\begin{thm}[Strong Small-Set Expansion Theorem]
\label{thm:sse} For $E_{1}\in[0,E_{1,\max}],E_{2}\in[0,E_{2,\max}]$,
the following hold. 
\begin{enumerate}
\item $\underline{\Theta}(E_{1},E_{2})=\underline{\Theta}^{*}(E_{1},E_{2}).$
Moreover, $\underline{\Theta}_{n}(E_{1},E_{2})\ge\underline{\Theta}^{*}(E_{1},E_{2})$
for any $n\ge1$. 
\item 
\begin{align*}
\overline{\Theta}(E_{1},E_{2}) & =\overline{\Theta}^{**}(E_{1},E_{2})\\
 & :=\begin{cases}
\overline{\Theta}^{*}(E_{1},E_{2}), & E_{1},E_{2}>0,\\
E_{1}, & E_{2}=0,\\
E_{2}, & E_{1}=0.
\end{cases}
\end{align*}
Moreover, $\overline{\Theta}_{n}(E_{1},E_{2})\leq\overline{\Theta}^{**}(E_{1},E_{2})$
for any $n\ge1$. 
\end{enumerate}
\end{thm}
\begin{rem}
We interpret $\overline{\Theta}^{*}(E_{1},E_{2})$ as $\infty$ if
$P_{XY}(x,y)=0$ for some $(x,y)$. It can be checked that this interpretation
is consistent with the definition in \eqref{eq:thetastar} because
$Q_{W}$, $Q_{X|W}$, and $Q_{Y|W}$ in the outer maximum can be chosen
so that $Q_{W}(w)>0$, $Q_{X|W}(x|w)>0$, and $Q_{Y|W}(y|w)>0$ for
some $w$. 
\end{rem}
\begin{rem}
By the convexity of $\underline{\Theta}^{*}$ and concavity of $\overline{\Theta}^{*}$,
Theorem \ref{thm:sse} implies $\underline{\Theta}_{n}(E_{1},E_{2})\ge\lim_{t\downarrow0}\frac{1}{t}\underline{\Theta}^{*}(tE_{1},tE_{2})$
and $\overline{\Theta}_{n}(E_{1},E_{2})\le\lim_{t\downarrow0}\frac{1}{t}\overline{\Theta}^{**}(tE_{1},tE_{2})$.
In particular, for the DSBS with correlation coefficient $\rho>0$,
these inequalities reduce to that 
\begin{align}
\underline{\Theta}_{n}(E_{1},E_{2}) & \ge\begin{cases}
\frac{E_{1}+E_{2}-2\rho\sqrt{E_{1}E_{2}}}{1-\rho^{2}}, & \rho^{2}E_{1}\le E_{2}\le E_{1}/\rho^{2},\\
E_{1}, & E_{2}<\rho^{2}E_{1},\\
E_{2}, & E_{2}>E_{1}/\rho^{2},
\end{cases}\\
\overline{\Theta}_{n}(E_{1},E_{2}) & \leq\frac{E_{1}+E_{2}+2\rho\sqrt{E_{1}E_{2}}}{1-\rho^{2}}.\label{eq:rsse2-1}
\end{align}
These inequalities correspond to the small-set expansion theorem given
in \cite[Lemma 1]{ahlswede1976spreading}\cite[Lemma 3.4]{kahn1988influence}\cite[Theorem 3.4]{mossel2006non}\cite[Generalized Small-Set Expansion Theorem on p. 285]{O'Donnell14analysisof}. 
\end{rem}
\begin{rem}
\label{rem:equality} We must have $\liminf_{n\to\infty}-\frac{1}{n}\log P_{X}^{n}({\calA}_{n})\ge E_{1}$
for any sequence $(\calA_{n},\calB_{n})$ attaining the asymptotic
exponent $\underline{\Theta}(E_{1},E_{2})$. If $\limsup_{n\to\infty}-\frac{1}{n}\log P_{X}^{n}({\calA}_{n})>E_{1}$
then it must be the case that $E_{1}<E_{1,\max}$. So in this case,
we can add sequences in $\mathcal{X}^{n}$ to $\calA_{n}$ to get
a new sequence $(\calA_{n},\calB_{n})$ such that $\lim_{n\to\infty}-\frac{1}{n}\log P_{X}^{n}({\calA}_{n})=E_{1}$.
This is possible since for each $E_{1}\in[0,E_{1,\max})$ there is
a sequence $\tilde{\calA}_{n}\subseteq\mathcal{X}^{n}$ such that
(1) each $\tilde{\calA}_{n}$ is a type class (the type can change
with $n$), (2) $P_{X}^{n}({\tilde{\calA}}_{n})\le2^{-nE_{1}}$, (3)
$\lim_{n\to\infty}|\tilde{\calA}_{n}|=\infty$, and (4) $\lim_{n\to\infty}-\frac{1}{n}\log P_{X}^{n}({\tilde{\calA}}_{n})=E_{1}$.
We can then simply replace $\calA_{n}$ by $\calA_{n}\cup\hat{\calA}_{n}$
where $\hat{\calA}_{n}$ is a maximal subset of $\tilde{\calA}_{n}$
among those that continue to satisfy $P_{X}^{n}(\calA_{n}\cup\hat{\calA}_{n})\le2^{-nE_{1}}$.
Now, the resulting new sequence $(\calA_{n},\calB_{n})$ continues
to attain the asymptotic exponent $\underline{\Theta}(E_{1},E_{2})$
(by the converse part in the theorem above). Similarly, if needed,
we can also add sequences to $\calB_{n}$ such that $\lim_{n\to\infty}-\frac{1}{n}\log P_{Y}^{n}({\calB}_{n})=E_{2}$.
This implies that there exists a sequence $(\calA_{n},\calB_{n})$
such that $\lim_{n\to\infty}-\frac{1}{n}\log P_{X}^{n}({\calA}_{n})=E_{1},\lim_{n\to\infty}-\frac{1}{n}\log P_{Y}^{n}({\calB}_{n})=E_{2}$,
and $\lim_{n\to\infty}-\frac{1}{n}\log P_{XY}^{n}({\calA}_{n}\times{\calB}_{n})=\underline{\Theta}(E_{1},E_{2}).$ 
\end{rem}
\begin{rem}
\label{rem:equality2} We define the effective region of $\overline{\Theta}^{*}$
as the set of $(E_{1},E_{2})$ for which $\overline{\Theta}^{*}(E_{1},E_{2})=\invbreve{\psi}(E_{1},E_{2})$,
i.e., there exists an optimal tuple $(Q_{W},Q_{X|W},Q_{Y|W})$ attaining
the maximum in the definition of $\overline{\Theta}^{*}(E_{1},E_{2})$
such that $D(Q_{X|W}\|P_{X}|Q_{W})=E_{1},D(Q_{Y|W}\|P_{Y}|Q_{W})=E_{2}$.
Note that $D(Q_{X|W}\|P_{X}|Q_{W})$ is the asymptotic exponent of
the probability of a conditional type class with type $Q_{X|W}$ given
$Q_{W}$, and $\DD(Q_{X|W},Q_{Y|W}\|P_{XY}|Q_{W})$ is the asymptotic
exponent of the probability of $\mathcal{T}_{Q_{X|W}}(\mathbf{w})\times\mathcal{T}_{Q_{Y|W}}(\mathbf{w})$
with $\mathbf{w}$ having type $Q_{W}$. Hence, for $(E_{1},E_{2})$
in the effective region of $\overline{\Theta}^{*}$, there exists
a sequence of $(\calA_{n},\calB_{n})$ such that $\lim_{n\to\infty}-\frac{1}{n}\log P_{X}^{n}({\calA}_{n})=E_{1},\lim_{n\to\infty}-\frac{1}{n}\log P_{Y}^{n}({\calB}_{n})=E_{2}$,
and $\lim_{n\to\infty}-\frac{1}{n}\log P_{XY}^{n}({\calA}_{n}\times{\calB}_{n})=\overline{\Theta}^{*}(E_{1},E_{2}).$
The strong small-set expansion theorem is improved in \cite{yu2021strong}
by proving asymptotically sharp bounds for equality constraints in
\eqref{eq:-72} and \eqref{eq:-71}. 
\end{rem}
\begin{rem}
The discrepancy between the noninteractive simulation problem with
a uniform source defined on a joint type and the one with a product
source was noted in \cite{liu2019smoothing}, and similar discrepancies
were also noted and exploited in some classic works on strong converses
in network information theory, e.g., \cite{Ahls76,Csi97}. Specifically,
for any joint distribution $P_{XY}$ and $0\le E_{1}\le H(X),0\le E_{2}\le H(Y)$,
we have $\underline{\Upsilon}^{*}(E_{1},E_{2})\ge\underline{\Theta}^{*}(E_{1},E_{2})$
(the inequality is strict in general), where $\underline{\Upsilon}^{*}$
and $\underline{\Theta}^{*}$ are both defined for $P_{XY}$. This
observation follows since for any distribution $Q_{XYW}$ with marginal
$Q_{XY}=P_{XY}$, it holds that $I_{Q}(XY;W)=D(Q_{XY|W}\|P_{XY}|Q_{W})=\mathbb{E}_{Q_{W}}D(Q_{XY|W}\|P_{XY})$.
Similar equalities hold for $I_{Q}(X;W)$ and $I_{Q}(Y;W)$. If we
drop the condition $Q_{XY}=P_{XY}$ in the definition of $\underline{\Upsilon}^{*}$,
then we obtain $\underline{\Theta}^{*}$. So, we have the observation.
Another observation is $\lim_{t\downarrow0}\frac{1}{t}\underline{\Upsilon}^{*}(tE_{1},tE_{2})=\lim_{t\downarrow0}\frac{1}{t}\underline{\Theta}^{*}(tE_{1},tE_{2})$.
This follows by the two equivalent information-theoretic characterizations
of hypercontractivity inequalities: one expressed in terms of relative
entropies and the other expressed in terms of mutual information \cite{nair2014equivalent}. 
\end{rem}
Since the alphabet size of $W$ in the definition of $\underline{\Theta}^{*}(E_{1},E_{2})$
or $\overline{\Theta}^{*}(E_{1},E_{2})$ can be taken to be at most
$3$, both $\underline{\Theta}(E_{1},E_{2})$ and $\overline{\Theta}(E_{1},E_{2})$
are achieved by a sequence of the time-sharing of at most \textit{three}
type codes (or equivalently, a conditional type class with conditional
random variable $W$ taking at most three values). Here a type code
refers to a code of the form $(\calA,\calB):=(\mathcal{T}_{T_{X}},\mathcal{T}_{T_{Y}})$
for a pair of types $(T_{X},T_{Y})$.

Define the optimal transport divergence\footnote{The reason for this name is due to its resemblance to the optimal
transport cost \cite{villani2008optimal}. In the latter, the objective
function is the expected cost, instead of the relative entropy. } (or the minimum relative entropy) of $(Q_{X},Q_{Y})$ with respect
to $P_{XY}$, as 
\[
\DD(Q_{X},Q_{Y}\|P_{XY}):=\min_{Q_{XY}\in\calC(Q_{X},Q_{Y})}D(Q_{XY}\|P_{XY}).
\]
For $s\in[0,E_{1,\max}],t\in[0,E_{2,\max}]$, define 
\begin{align}
\varphi(s,t) & :=\min_{\substack{Q_{XY}:D(Q_{X}\|P_{X})=s,\\
D(Q_{Y}\|P_{Y})=t
}
}D(Q_{XY}\|P_{XY})\label{eq:-11-3}\\
 & =\min_{\substack{Q_{X},Q_{Y}:D(Q_{X}\|P_{X})=s,\\
D(Q_{Y}\|P_{Y})=t
}
}\DD(Q_{X},Q_{Y}\|P_{XY}),
\end{align}
and 
\begin{align}
\psi(s,t) & :=\max_{\substack{Q_{X},Q_{Y}:D(Q_{X}\|P_{X})=s,\\
D(Q_{Y}\|P_{Y})=t
}
}\DD(Q_{X},Q_{Y}\|P_{XY}).\label{eq:-11-3-1}
\end{align}
Define $\breve{f}$ as the lower convex envelope of a function $f$,
and $\invbreve{f}$ as its upper concave envelope. Then, by definition,
we have 
\begin{align}
 & \underline{\Theta}^{*}(E_{1},E_{2})\nonumber \\
 & =\min_{\substack{(q_{i},s_{i},t_{i})_{i\in[3]}:\\
\sum_{i}q_{i}=1,\;q_{i}\ge0,\forall i\in[3]\\
\sum_{i}q_{i}s_{i}\ge E_{1},\sum_{i}q_{i}t_{i}\ge E_{2}
}
}\sum_{i}q_{i}\varphi(s_{i},t_{i})\\
 & =\min_{s\ge E_{1},t\ge E_{2}}\min_{\substack{(q_{i},s_{i},t_{i})_{i\in[3]}:\\
\sum_{i}q_{i}=1,\;q_{i}\ge0,\forall i\in[3]\\
\sum_{i}q_{i}s_{i}=s,\sum_{i}q_{i}t_{i}=t
}
}\sum_{i}q_{i}\varphi(s_{i},t_{i})\nonumber \\
 & =\min_{s\ge E_{1},t\ge E_{2}}\breve{\varphi}(s,t),\label{eq:lce}
\end{align}
and similarly, 
\begin{align}
\overline{\Theta}^{*}(E_{1},E_{2}) & =\max_{s\leq E_{1},t\leq E_{2}}\invbreve{\psi}(s,t).\label{eq:uce}
\end{align}
Hence $\underline{\Theta}^{*}(E_{1},E_{2})$ is convex in $(E_{1},E_{2})$,
and $\overline{\Theta}^{*}(E_{1},E_{2})$ is concave in $(E_{1},E_{2})$.
By definition, it holds that for $s\in[0,E_{1,\max}],t\in[0,E_{2,\max}]$,
\begin{align}
 & \underline{\Theta}^{*}(s,t)\le\breve{\varphi}(s,t)\le\varphi(s,t)\nonumber \\
 & \qquad\le\psi(s,t)\le\invbreve{\psi}(s,t)\le\overline{\Theta}^{*}(s,t).\label{eq:ordered}
\end{align}

\begin{prop}
Both $\underline{\Theta}^{*}(E_{1},E_{2})$ and $\overline{\Theta}^{*}(E_{1},E_{2})$
are continuous over $E_{1}\in[0,E_{1,\max}],E_{2}\in[0,E_{2,\max}]$. 
\end{prop}
\begin{IEEEproof}
By the convexity and concavity, $\underline{\Theta}^{*}(E_{1},E_{2})$
and $\overline{\Theta}^{*}(E_{1},E_{2})$ are continuous over $E_{1}\in(0,E_{1,\max}),E_{2}\in(0,E_{2,\max})$.
On the boundary, the continuity of these two functions follows by
the continuity of the constraint functions and the continuity of the
objective function, i.e., the continuity of $D(Q_{X|W}\|P_{X}|Q_{W}),D(Q_{Y|W}\|P_{Y}|Q_{W})$
and $\DD(Q_{X|W},Q_{Y|W}\|P_{XY}|Q_{W})$ in $(Q_{W},Q_{X|W},Q_{Y|W})$.

The continuity of $D(Q_{X|W}\|P_{X}|Q_{W}),D(Q_{Y|W}\|P_{Y}|Q_{W})$
in $(Q_{W},Q_{X|W},Q_{Y|W})$ is obvious, since as assumed, $P_{X}$
and $P_{Y}$ have full support. We claim that 
\begin{align}
f(Q_{W},Q_{X|W},Q_{Y|W}) & :=\DD(Q_{X|W},Q_{Y|W}\|P_{XY}|Q_{W})\label{eq:fun_f}
\end{align}
is continuous in $(Q_{W},Q_{X|W},Q_{Y|W})$, which follows by the
following lemma. 
\begin{lem}
\label{lem:marginalbound} \cite[Lemma 13]{yu2020asymptotics} Let
$P_{X},Q_{X}$ be distributions on $\mathcal{X}$, and $P_{Y},Q_{Y}$
distributions on $\mathcal{Y}$. Then for any $Q_{XY}\in\calC(Q_{X},Q_{Y})$,
there exists $P_{XY}\in\calC(P_{X},P_{Y})$ such that 
\begin{equation}
\Vert P_{XY}-Q_{XY}\Vert\leq\Vert P_{X}-Q_{X}\Vert+\Vert P_{Y}-Q_{Y}\Vert,\label{eq:-81}
\end{equation}
where $\Vert P-Q\Vert:=\sup_{\calA}P(\calA)-Q(\calA)$ denotes the
total variation distance between $P$ and $Q$. 
\end{lem}
By Lemma \ref{lem:marginalbound}, given $(Q_{W},Q_{X|W},Q_{Y|W},P_{W},P_{X|W},P_{Y|W})$,
for any $Q_{XY|W}\in\calC(Q_{X|W},Q_{Y|W})$, there exists $P_{XY|W}\in\calC(P_{X|W},P_{Y|W})$
such that 
\begin{align*}
 & \Vert P_{WXY}-Q_{WXY}\Vert\\
 & \leq\Vert P_{W}-Q_{W}\Vert+\max_{w}\Vert P_{X|W=w}-Q_{X|W=w}\Vert\\
 & \qquad+\max_{w}\Vert P_{Y|W=w}-Q_{Y|W=w}\Vert.
\end{align*}
Hence, for any sequence $(P_{W}^{(k)},P_{X|W}^{(k)},P_{Y|W}^{(k)})$
convergent to $(Q_{W},Q_{X|W},Q_{Y|W})$, $\limsup_{k\to\infty}f(P_{W}^{(k)},P_{X|W}^{(k)},P_{Y|W}^{(k)})\le f(Q_{W},Q_{X|W},Q_{Y|W})$,
and $f(Q_{W},Q_{X|W},Q_{Y|W})\le\liminf_{k\to\infty}f(P_{W}^{(k)},P_{X|W}^{(k)},P_{Y|W}^{(k)})$.
Hence $f(Q_{W},Q_{X|W},Q_{Y|W})$ is continuous in $(Q_{W},Q_{X|W},Q_{Y|W})$. 
\end{IEEEproof}

\subsection{\label{subsec:Example:-DSBS}Example: DSBS}

Consider a DSBS with correlation coefficient $\rho$, whose distribution
$P_{XY}$ is given in Table \ref{tab:The-distribution-of}. We assume
that $0<\rho<1$. Denote by $h:t\mapsto-t\log t-(1-t)\log(1-t)$ the
binary entropy function, and $h^{-1}$ as the inverse of the restriction
of $h$ to the set $[0,\frac{1}{2}]$.

\begin{figure}
\centering %
\includegraphics[width=1\columnwidth]{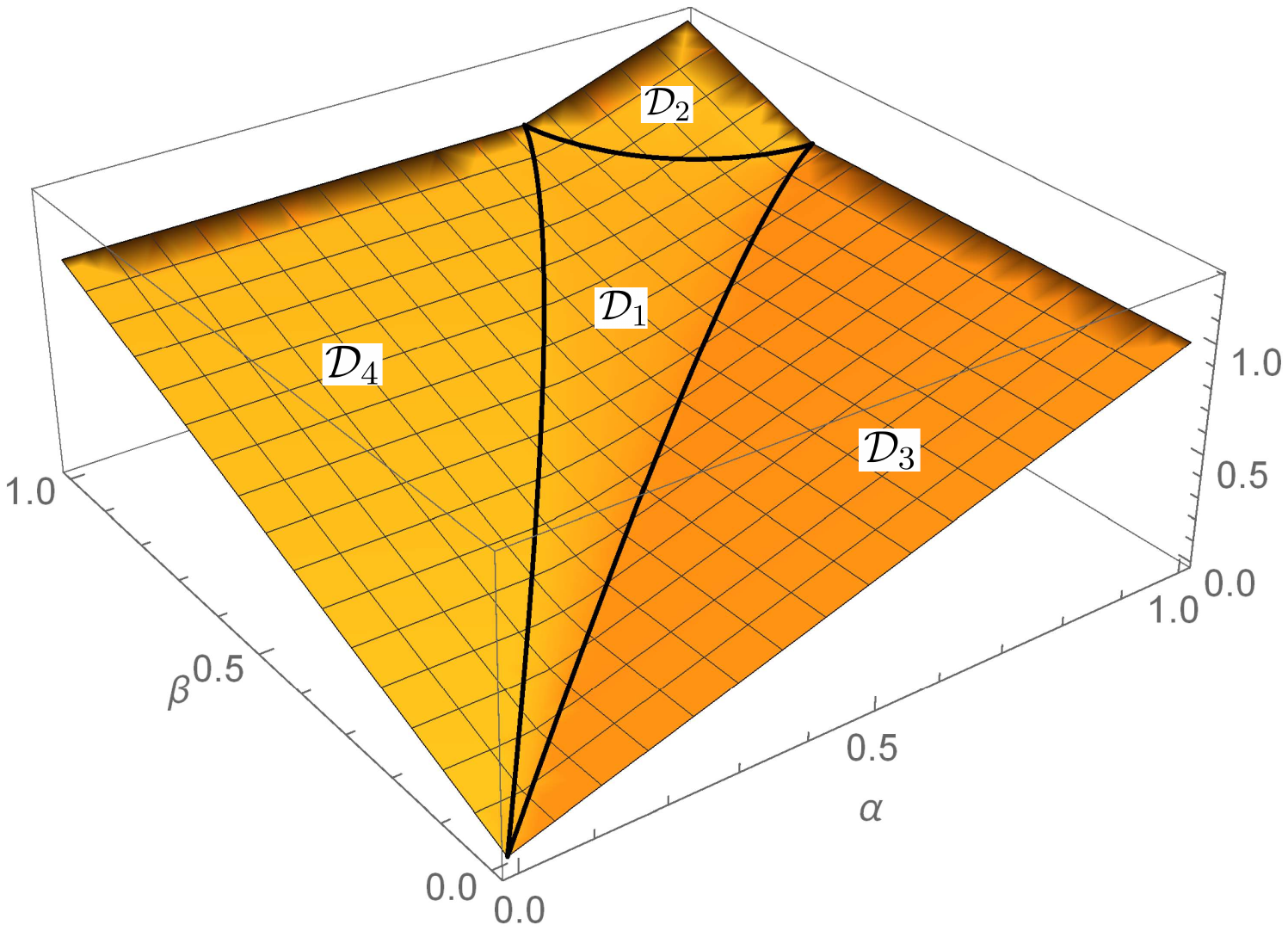}

\caption{\label{fig:upsilon1}Illustration of $\underline{\Upsilon}^{*}$ for
$\rho=0.9$. }
\end{figure}

The following explicit expression for $\underline{\Upsilon}^{*}$
(or $E^{*}$ given in \eqref{eq:E*}, or equivalently, the Gray--Wyner
coding region or the mutual information region \cite{gray1974source})
for the DSBS was conjectured by Gray and Wyner \cite{gray1974source,wyner1974recent},
and recently confirmed positively by the first author \cite{yu2022gray}.
For $(s,t)\in[0,1]^{2}$, it holds that 
\begin{equation}
\underline{\Upsilon}^{*}(s,t)=\begin{cases}
1-(1-q)h(\frac{a+b-q}{2(1-q)})\\
\qquad-qh(\frac{a-b+q}{2q}), & (s,t)\in\mathcal{D}_{1},\\
1+h(q)-h(a)-h(b), & (s,t)\in\mathcal{D}_{2},\\
1-h(a), & (s,t)\in\mathcal{D}_{3},\\
1-h(b), & (s,t)\in\mathcal{D}_{4},
\end{cases}\label{eq:mi}
\end{equation}
where $q=\frac{1-\rho}{2},a=h^{-1}(1-s),b=h^{-1}(1-t)$, and 
\begin{align*}
\mathcal{D}_{1} & :=\{(s,t)\in[0,1]^{2}:a*q\ge b,\\
 & \qquad\qquad\;b*q\ge a,\;a*b\ge q\},\\
\mathcal{D}_{2} & :=\{(s,t)\in[0,1]^{2}:a*b<q\},\\
\mathcal{D}_{3} & :=\{(s,t)\in[0,1]^{2}:a*q<b\},\\
\mathcal{D}_{4} & :=\{(s,t)\in[0,1]^{2}:b*q<a\}.
\end{align*}
The function $\underline{\Upsilon}^{*}$ is plotted in Fig. \ref{fig:upsilon1}.

\begin{figure}
\label{fig:DSBS} \centering %
\begin{tabular}{c}
\includegraphics[width=1\columnwidth]{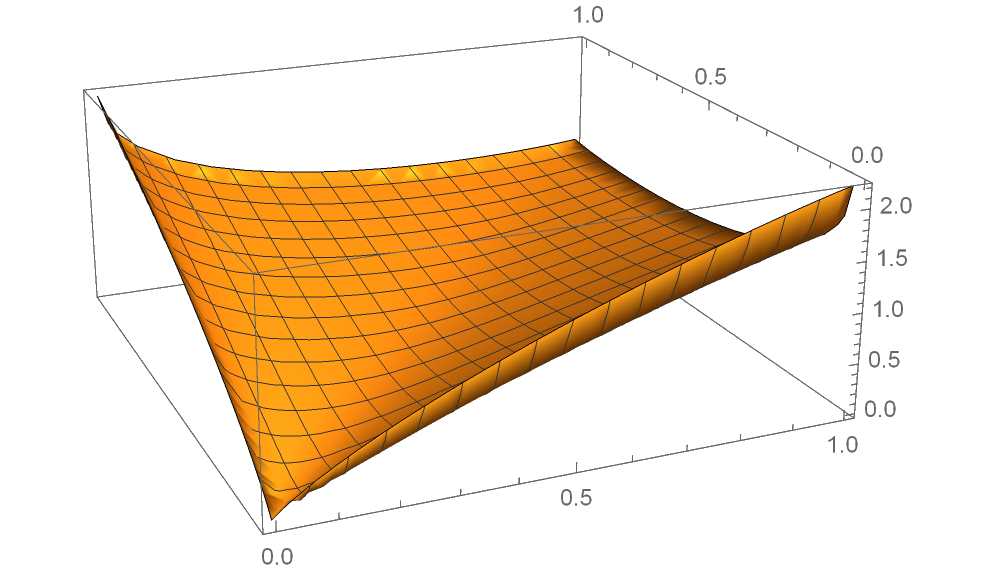} \tabularnewline
{\footnotesize{}{}{}{}$\varphi(s,t)$} \tabularnewline
\end{tabular}

\begin{tabular}{c}
\includegraphics[width=0.8\columnwidth]{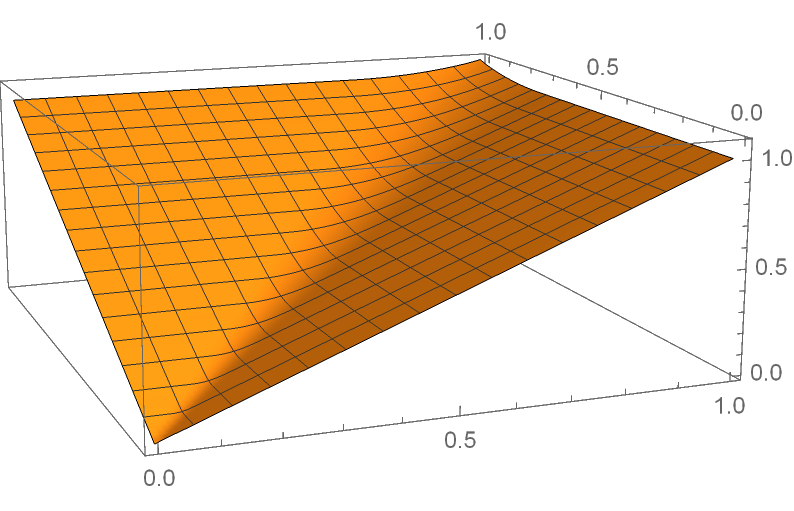}\tabularnewline
{\footnotesize{}{}{}{}$\underline{\Theta}^{*}(s,t)$}\tabularnewline
\end{tabular}

\begin{tabular}{c}
\includegraphics[width=1\columnwidth]{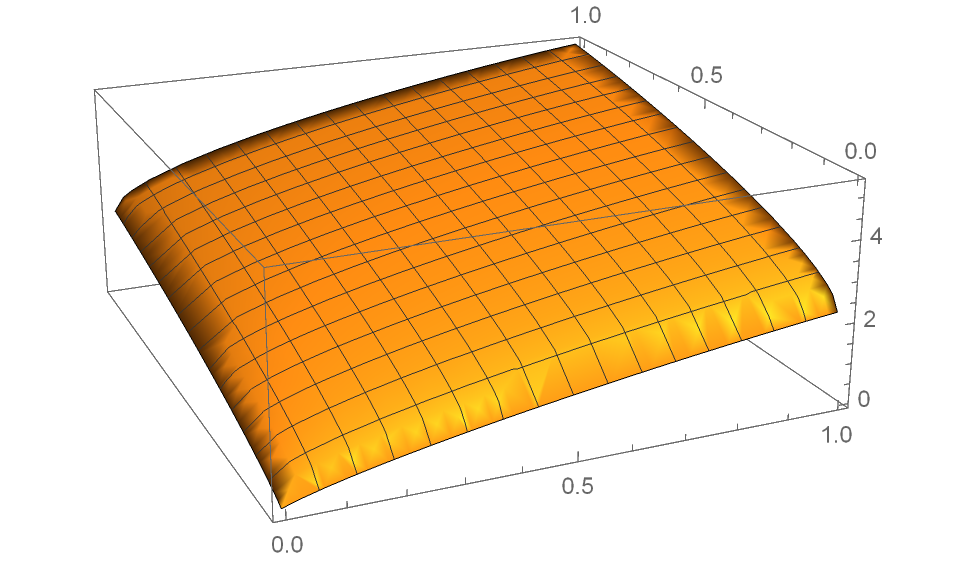}\tabularnewline
{\footnotesize{}{}{}{}$\overline{\Theta}^{*}(s,t)=\psi(s,t)$}\tabularnewline
\end{tabular}\caption{\label{fig:upsilon}Illustration of $\varphi(s,t)$, $\psi(s,t)$,
$\underline{\Theta}^{*}(E_{1},E_{2})$, and $\overline{\Theta}^{*}(E_{1},E_{2})$
for the DSBS for $\rho=0.9$. Note that $\underline{\Theta}^{*}(E_{1},E_{2})$
and $\overline{\Theta}^{*}(E_{1},E_{2})$ are expressed in terms of
$\varphi(s,t)$ and $\psi(s,t)$ in (\ref{eq:lce}) and (\ref{eq:uce}).
All the bases of logarithms are 2 for these figures. }
\end{figure}



We next provide explicit expressions for $\underline{\Theta}^{*}$
and $\overline{\Theta}^{*}$. Suppose $Q_{X}=(a,1-a)$ and $Q_{Y}=(b,1-b)$.
For the DSBS, we have $D(Q_{X}\|P_{X})=1-h(a)$. Hence, if $D(Q_{X}\|P_{X})=s$,
then we have $a=h^{-1}(1-s)$ or $1-h^{-1}(1-s)$. Similarly, for
$Q_{Y}$ such that $D(Q_{Y}\|P_{Y})=t$, we have $b=h^{-1}(1-t)$
or $1-h^{-1}(1-t)$.

Define $\kappa:=(\frac{1+\rho}{1-\rho})^{2}$. For $\max\{0,a+b-1\}\le p\le\min\{a,b\}$,
define 
\begin{align}
D_{a,b}(p) & :=D((p,a-p,b-p,1+p-a-b)\|\nonumber \\
 & \qquad(\frac{1+\rho}{4},\frac{1-\rho}{4},\frac{1-\rho}{4},\frac{1+\rho}{4})),\\
\DD(a,b) & :=\min_{0,a+b-1\le p\le a,b}D_{a,b}(p)\\
 & =D_{a,b}(p^{*}),\label{eq:Dab}
\end{align}
where 
\begin{align*}
p^{*} & =\frac{1}{2(\kappa-1)}\Big((\kappa-1)(a+b)+1\\
 & \quad-\sqrt{((\kappa-1)(a+b)+1)^{2}-4\kappa(\kappa-1)ab}\Big).
\end{align*}
Equation \eqref{eq:Dab} follows from the facts that $p\mapsto D_{a,b}(p)$
is convex (due to the convexity of the relative entropy), ${\max\{0,a+b-1\}\le p^{*}\le\min\{a,b\}}$,
and the extreme value is taken at $p^{*}$. Furthermore, we have the
following lemma, whose proof is provided in Appendix \ref{sec:Proof-of-Lemma-D}. 
\begin{lem}
\label{lem:InequalityD} For $0\le a,b\le\frac{1}{2}$, it holds that
\begin{align}
\DD(a,b) & =\DD(1-a,1-b)\nonumber \\
 & \le\DD(a,1-b)=\DD(1-a,b).\label{eq:-5}
\end{align}
\end{lem}
By Lemma \ref{lem:InequalityD}, we have 
\begin{align*}
\varphi(s,t) & =\DD(h^{-1}(1-s),h^{-1}(1-t)),\\
\psi(s,t) & =\DD(h^{-1}(1-s),1-h^{-1}(1-t)),
\end{align*}
Then $\underline{\Theta}^{*}(E_{1},E_{2})$ and $\overline{\Theta}^{*}(E_{1},E_{2})$
are determined by $\varphi(s,t)$ and $\psi(s,t)$ via (\ref{eq:lce})
and (\ref{eq:uce}). Moreover, by Theorem \ref{thm:sse}, $\underline{\Theta}(E_{1},E_{2})$
is attained by a sequence involving the time-sharing of at most three
pairs of concentric\footnote{
Here we call two Hamming spheres concentric if they have the same
center and the radiuses are both not larger than or both not smaller
than $n/2$. Similarly, two spheres are called anti-concentric if
they have the same center and one of the two radiuses is not larger
than $n/2$ while the other one is not smaller than $n/2$.} Hamming spheres, and $\overline{\Theta}(E_{1},E_{2})$ is attained
by a sequence involving the time-sharing of at most three pairs of
anti-concentric Hamming spheres. 
\begin{prop}[DSBS]
\label{prop:OPSconjecture} For the DSBS, the following hold.\\
 1) $\underline{\Theta}(E_{1},E_{2})$ is achieved by a sequence of
pairs of concentric Hamming spheres if $\tilde{\varphi}(E_{1},E_{2}):=\min_{s\ge E_{1},t\ge E_{2}}\varphi(s,t)$
is convex in $(E_{1},E_{2})$.\\
 2) $\overline{\Theta}(E_{1},E_{2})$ is achieved by a sequence of
pairs of anti-concentric Hamming spheres if $\psi(s,t)$ is concave
in $(s,t)$. 
\end{prop}
\begin{IEEEproof}
We prove Proposition \ref{prop:OPSconjecture}. For a function $f:\mathbb{R}_{\ge0}\times\mathbb{R}_{\ge0}\to\mathbb{R}_{\ge0}$,
we use $\tilde{f}$ to denote $(x,y)\mapsto\inf_{s\ge x,t\ge y}f(s,t)$.
Then, by assumption, $\tilde{\varphi}$ is convex. We now prove that
$\tilde{\varphi}=\underline{\Theta}^{*}$, where by definition, $\underline{\Theta}^{*}=\tilde{\breve{\varphi}}$.
On one hand, $\tilde{\breve{\varphi}}\le\tilde{\varphi}$. On the
other hand, $\tilde{\breve{\varphi}}\ge\tilde{\breve{\tilde{\varphi}}}=\tilde{{\tilde{\varphi}}}={\tilde{\varphi}}$.
Hence, $\tilde{\varphi}=\underline{\Theta}^{*}$. This implies that
the time-sharing random variable $W$ can be removed. By Theorem \ref{thm:sse},
time-sharing is not needed to attain $\underline{\Theta}$, completing
the proof of Statement 1).

Statement 2) follows similarly, but we need to prove that $\psi(s,t)=\max_{s\le E_{1},t\le E_{2}}\psi(s,t)$.
This equality is equivalent to that $\max_{Q_{X},Q_{Y}:D(Q_{X}\|P_{X})\le s,D(Q_{Y}\|P_{Y})\le t}\DD(Q_{X},Q_{Y}\|P_{XY})$
is always attained by some $Q_{X},Q_{Y}$ satisfying that both the
equalities in the constraints hold. Observe that in this maximization
both the objective function and the constraint functions are convex
and, moreover, the set of feasible solutions is compact. By the Krein--Milman
theorem, the set of feasible solutions is the closed convex hull of
its extreme points. Hence, the maximization is attained by an extreme
point. An extreme point here is a pair $(Q_{X},Q_{Y})$ such that
$Q_{X}$ is either a Dirac distribution or a distribution satisfying
$D(Q_{X}\|P_{X})=s$, and so is $Q_{Y}$. For the DSBS considered
here, when $s<1$, there is no Dirac distribution in the set of feasible
solutions, which means any extreme points must satisfy $D(Q_{X}\|P_{X})=s$.
Similarly, they must also satisfy $D(Q_{Y}\|P_{Y})=t$. These are
the desired, which imply Statement 2). 
\end{IEEEproof}
Ordentlich, Polyanskiy, and Shayevitz \cite{ordentlich2020note} conjectured
that $\underline{\Theta}(E_{1},E_{2})$ is achieved by a sequence
of pairs of concentric Hamming spheres, and $\overline{\Theta}(E_{1},E_{2})$
is achieved by a sequence of pairs of anti-concentric Hamming spheres.
Hence their conjecture is true under the assumptions in Proposition
\ref{prop:OPSconjecture}. Given Theorem \ref{thm:sse}, Ordentlich--Polyanskiy--Shayevitz's
conjecture boils down to proving the convexity of $\tilde{\varphi}$
and concavity of $\psi$. In other words, the essence is to remove
the time-sharing random variable $W$ in both $\underline{\Theta}^{*}$
and $\overline{\Theta}^{*}$. Subsequent to the completion of this
paper, the first author proved this point, and hence confirmed positively
the Ordentlich--Polyanskiy--Shayevitz conjecture \cite{leiyu2021june,yu2021strong}.
Furthermore, noninteractive simulation in the exponential regime was
also studied by Kirshner and Samorodnitsky \cite{kirshner2021moment}
who solved the symmetric case $E_{1}=E_{2}$.

The functions $\varphi,\psi,\underline{\Theta}^{*}$ and $\overline{\Theta}^{*}$
for $\rho=0.9$ are plotted in Fig. \ref{fig:upsilon}. This figure
numerically verifies the assumptions in Proposition \ref{prop:OPSconjecture}.

\subsection{\label{subsec:Applications-to-Zero-Error}Applications to Zero-Error
Coding}

As mentioned in \cite{ordentlich2020note}, the minimization part
of the conjecture of Ordentlich, Polyanskiy, and Shayevitz implies
a sharper outer bound for the zero-error capacity region of the binary
adder channel.

Consider the two-user binary adder channel (BAC) $(\mathbf{a},\mathbf{b})\in\{0,1\}^{2n}\mapsto\mathbf{a}+\mathbf{b}\in\{0,1,2\}^{n}$
and a code $(\calA_{n},\calB_{n})$ with $\calA_{n},\calB_{n}\subseteq\{0,1\}^{n}$
for this channel. Here $\mathbf{a}+\mathbf{b}$ denotes addition over
$\mathbb{Z}^{n}$. When the code $(\calA_{n},\calB_{n})$ is used
to transmit messages over the BAC, the receiver is able to decode
the messages without any error if and only if any pair $(\mathbf{a},\mathbf{b})\in\calA_{n}\times\calB_{n}$
is mapped to a unique sequence in $\{0,1,2\}^{n}$, i.e., $|\calA_{n}+\calB_{n}|=|\calA_{n}|\cdot|\calB_{n}|$,
where $\calA_{n}+\calB_{n}$ denotes the sumset $\{\mathbf{a}+\mathbf{b}:\mathbf{a}\in\calA_{n},\mathbf{b}\in\calB_{n}\}$.
The zero-error capacity region $\calC$ of the BAC (or the rate region
of uniquely decodable code pairs) is defined as the set of $(R_{1},R_{2})$
for which there is a sequence of pairs $\calA_{n},\calB_{n}\subseteq\{0,1\}^{n}$
with $|\calA_{n}|=2^{n(R_{1}+o(1))},|\calB_{n}|=2^{n(R_{2}+o(1))}$
such that $|\calA_{n}+\calB_{n}|=|\calA_{n}|\cdot|\calB_{n}|$ for
every $n$.

Finding the capacity region of the BAC is a long standing open problem;
refer to \cite{lindstrom1969determination,van1978upper,kasami1978bounds,weldon1978coding,kasami1983graph,urbanke1998zero,ajjanagadde2015adder,ordentlich2016upper,austrin2017sharper,ordentlich2020note}
for details. The current progress on this topic is rather unsatisfactory.
The upper bound on the sum rate $R_{1}+R_{2}$ is still the simple
bound $3/2$, which corresponds to the maximum sum rate in the Shannon
capacity. However, Urbanke and Li \cite{urbanke1998zero} broke through
the $3/2$ bound in the unbalanced case, in which it is assumed that
$R_{1}=1$ (note that it does not mean $\calA_{n}=\{0,1\}^{n}$) and
they showed that $R_{2}\le0.4921$. Later, this result was improved
to $R_{2}\le0.4798$ in \cite{ordentlich2016upper} and $R_{2}\le0.4228$
respectively in \cite{austrin2017sharper}. The latter is the best
known upper bound until now. The best known lower bound for this case
is $R_{2}\ge1/4$ given in \cite{kasami1983graph}.

In particular, the reverse small-set expansion inequality given in
\eqref{eq:rsse2-1} for the DSBS was used by Austrin, Kaski, Koivisto,
and Nederlof \cite{austrin2017sharper} to prove the best known upper
bound. As mentioned by Ordentlich, Polyanskiy, and Shayevitz \cite{ordentlich2020note},
repeating the arguments in \cite{austrin2017sharper} with improved
bounds on $\overline{\Theta}(E_{1},E_{2})$ will yield tighter bounds
on $R_{2}$ when $R_{1}=1$. Replacing the reverse small-set expansion
inequality in the proof given in \cite{austrin2017sharper} with the
characterization of $\overline{\Theta}(E_{1},E_{2})$ in Theorem \ref{thm:sse},
we obtain the following result. 
\begin{thm}
\label{thm:zeroerror} If $(1-\epsilon,R_{2})\in\calC$, then for
any $\rho\in(0,1)$ there exists some $\lambda\in\frac{1}{2}\pm\sqrt{\frac{\ln(2)\epsilon}{2}}$
such that 
\begin{align*}
 & \lambda\overline{\Theta}^{*}(\frac{\epsilon}{\lambda},\frac{\lambda+\epsilon-R_{2}}{\lambda})\\
 & \ge\lambda(\frac{5}{2}-\log(3-\rho))-\frac{1}{2}-\epsilon-\sqrt{\frac{\ln(2)\epsilon}{2}},
\end{align*}
where $\overline{\Theta}^{*}$ is defined for the DSBS with correlation
coefficient $\rho$. In particular, if $\epsilon=0$, we obtain for
any $\rho\in(0,1)$, 
\begin{equation}
\frac{1}{2}\overline{\Theta}^{*}(0,1-2R_{2})\geq\frac{1}{2}(\frac{3}{2}-\log(3-\rho)).\label{eq:-57}
\end{equation}
\end{thm}
Numerical results show that if $R_{1}=1$ (i.e., $\epsilon=0$), by
choosing the almost best $\rho=0.6933$, (\ref{eq:-57}) implies $R_{2}\leq0.4177$,
which improves the previously best known bound $R_{2}\leq0.4228$
established in \cite{austrin2017sharper}. Note that the upper bound
$R_{2}\leq0.4177$ was first calculated in \cite{ordentlich2020note}.
Subsequent to the completion of this paper, $\overline{\Theta}^{*}=\psi$
was proven by the first author \cite{leiyu2021june,yu2021strong},
which further simplifies the inequalities in Theorem \ref{thm:zeroerror}.

\section{Brascamp--Lieb and Hypercontractivity Inequalities}

In this section, we relax Boolean functions in noninteractive simulation
problems to any nonnegative functions, but still restrict their supports
to be exponentially small. Let $(\mathbf{X},\mathbf{Y})\sim P_{XY}^{n}$,
where $P_{XY}$ is a joint distribution defined on $\mathcal{X}\times\mathcal{Y}$.
Recall the notation $\langle f,g\rangle=\mathbb{E}[f(\mathbf{X})g(\mathbf{Y})]$
and $\Vert f\Vert_{p}=\big(\mathbb{E}[f(\mathbf{X})^{p}]\big)^{1/p}$.
We continue to assume that $\mathcal{X},\mathcal{Y}$ are finite sets,
each with cardinality at least $2$, and with $P_{X}(x)>0$, $P_{Y}(y)>0$
for all $(x,y)\in\mathcal{X}\times\mathcal{Y}$. We next derive strengthened
versions of (forward and reverse) Brascamp--Lieb and hypercontractivity
inequalities by using Theorem \ref{thm:sse}. Our inequalities reduce
to the usual ones when $\alpha=\beta=0$. For $\alpha\in[0,E_{1,\max}],\beta\in[0,E_{2,\max}]$
and $p,q\in(0,\infty)$, define 
\begin{align*}
\underline{\Lambda}_{p,q}^{*}(\alpha,\beta) & :=\min_{\substack{\alpha\le s\le E_{1,\max}\\
\beta\le t\le E_{2,\max}
}
}(\underline{\Theta}^{*}(s,t)-\frac{s}{p}-\frac{t}{q}),
\end{align*}
and 
\begin{align*}
\overline{\Lambda}_{p,q}^{*}(\alpha,\beta) & :=\min_{\substack{\alpha\le s\le E_{1,\max}\\
\beta\le t\le E_{2,\max}
}
}(\frac{s}{p}+\frac{t}{q}-\overline{\Theta}^{*}(s,t)).
\end{align*}

\begin{rem}
Subsequent to the completion of this paper, the first author proved
that $\underline{\Theta}^{*}=\tilde{\varphi}$ and $\overline{\Theta}^{*}=\psi$
for the DSBS in \cite{leiyu2021june}. 
\end{rem}
The strengthened (forward and reverse) Brascamp--Lieb inequalities
are given in the following theorem, whose proof is provided in Appendix
\ref{sec:Proof-of-Theorem-1}. 
\begin{thm}
\label{thm:stronghypercontractivity-2} Let{} $p,q>0$ and $\alpha\in[0,E_{1,\max}],\beta\in[0,E_{2,\max}]$.
Let $f,g$ be nonnegative functions on $\mathcal{X}^{n}$ and $\mathcal{Y}^{n}$
respectively such that $P_{X}^{n}(\mathrm{supp}(f))\leq2^{-n\alpha},P_{Y}^{n}(\mathrm{supp}(g))\leq2^{-n\beta}$.
Then 
\begin{align}
\langle f,g\rangle & \leq2^{-n\underline{\Lambda}_{p,q}^{*}(\alpha,\beta)}\Vert f\Vert_{p}\Vert g\Vert_{q},\label{eq:fh}\\
\langle f,g\rangle & \geq2^{n\overline{\Lambda}_{p,q}^{*}(\alpha,\beta)}\Vert f\Vert_{p}\Vert g\Vert_{q}.\label{eq:rh}
\end{align}
\end{thm}
\begin{rem}
\label{rem:HCsharpness} Given $\alpha\in[0,E_{1,\max}],\beta\in[0,E_{2,\max}]$
and $p,q>0$, the inequality (\ref{eq:fh}) is exponentially sharp,
in the sense that the exponents on the two sides of (\ref{eq:fh})
are asymptotically equal as $n\to\infty$, for a sequence of Boolean
functions $f_{n}=1_{\calA_{n}},g_{n}=1_{\calB_{n}}$ with $(\calA_{n},\calB_{n})$
denoting the sets given in Remark \ref{rem:equality} but with $(E_{1},E_{2})$
there replaced by the optimal $(s^{*},t^{*})$ attaining the minimum
in the definition of $\underline{\Lambda}_{p,q}^{*}(\alpha,\beta)$.
Note that if $(\alpha,\beta)$ is in the effective region of $\overline{\Theta}^{*}$,
then the optimal $(s^{*},t^{*})$ attaining the minimum in the definition
of $\overline{\Lambda}_{p,q}^{*}(\alpha,\beta)$ is still in the effective
region of $\overline{\Theta}^{*}$. Given $(\alpha,\beta)$ in the
effective region of $\overline{\Theta}^{*}$ and $p,q>0$, the inequality
(\ref{eq:rh}) is exponentially sharp, in the sense that the exponents
on the two sides of (\ref{eq:rh}) are asymptotically equal as $n\to\infty$,
for a sequence of Boolean functions $f_{n}=1_{\calA_{n}},g_{n}=1_{\calB_{n}}$
with some sequence $(\calA_{n},\calB_{n})$; see Remark \ref{rem:equality2}. 
\end{rem}
\begin{rem}
A special case of Theorem \ref{thm:stronghypercontractivity-2} with
$p\ge p_{0},q\ge q_{0}$ for (\ref{eq:fh}) and $p\le p_{1},q\le q_{1}$
for (\ref{eq:fh}){} can be recovered by the information-theoretic
characterization of classic Brascamp--Lieb inequalities, where $(\frac{1}{p_{0}},\frac{1}{q_{0}})$
is a subgradient of $\underline{\Theta}^{*}$ and $(\frac{1}{p_{1}},\frac{1}{q_{1}})$
is a subgradient of $\overline{\Theta}^{*}$. See Corollary 3 in \cite{yu2021strong}
which is a consequence of Theorem 2 therein, and also the simple proof
of Theorem 2 therein given in Appendix C in \cite{yu2021strong}. 
\end{rem}
For $\alpha\in[0,E_{1,\max}],\beta\in[0,E_{2,\max}]$, define the
forward and reverse $(\alpha,\beta)$-hypercontractivity regions as
\begin{align*}
\mathcal{R}_{\alpha,\beta}^{+}(P_{XY}) & :=\{(p,q)\in(0,\infty)^{2}:\\
 & \qquad\underline{\Theta}^{*}(E_{1},E_{2})\ge\frac{1}{p}E_{1}+\frac{1}{q}E_{2},\\
 & \qquad\forall E_{1}\in[\alpha,E_{1,\max}],E_{2}\in[\beta,E_{2,\max}]\},\\
\mathcal{R}_{\alpha,\beta}^{-}(P_{XY}) & :=\{(p,q)\in(0,\infty)^{2}:\\
 & \qquad\overline{\Theta}^{*}(E_{1},E_{2})\le\frac{1}{p}E_{1}+\frac{1}{q}E_{2},\\
 & \qquad\forall E_{1}\in[\alpha,E_{1,\max}],E_{2}\in[\beta,E_{2,\max}]\}.
\end{align*}
For $\alpha=\beta=0$, $\mathcal{R}_{0,0}^{+}(P_{XY})$ and $\mathcal{R}_{0,0}^{-}(P_{XY})$
correspond to the classic hypercontractivity regions in \cite{ahlswede1976spreading,carlen2009subadditivity,nair2014equivalent,liu2018information}
for the forward one and \cite{kamath2015reverse,beigi2016equivalent,liu2016brascamp,liu2018information,yu2021strong}
for the reverse one.

As a consequence of Theorem \ref{thm:stronghypercontractivity-2},
we obtain the following new version of hypercontractivity. 
\begin{thm}
\label{thm:stronghypercontractivity-1} Under the assumption in Theorem
\ref{thm:stronghypercontractivity-2}, it holds that 
\begin{align}
\langle f,g\rangle & \leq\Vert f\Vert_{p}\Vert g\Vert_{q},\forall(p,q)\in\mathcal{R}_{\alpha,\beta}^{+}(P_{XY}),\label{eq:fh-1}\\
\langle f,g\rangle & \geq\Vert f\Vert_{p}\Vert g\Vert_{q},\forall(p,q)\in\mathcal{R}_{\alpha,\beta}^{-}(P_{XY}).\label{eq:rh-1}
\end{align}
\end{thm}
\begin{rem}
\label{rem:hypsharpness} These two inequalities are exponentially
sharp in the same sense as \eqref{eq:fh} and \eqref{eq:rh}; see
Remark \ref{rem:HCsharpness}. That is, given $\alpha\in[0,E_{1,\max}],\beta\in[0,E_{2,\max}]$
and $(p,q)$ in the boundary of $\mathcal{R}_{\alpha,\beta}^{+}(P_{XY})$,
the inequality (\ref{eq:fh}) is exponentially sharp, in the sense
that the exponents on the two sides of (\ref{eq:fh}) are asymptotically
equal as $n\to\infty$, for a sequence of Boolean functions $f_{n}=1_{\calA_{n}},g_{n}=1_{\calB_{n}}$
with some sequence $(\calA_{n},\calB_{n})$. Given $(\alpha,\beta)$
in the effective region of $\overline{\Theta}^{*}$ and $(p,q)$ in
the boundary of $\mathcal{R}_{\alpha,\beta}^{-}(P_{XY})$, the inequality
(\ref{eq:rh}) is exponentially sharp, in the sense that the exponents
on the two sides of (\ref{eq:rh}) are asymptotically equal as $n\to\infty$,
for a sequence of Boolean functions $f_{n}=1_{\calA_{n}},g_{n}=1_{\calB_{n}}$
with some sequence $(\calA_{n},\calB_{n})$. 
\end{rem}
Note that the hypercontractivity inequalities in Theorem \ref{thm:stronghypercontractivity-2}
differ from the common ones in the factors $2^{-n\underline{\Lambda}_{p,q}^{*}(\alpha,\beta)}$
and $2^{n\overline{\Lambda}_{p,q}^{*}(\alpha,\beta)}$; while the
ones in Theorem \ref{thm:stronghypercontractivity-1} differ from
the common ones in the region of parameters $p,q$. Strengthening
the forward hypercontractivity was previously studied in \cite{polyanskiy2019improved,kirshner2021moment}.
Polyanskiy and Samorodnitsky \cite{polyanskiy2019improved} strengthened
the hypercontractivity inequalities in a similar sense to Theorem
\ref{thm:stronghypercontractivity-2}; while Kirshner and Samorodnitsky
\cite{kirshner2021moment} strengthened the hypercontractivity inequalities
in a similar sense to Theorem \ref{thm:stronghypercontractivity-1}.
However, both works in \cite{polyanskiy2019improved,kirshner2021moment}
focused on strengthening the single-function version of forward hypercontractivity.
Moreover, the hypercontractivity inequalities in \cite{polyanskiy2019improved}
are only sharp at extreme cases, and only DSBSes were considered in
\cite{kirshner2021moment}. A systematic investigation of the exponentially
sharp version of Brascamp--Lieb and hypercontractivity inequalities
in Polish spaces and under a general measure of the ``sizes'' of
functions (termed the two-parameter entropy) was done by the first
author in \cite{yu2021strong}.

\section{\label{sec:Concluding-Remarks}Concluding Remarks}

The maximal density of subgraphs of a type graph and the biclique
rate region have been studied in this paper. One may be also interested
in their counterparts---the minimal density of subgraphs of a type
graph and the independent-set rate region. Here, given a joint $n$-type
$T_{XY}$, $1\le M_{1}\le|\mathcal{T}_{T_{X}}|$, and $1\le M_{2}\le|\mathcal{T}_{T_{Y}}|$,
we define the minimal density of subgraphs of the type graph of $T_{XY}$
with size $(M_{1},M_{2})$ as 
\begin{align*}
\underline{\Gamma}_{n}(M_{1},M_{2}) & :=\min_{\substack{\calA\subseteq\mathcal{T}_{T_{X}},\calB\subseteq\mathcal{T}_{T_{Y}}:\\
|\calA|=M_{1},|\calB|=M_{2}
}
}\rho(G[\calA,\calB]).
\end{align*}
Similar to the biclique rate region, we define the independent-set
rate region as 
\begin{align*}
\underline{\mathcal{R}}_{n}(T_{XY}) & :=\{(R_{1},R_{2})\in\mathcal{R}_{X}^{(n)}\times\mathcal{R}_{Y}^{(n)}:\\
 & \qquad\qquad\underline{\Gamma}_{n}(2^{nR_{1}},2^{nR_{2}})=0\}.
\end{align*}
Then one can easily obtain the following inner bound and outer bound
on $\underline{\mathcal{R}}_{n}(T_{XY})$. 
\begin{prop}
\label{thm:bicliquerateregion-1} For any $n$ and $T_{XY}$, 
\begin{align*}
 & (\underline{\mathcal{R}}^{(i)}(T_{XY})-[0,\varepsilon_{1,n}]\times[0,\varepsilon_{2,n}])\cap(\mathcal{R}_{X}^{(n)}\times\mathcal{R}_{Y}^{(n)})\\
 & \subseteq\underline{\mathcal{R}}_{n}(T_{XY})\\
 & \subseteq\underline{\mathcal{R}}^{(o)}(T_{XY})\cap(\mathcal{R}_{X}^{(n)}\times\mathcal{R}_{Y}^{(n)})
\end{align*}
for some positive sequences $\{\varepsilon_{1,n}\}$ and $\{\varepsilon_{2,n}\}$
which both vanish as $n\to\infty$, where 
\begin{align*}
 & \underline{\mathcal{R}}^{(o)}(T_{XY}):=\{(R_{1},R_{2}):R_{1}\le H(X),R_{2}\le H(Y)\},\\
 & \underline{\mathcal{R}}^{(i)}(T_{XY}):=\bigcup_{\substack{P_{W},P_{X|W},P_{Y|W}:\\
P_{W}P_{X|W},P_{W}P_{Y|W}\textrm{ are }n\textrm{-types},\\
P_{X}=T_{X},P_{Y}=T_{Y},\\
Q_{XY}\neq T_{XY},\forall Q_{XY|W}\in\calC(P_{X|W},P_{Y|W})
}
}\\
 & \qquad\{(R_{1},R_{2}):R_{1}\le H(X|W),R_{2}\le H(Y|W)\}.
\end{align*}
\end{prop}
The inner bound above can be proven by using the codes used in proving
the achievability part of Theorem \ref{thm:exponentoftypegraph}.
The outer bound above is trivial. Determining the asymptotics of $\underline{\mathcal{R}}_{n}(T_{XY})$
could be of interest. However, currently, we have no idea how to tackle
it. In addition, if $\underline{\mathcal{R}}_{n}(T_{XY})$ is not
asymptotically equal to $\underline{\mathcal{R}}^{(o)}(T_{XY})$,
then determining the exponent of the minimal density is also interesting.

Furthermore, many other fundamental properties of type graphs remain
to be investigated, including graph coloring, graph circuits, graph
embedding, graph connectivity, covering and packing, etc. \cite{west2001introduction}.
Thanks to good structures enjoyed by type graphs, it seems not hopeless
to characterize them.

\section*{Acknowledgement}

The authors are grateful to Or Ordentlich, Yury Polyanskiy, and Ofer
Shayevitz for sharing their code and the details about their calculation
in \cite{ordentlich2020note} to help us find out and fix an error
in the previous version of Theorem \ref{thm:zeroerror}. We also thank
Amin Gohari and Sandeep Pradhan for pointing out the related references
\cite{krithivasan2007large,nazari2009new,nazari2010typicality} to
us. We would like to thank anonymous reviewers for pointing out related
references, and especially thank one of reviewers for pointing out
that the strong small-set theorem is not new, and in fact it is a
direct consequence of the information-theoretic characterization of
Brascamp--Lieb inequalities.

\appendices{}

\section{\label{sec:Proof-of-Lemma-F}Proof of Lemma \ref{lem:F}}

Statements 1) and 2) follow directly from the definition of $F^{*}(R_{1},R_{2})$.
Note that in Statement 2), the maximum $H_{T}(X,Y)$ is attained by
$P_{XYW}=T_{XY}P_{W}$ (or set $W$ to constant) when $R_{1}=H_{T}(X),R_{2}=H_{T}(Y)$.

Statement 3): By symmetry, it suffices to only consider the case $R_{1}=0$.
By Statement 2), $F^{*}(0,R_{2})\le\min\{R_{2},H_{T}(Y|X)\}$. On
the other hand, if $R_{2}\ge H_{T}(Y|X)$, then we choose $W=X$,
which leads to 
$H(X|W)=0$ and $H(Y|W)=H(X,Y|W)=H_{T}(Y|X)$. Hence we have $F^{*}(0,R_{2})=H_{T}(Y|X)$
for $R_{2}\ge H_{T}(Y|X)$. If $R_{2}\le H_{T}(Y|X)$, then one can
find a random variable $U$ such that $H(Y|X,U)=R_{2}$. For example,
we choose $U=(V,J)$ with $V$ defined on $\mathcal{X}\cup\mathcal{Y}$
and $J$ defined on $\{0,1\}$ such that $V=X$ if $J=0$ and $V=Y$
if $J=1$, where $J\sim\mathrm{Bern}(\alpha)$ 
for $\alpha:=R_{2}/H_{T}(Y|X)$ is independent of $(X,Y)$. 
Set $W=(X,U)$. We have 
$H(X|W)=0$ and $H(Y|W)=H(X,Y|W)=H(X,Y|W,J)=R_{2}$. 
Hence we have $F^{*}(0,R_{2})=R_{2}$ for $R_{2}\le H_{T}(Y|X)$.

Statement 4): Let $P_{XYW_{0}}$ attain $F^{*}(R_{1},R_{2})$, and
$P_{XYW_{1}}$ attain $F^{*}(\hat{R}_{1},\hat{R}_{2})$. For $0<\alpha<1$,
define $J\sim\mathrm{Bern}(\alpha)$ independent of $(X,Y,W_{0},W_{1})$
and let 
$W:=W_{J}$, taking values in $\mathcal{W}_{0}\cup\mathcal{W}_{1}$,
where $\mathcal{W}_{j}$ denotes the alphabet of $W_{j}$ for $j=0,1$.
Note that $J$ is a deterministic function of $W$. Then $P_{XYW}$
induces 
\begin{align*}
H(X,Y|W) & =\alpha H(X,Y|W_{0})+(1-\alpha)H(X,Y|W_{1}),\\
H(X|W) & =\alpha H(X|W_{0})+(1-\alpha)H(X|W_{1}),\\
H(Y|W) & =\alpha H(Y|W_{0})+(1-\alpha)H(Y|W_{1}).
\end{align*}
Therefore, 
\begin{align*}
 & F^{*}(\alpha R_{1}+(1-\alpha)\hat{R}_{1},\alpha R_{2}+(1-\alpha)\hat{R}_{2})\\
 & \ge\alpha F^{*}(R_{1},R_{2})+(1-\alpha)F^{*}(\hat{R}_{1},\hat{R}_{2}).
\end{align*}

Statement 5): If $\delta_{1}=\delta_{2}=0$, there is nothing to prove.
If $\delta_{2}>\delta_{1}=0$, then, for $t\ge0$, 
\[
f(t):=F^{*}(R_{1},t)
\]
is nondecreasing and concave, by Statements 1) and 4). Hence, for
fixed $\delta_{2}$, 
\[
\frac{f(t+\delta_{2})-f(t)}{\delta_{2}}
\]
is nonincreasing in $t$. Combining this with Statements 2) and 3)
yields 
\begin{align*}
 & \frac{f(t+\delta_{2})-f(t)}{\delta_{2}}\le\frac{f(\delta_{2})-f(0)}{\delta_{2}}\\
 & \leq\frac{\delta_{2}+\min\{R_{1},H_{T}(X|Y)\}-\min\{R_{1},H_{T}(X|Y)\}}{\delta_{2}}\\
 & =1.
\end{align*}
Setting $t=R_{2}$, we obtain $F^{*}(R_{1},R_{2}+\delta_{2})-F^{*}(R_{1},R_{2})\leq\delta_{2}$,
as desired. 

By symmetry, 
the claim also holds in the case $\delta_{1}>\delta_{2}=0$. Now we
consider the case $\delta_{1},\delta_{2}>0$. 
Without loss of generality, we assume $\frac{R_{1}}{\delta_{1}}\ge\frac{R_{2}}{\delta_{2}}$.
For $t\ge-\frac{R_{2}}{\delta_{2}}$, define 
\[
g(t):=F^{*}(R_{1}+\delta_{1}t,R_{2}+\delta_{2}t).
\]
By Statements 1) and 4), $g(t)$ is nondecreasing and concave. Hence,
for fixed $\delta_{2}$, 
\[
g(t+1)-g(t)
\]
is nonincreasing in $t$. Combining this with Statements 2) and 3)
yields that for $t\ge-\frac{R_{2}}{\delta_{2}}$ we have 
\begin{align*}
 & g(t+1)-g(t)\le g(-\frac{R_{2}}{\delta_{2}}+1)-g(-\frac{R_{2}}{\delta_{2}})\\
 & =F^{*}(R_{1}-\frac{\delta_{1}R_{2}}{\delta_{2}}+\delta_{1},\delta_{2})-F^{*}(R_{1}-\frac{\delta_{1}R_{2}}{\delta_{2}},0)\\
 & \leq\min\{R_{1}-\frac{\delta_{1}R_{2}}{\delta_{2}}+\delta_{1},H_{T}(X|Y)\}+\delta_{2}\\
 & \qquad-\min\{R_{1}-\frac{\delta_{1}R_{2}}{\delta_{2}},H_{T}(X|Y)\}\\
 & \leq\delta_{1}+\delta_{2}.
\end{align*}
Setting $t=0$, we obtain $F^{*}(R_{1}+\delta_{1},R_{2}+\delta_{2})-F^{*}(R_{1},R_{2})\leq\delta_{1}+\delta_{2}$,
as desired. 

\section{\label{sec:Proof-of-Theorem-exponent}Proof of Theorem \ref{thm:exponentoftypegraph}}

The claim that we can restrict attention to the case $|\mathcal{W}|\le|\mathcal{X}||\mathcal{Y}|+2$
in the definition of $F^{*}(R_{1},R_{2})$ comes from the support
lemma in \cite{Gamal}. We next prove \eqref{eq:-10}.

Lower bound: Let $\calC:=(\calA\times\calB)\cap\mathcal{T}_{T_{XY}}$
for some optimal $(\calA,\calB)$ attaining $\Gamma_{n}(2^{nR_{1}},2^{nR_{2}})$.
Let $(\mathbf{X},\mathbf{Y})\sim\mathrm{Unif}(\calC)$. Then, 
\begin{align*}
\Gamma_{n}(2^{nR_{1}},2^{nR_{2}}) & =\frac{|\calC|}{|\calA||\calB|}=\frac{2^{H(\mathbf{X},\mathbf{Y})}}{2^{nR_{1}}2^{nR_{2}}},\\
\frac{1}{n}H(\mathbf{X}) & \leq R_{1},\\
\frac{1}{n}H(\mathbf{Y}) & \leq R_{2},
\end{align*}
which follow by the fact that the entropy of a random variable is
no larger than the logarithm of its support size, and they are equal
if the random variable is uniformly distributed over its support.
Therefore, 
\begin{align*}
E_{n}(R_{1},R_{2}) & =R_{1}+R_{2}-\frac{1}{n}H(\mathbf{X},\mathbf{Y})\\
 & =R_{1}+R_{2}-\frac{1}{n}\sum_{i=1}^{n}H(X_{i},Y_{i}|X^{i-1},Y^{i-1})\\
 & =R_{1}+R_{2}-H(X_{J},Y_{J}|X^{J-1},Y^{J-1},J),
\end{align*}
where $J\sim\mathrm{Unif}[n]$ is a random time index independent
of $(X^{n},Y^{n})$ and $X^{J-1}$ denotes a ``random vector''\footnote{Rigorously speaking, the ``random vector'' $X^{J-1}$ is not well
defined since for different $i$, the random vectors $X^{i-1}$ are
defined on different spaces. (The space of $X^{i-1}$ is $\mathcal{X}^{i-1}$
for each $i$.) One way to address this issue is to map $X^{i-1}$
to a common (measurable) space via one-to-one functions. Another simpler
way is to concatenate $X^{i-1}$ with a length-$(n-i+1)$ of constant
symbols, e.g., $\hat{X}_{(i-1)}^{n}:=(X^{i-1},x_{0},\dots,x_{0})$
where $x_{0}$ is a fixed symbol and appears $n-i+1$ times here.
In this case, $X^{J-1}$ denotes $\hat{X}_{(J-1)}^{n}$. This convention
applies throughout this paper. } induced by $(J,X^{n})$. On the other hand, 
\begin{align*}
H(X_{J}|X^{J-1},Y^{J-1},J) & \leq H(X_{J}|X^{J-1}J)=\frac{1}{n}H(\mathbf{X})\leq R_{1},\\
H(Y_{J}|X^{J-1},Y^{J-1},J) & \leq R_{2}.
\end{align*}
Using the notation 
\[
X:=X_{J},Y:=Y_{J},W:=(X^{J-1},Y^{J-1},J),
\]
we obtain $(X,Y)\sim T_{XY},$ and 
\begin{align*}
E_{n}(R_{1},R_{2}) & \geq\inf_{\substack{P_{XYW}:P_{XY}=T_{XY},\\
H(X|W)\leq R_{1},\\
H(Y|W)\leq R_{2}
}
}R_{1}+R_{2}-H(X,Y|W)\\
 & =E^{*}(R_{1},R_{2}).
\end{align*}

Upper bound: In this part, we assume that $W$ is a random variable
defined on an alphabet $\mathcal{W}$ such that $|\mathcal{W}|\le|\mathcal{X}||\mathcal{Y}|+2$.
For a joint $n$-type $P_{XYW}$ such that $P_{XY}=T_{XY},H(X|W)\le R_{1},H(Y|W)\le R_{2}$
and for a fixed sequence $\mathbf{w}$ with type $P_{W}$, we choose
$\calA$ as the union of $\mathcal{T}_{P_{X|W}}(\mathbf{w})$ and
$2^{nR_{1}}-|\mathcal{T}_{P_{X|W}}(\mathbf{w})|$ of arbitrary sequences
outside $\mathcal{T}_{P_{X|W}}(\mathbf{w})$, which is possible because
$\mathcal{T}_{P_{X|W}}(\mathbf{w})\le2^{nH(X|W)}$, see \cite[Lemma 2.5]{Csiszar},
and choose $\calB$ in a similar way, but with $\mathcal{T}_{P_{X|W}}(\mathbf{w})$
replaced by $\mathcal{T}_{P_{Y|W}}(\mathbf{w})$. Then $|\calA|=2^{nR_{1}}$
and $|\calB|=2^{nR_{2}}$. Observe that 
\begin{align*}
|(\calA\times\calB)\cap\mathcal{T}_{T_{XY}}| & \geq|\mathcal{T}_{P_{XY|W}}(\mathbf{w})|\\
 & \geq2^{n(H(X,Y|W)-\frac{|\mathcal{W}||\mathcal{X}||\mathcal{Y}|\log(n+1)}{n})},
\end{align*}
where 
\begin{itemize}
\item the first inequality follows since for any pair $(\mathbf{x},\mathbf{y})\in\mathcal{T}_{P_{XY|W}}(\mathbf{w})$,
the tuple $(\mathbf{w},\mathbf{x},\mathbf{y})$ must have joint type
$P_{WXY}$, and hence, $(\mathbf{w},\mathbf{x})$ has joint type $P_{WX}$,
$(\mathbf{w},\mathbf{y})$ has joint type $P_{WY}$, and $(\mathbf{x},\mathbf{y})$
has joint type $T_{XY}$; 
\item the second inequality follows from \cite[Lemma 2.5]{Csiszar}. 
\end{itemize}
Thus we have 
\begin{align}
\rho(G[\calA,\calB]) & =\frac{|(\calA\times\calB)\cap\mathcal{T}_{T_{XY}}|}{2^{nR_{1}}2^{nR_{2}}}\nonumber \\
 & \geq2^{-n(R_{1}+R_{2}-H(X,Y|W)+\frac{|\mathcal{W}||\mathcal{X}||\mathcal{Y}|\log(n+1)}{n})}.\label{eq:-8}
\end{align}
Optimizing the exponent in (\ref{eq:-8}) over all joint $n$-types
$P_{XYW}$ such that $P_{XY}=T_{XY},H(X|W)\le R_{1},H(Y|W)\le R_{2}$
yields the upper bound 
\begin{align}
E_{n}(R_{1},R_{2}) & \leq R_{1}+R_{2}-F_{n}(R_{1},R_{2})\nonumber \\
 & \qquad+\frac{|\mathcal{W}||\mathcal{X}||\mathcal{Y}|\log(n+1)}{n},\label{eq:-51}
\end{align}
where $F_{n}(R_{1},R_{2})$ is defined similarly as $F^{*}(R_{1},R_{2})$
in (\ref{eq:F}) but with the $P_{XYW}$ in (\ref{eq:F}) restricted
to be a joint $n$-type and $\mathcal{W}$ assumed to satisfy $|\mathcal{W}|\le|\mathcal{X}||\mathcal{Y}|+2$.

We next show that the values of $F_{n}(R_{1},R_{2})$ and $F^{*}(R_{1},R_{2})$
do not differ too much. For a joint $n$-type $T_{XY}$ and a distribution
$P_{XYW}$ with $P_{XY}=T_{XY}$, one can find a $n$-type $Q_{XYW}$
with $Q_{XY}=T_{XY}$ such that $\Vert P_{XYW}-Q_{XYW}\Vert\le\frac{|\mathcal{W}||\mathcal{X}||\mathcal{Y}|}{2n}$,
where $\Vert\cdot\Vert$ denotes the TV distance, see \cite[Lemma 3]{yu2018renyi}.
Combining this with \cite[Lemma 2.7]{Csiszar} (i.e., if $\Vert P_{X}-Q_{X}\Vert\le\Theta\le\frac{1}{4}$,
then $|H_{P}(X)-H_{Q}(X)|\le-2\Theta\log\frac{2\Theta}{|\mathcal{X}|}$),
we have for $\frac{|\mathcal{W}||\mathcal{X}||\mathcal{Y}|}{2n}\le\frac{1}{4}$
that 
\begin{align}
 & |H_{P}(X|W)-H_{Q}(X|W)|\nonumber \\
 & \leq|H_{P}(X,W)-H_{Q}(X,W)|+|H_{P}(W)-H_{Q}(W)|\nonumber \\
 & \leq-2\frac{|\mathcal{W}||\mathcal{X}||\mathcal{Y}|}{2n}\log\frac{2\frac{|\mathcal{W}||\mathcal{X}||\mathcal{Y}|}{2n}}{|\mathcal{X}||\mathcal{W}|}\nonumber \\
 & \qquad-2\frac{|\mathcal{W}||\mathcal{X}||\mathcal{Y}|}{2n}\log\frac{2\frac{|\mathcal{W}||\mathcal{X}||\mathcal{Y}|}{2n}}{|\mathcal{W}|}\\
 & =-\frac{|\mathcal{W}||\mathcal{X}||\mathcal{Y}|}{n}\log\frac{|\mathcal{X}||\mathcal{Y}|^{2}}{n^{2}},\label{eq:-73}
\end{align}
and similarly, 
\begin{align}
|H_{P}(Y|W)-H_{Q}(Y|W)| & \leq-\frac{|\mathcal{W}||\mathcal{X}||\mathcal{Y}|}{n}\log\frac{|\mathcal{X}|^{2}|\mathcal{Y}|}{n^{2}},\\
|H_{P}(XY|W)-H_{Q}(XY|W)| & \leq-\frac{|\mathcal{W}||\mathcal{X}||\mathcal{Y}|}{n}\log\frac{|\mathcal{X}||\mathcal{Y}|}{n^{2}}.\label{eq:-79}
\end{align}
Combining (\ref{eq:-73})-(\ref{eq:-79}) yields that 
\begin{align*}
F_{n}(R_{1},R_{2}) & \geq F^{*}(R_{1}+\frac{|\mathcal{W}||\mathcal{X}||\mathcal{Y}|}{n}\log\frac{|\mathcal{X}||\mathcal{Y}|^{2}}{n^{2}},\\
 & \qquad R_{2}+\frac{|\mathcal{W}||\mathcal{X}||\mathcal{Y}|}{n}\log\frac{|\mathcal{X}|^{2}|\mathcal{Y}|}{n^{2}})\\
 & \qquad+\frac{|\mathcal{W}||\mathcal{X}||\mathcal{Y}|}{n}\log\frac{|\mathcal{X}||\mathcal{Y}|}{n^{2}}.
\end{align*}

Applying Statement 5) of Lemma \ref{lem:F}, we obtain 
\begin{align*}
F_{n}(R_{1},R_{2}) & \geq F^{*}(R_{1},R_{2})+\frac{|\mathcal{W}||\mathcal{X}||\mathcal{Y}|}{n}\log\frac{|\mathcal{X}|^{4}|\mathcal{Y}|^{4}}{n^{6}}.
\end{align*}

Substituting this into the upper bound in (\ref{eq:-51}) and combining
with the assumption $|\mathcal{W}|\le|\mathcal{X}||\mathcal{Y}|+2$
yields the desired upper bound.

\section{\label{sec:Proof-of-Theorem-exponent-1}Proof of Theorem \ref{thm:bicliquerateregion}}

We now prove Theorem~\ref{thm:bicliquerateregion}. Since \eqref{eq:-3}
follows from \eqref{eq:bicliquerateregion}, it suffices to prove
\eqref{eq:bicliquerateregion}.

Inner Bound: The inner bound proof here uses a standard time-sharing
argument. Let $d$ be an integer such that $1\le d\le n-1$. Let $(P_{XY},Q_{XY})$
be a pair comprised of a $d$-joint type and an $(n-d)$-joint type
on $\mathcal{X}\times\mathcal{Y}$ such that $\frac{d}{n}P_{XY}+(1-\frac{d}{n})Q_{XY}=T_{XY}$.
For a fixed length-$d$ sequence $\mathbf{y}$ with type $P_{Y}$
an a fixed length-$(n-d)$ sequence $\mathbf{x}$ with type $Q_{X}$,
we choose $\calA=\mathcal{T}_{P_{X|Y}}(\mathbf{y})\times\{\mathbf{x}\}$
and $\calB=\{\mathbf{y}\}\times\mathcal{T}_{Q_{Y|X}}(\mathbf{x})$.
Then, from \cite[Lemma 2.5]{Csiszar}, we have $|\calA|\geq2^{d(H_{P}(X|Y)-\frac{|\mathcal{X}||\mathcal{Y}|\log(d+1)}{d})}$
and similarly $|\calB|\geq2^{(n-d)(H_{Q}(Y|X)-\frac{|\mathcal{X}||\mathcal{Y}|\log(n-d+1)}{n-d})}$.
On the other hand, for this code we have $\calA\times\calB\subseteq\mathcal{T}_{T_{XY}}$.
Hence any rate pair $(R_{1},R_{2})\in(\mathcal{R}_{X}^{(n)}\times\mathcal{R}_{Y}^{(n)})$
with 
\begin{align*}
R_{1} & \le\frac{d}{n}(H_{P}(X|Y)-\frac{|\mathcal{X}||\mathcal{Y}|\log(d+1)}{d}),\\
R_{2} & \le(1-\frac{d}{n})(H_{Q}(Y|X)-\frac{|\mathcal{X}||\mathcal{Y}|\log(n-d+1)}{n-d}),
\end{align*}
is achievable (i.e., it is in $\mathcal{R}_{n}(T_{XY})$), which in
turn implies that a pair of smaller rates $(R_{1},R_{2})\in(\mathcal{R}_{X}^{(n)}\times\mathcal{R}_{Y}^{(n)})$
with 
\begin{align}
R_{1} & \le\frac{d}{n}H_{P}(X|Y)-\frac{|\mathcal{X}||\mathcal{Y}|\log(n+1)}{n},\label{eq:-53}\\
R_{2} & \le(1-\frac{d}{n})H_{Q}(Y|X)-\frac{|\mathcal{X}||\mathcal{Y}|\log(n+1)}{n},\label{eq:-52}
\end{align}
is achievable.

We next remove the constraint that $(P_{XY},Q_{XY})$ are joint types.
For $0\le\alpha\le1$, let $(\hat{P}_{XY},\hat{Q}_{XY})$ be a pair
of distributions such that $\alpha\hat{P}_{XY}+(1-\alpha)\hat{Q}_{XY}=T_{XY}$.
Define $d:=\Vert\lfloor n\alpha\hat{P}_{XY}\rfloor\Vert_{1}$. Note
that we have 
\begin{equation}
n\alpha-|\mathcal{X}||\mathcal{Y}|\le d\le n\alpha.\label{eq:-11}
\end{equation}

We first consider the case 
\begin{equation}
4|\mathcal{X}||\mathcal{Y}|\le d\le n-4|\mathcal{X}||\mathcal{Y}|.\label{eq:d-condition}
\end{equation}
Define $P_{XY}:=\frac{\lfloor n\alpha\hat{P}_{XY}\rfloor}{d}$. Then
$P_{XY}$ is a joint $d$-type and $\Vert P_{XY}-\hat{P}_{XY}\Vert\le\frac{|\mathcal{X}||\mathcal{Y}|}{d}\le\frac{1}{4}$.
Define $Q_{XY}:=\frac{nT_{XY}-dP_{XY}}{n-d}$, which is a joint $(n-d)$-type
and satisfies $\Vert Q_{XY}-\hat{Q}_{XY}\Vert\le\frac{|\mathcal{X}||\mathcal{Y}|}{n-d}\le\frac{1}{4}$.
Combining \cite[Lemma 2.7]{Csiszar} with the equality $H(X|Y)=H(X,Y)-H(Y)$,
we have 
\begin{align*}
H_{P}(X|Y) & \geq H_{\hat{P}}(X|Y)+\frac{2|\mathcal{X}||\mathcal{Y}|}{d}\log\frac{4|\mathcal{X}|}{d^{2}},\\
H_{Q}(Y|X) & \geq H_{\hat{Q}}(Y|X)+\frac{2|\mathcal{X}||\mathcal{Y}|}{n-d}\log\frac{4|\mathcal{Y}|}{(n-d)^{2}}.
\end{align*}
These inequalities, together with \eqref{eq:-53} and \eqref{eq:-52},
imply that 
\begin{align}
\textrm{RHS of }\eqref{eq:-53} & \geq\frac{d}{n}H_{\hat{P}}(X|Y)-\frac{2|\mathcal{X}||\mathcal{Y}|}{n}\log\frac{n^{2}}{4|\mathcal{X}|}\nonumber \\
 & \qquad-\frac{|\mathcal{X}||\mathcal{Y}|\log(n+1)}{n}\\
 & \geq\alpha H_{\hat{P}}(X|Y)-\frac{|\mathcal{X}||\mathcal{Y}|}{n}\log|\mathcal{X}|\nonumber \\
 & \qquad-\frac{2|\mathcal{X}||\mathcal{Y}|}{n}\log\frac{n^{2}}{4|\mathcal{X}|}\nonumber \\
 & \qquad-\frac{|\mathcal{X}||\mathcal{Y}|\log(n+1)}{n}\\
 & =\alpha H_{\hat{P}}(X|Y)-\epsilon_{1,n};\label{eq:-54}\\
\textrm{RHS of }\eqref{eq:-52} & \geq(1-\frac{d}{n})H_{\hat{Q}}(Y|X)-\frac{2|\mathcal{X}||\mathcal{Y}|}{n}\log\frac{n^{2}}{4|\mathcal{Y}|}\nonumber \\
 & \qquad-\frac{|\mathcal{X}||\mathcal{Y}|\log(n+1)}{n}\\
 & \ge(1-\alpha)H_{\hat{Q}}(Y|X)-\frac{2|\mathcal{X}||\mathcal{Y}|}{n}\log\frac{n^{2}}{4|\mathcal{Y}|}\nonumber \\
 & \qquad-\frac{|\mathcal{X}||\mathcal{Y}|\log(n+1)}{n}\\
 & =(1-\alpha)H_{\hat{Q}}(Y|X)-\epsilon_{2,n}.\label{eq:-55}
\end{align}
Recall the definitions of $\epsilon_{1,n}$ and $\epsilon_{2,n}$
in Theorem \ref{thm:bicliquerateregion}.

Combining \eqref{eq:-53}, \eqref{eq:-52}, \eqref{eq:-54}, and \eqref{eq:-55}
yields that any rate pair $(R_{1},R_{2})\in(\mathcal{R}_{X}^{(n)}\times\mathcal{R}_{Y}^{(n)})$
with $R_{1}\le\alpha H_{\hat{P}}(X|Y)-\epsilon_{1,n}$ and $R_{2}\le(1-\alpha)H_{\hat{Q}}(Y|X)-\epsilon_{2,n}$,
for any $0\le\alpha\le1$ and $(\hat{P}_{XY},\hat{Q}_{XY})$ a pair
of distributions such that $\alpha\hat{P}_{XY}+(1-\alpha)\hat{Q}_{XY}=T_{XY}$,
is achievable as long as the condition in \eqref{eq:d-condition}
holds.

We next consider the case $0\le d<4|\mathcal{X}||\mathcal{Y}|$. For
this case, we have 
\begin{align*}
\alpha H_{\hat{P}}(X|Y) & \le\frac{d+|\mathcal{X}||\mathcal{Y}|}{n}\log|\mathcal{X}|\\
 & \le\frac{5|\mathcal{X}||\mathcal{Y}|}{n}\log|\mathcal{X}|\le\varepsilon_{1,n},
\end{align*}
where the first inequality follows by \eqref{eq:-11} and the fact
that $H_{\hat{P}}(X|Y)\le\log|\mathcal{X}|$. Hence 
\begin{align*}
 & \{(R_{1},R_{2}):R_{1}\le\alpha H_{\hat{P}}(X|Y),R_{2}\le(1-\alpha)H_{\hat{Q}}(Y|X)\}\\
 & \qquad-[0,\varepsilon_{1,n}]\times[0,\varepsilon_{2,n}]
\end{align*}
is empty, and so its intersection with $(\mathcal{R}_{X}^{(n)}\times\mathcal{R}_{Y}^{(n)})$
is also empty. 
Therefore, there is nothing to prove 
in this case. 
The case when $n-4|\mathcal{X}||\mathcal{Y}|<d\le n$ can be handled
similarly. 
This completes the proof for the inner bound.

Outer Bound: We next prove the outer bound by combining information-theoretic
methods and linear algebra. Observe that the biclique rate region
only depends on the probability values of $T_{XY}$, rather than the
alphabets $\mathcal{X},\mathcal{Y}$. With this in mind, we observe
that we can identify $\mathcal{X}$ and $\mathcal{Y}$ with subsets
of $\mathbb{R}$ by one-to-one mappings such that, for any probability
distribution $P_{XY}$, if $(X,Y)\in\mathcal{X}\times\mathcal{Y}$
satisfies $(X,Y)\sim P_{XY}$ we can talk about the expectations $\mathbb{E}_{P}[X]$,
$\mathbb{E}_{P}[Y]$, the covariance $\mbox{Cov}_{P}(X,Y)$, and the
correlation $\mathbb{E}_{P}[XY]$. Translating the choices of $\mathcal{X}$
and/or $\mathcal{Y}$ (as subsets of $\mathbb{R}$) does not change
$\mbox{Cov}_{P}(X,Y)$, so we can ensure that we make these choices
in such a way that $\mathbb{E}_{P}[XY]=\mbox{Cov}_{P}(X,Y)+\mathbb{E}_{P}[X]\mathbb{E}_{P}[Y]=0$.

Let us now choose $\mathcal{X},\mathcal{Y}\subseteq\mathbb{R}$ in
this way, such that for the given joint $n$-type $T_{XY}$ we have
$E_{T}[XY]=0$. 
Then, for $\calA\times\calB\subseteq\mathcal{T}_{T_{XY}}$, we will
have $\langle\mathbf{x},\mathbf{y}\rangle=0$ for any $(\mathbf{x},\mathbf{y})\in\calA\times\calB$,
where $\mathbf{x},\mathbf{y}$ are now viewed as row vectors in $\mathbb{R}^{n}$.
Let $\overline{\calA}$ denote the linear space spanned by all the
vectors in $\calA$, and let $\overline{\calB}$ denote the linear
space spanned by all the vectors in $\calB$. Hence $\overline{\calB}\subseteq\overline{\calA}^{\perp}$,
where $\overline{\calA}^{\perp}$ denotes the orthogonal complement
of a subspace $\overline{\calA}$. As an important property of the
orthogonal complement, $\dim(\overline{\calA})+\dim(\overline{\calA}^{\bot})=n.$
Hence $\dim(\overline{\calA})+\dim(\overline{\calB})\leq n.$

We next establish the following exchange lemma. The proof is provided
in Appendix \ref{sec:Proof-of-Lemma}, and is based on the well-known
exchange lemma in linear algebra. 
\begin{lem}
\label{lem:probexchangelemma} Let $\calV_{1},\calV_{2}$ be mutually
orthogonal linear subspaces of $\mathbb{R}^{n}$ with dimensions,
denoted as $n_{1},n_{2}$, satisfying $n_{1}+n_{2}=n$. Then there
always exists a partition $\{\mathcal{J}_{1},\mathcal{J}_{2}\}$ of
$[n]$ such that $|\mathcal{J}_{i}|=n_{i}$ and $\mathbf{x}=f_{i}(\mathbf{x}_{\mathcal{J}_{i}}),\forall\mathbf{x}\in\calV_{i},i=1,2$
for some deterministic linear functions $f_{i}:\mathbb{R}^{n_{i}}\to\mathbb{R}^{n}$,
where $\mathbf{x}_{\mathcal{J}_{i}}:=(x_{j})_{j\in\mathcal{J}_{i}}$. 
\end{lem}
\begin{rem}
The natural generalization of this lemma also holds for $k$ mutually
orthogonal linear subspaces of $\mathbb{R}^{n}$ with total dimensions
equal to $n$, and can be proved using Lemma \ref{lem:exchangek}
in Appendix \ref{sec:Proof-of-Lemma}. Furthermore, 
the condition ``mutually orthogonal linear subspaces of $\mathbb{R}^{n}$''
can be replaced by ``mutually (linearly) independent linear subspaces
of $\mathbb{R}^{n}$'' 
(i.e., such that the dimension of the span of the subspaces equals
the sum of the dimensions of the subspaces), 
or, more generally, 
affine subspaces each of which is a 
translate of one of a mutually independent family of linear subspaces
of $\mathbb{R}^{n}$. 
\end{rem}
\begin{rem}
In other words, under the assumption in this lemma there always exists
a permutation $\sigma$ of $[n]$ such that $\mathbf{x}^{(\sigma)}=f_{1}(\mathbf{x}_{[1:n_{1}]}^{(\sigma)}),\forall\mathbf{x}\in V_{1}$
and $\mathbf{x}^{(\sigma)}=f_{2}(\mathbf{x}_{[n_{1}+1:n]}^{(\sigma)}),\forall\mathbf{x}\in V_{2}$
for some deterministic functions $f_{i}:\mathbb{R}^{n_{i}}\to\mathbb{R}^{n}$,
where $\mathbf{x}^{(\sigma)}$ is obtained by permuting the components
of $\mathbf{x}$ using $\sigma$. 
\end{rem}
Let $d$ denote $\dim(\overline{\calA})$, so we have $\dim(\overline{\calA}^{\bot})=n-d$.
Let $\mathbf{X}\sim\mathrm{Unif}(\calA),\mathbf{Y}\sim\mathrm{Unif}(\calB)$
be two independent random vectors, i.e., $(\mathbf{X},\mathbf{Y})\sim P_{\mathbf{X},\mathbf{Y}}:=\mathrm{Unif}(\calA)\mathrm{Unif}(\calB)$.
Now we choose $V_{1}=\overline{\calA},V_{2}=\overline{\calA}^{\bot},\mathbf{X}_{1}=\mathbf{X},\mathbf{X}_{2}=\mathbf{Y}$
in Lemma \ref{lem:probexchangelemma}. Then there exists a partition
$\{\mathcal{J},\mathcal{J}^{c}\}$ of $[n]$ such that $|\mathcal{J}|=d$
and $\mathbf{X}=f_{1}(\mathbf{X}_{\mathcal{J}}),\mathbf{Y}=f_{2}(\mathbf{Y}_{\mathcal{J}^{c}})$
for some deterministic functions $f_{1}:\mathbb{R}^{d}\to\mathbb{R}^{n},f_{2}:\mathbb{R}^{n-d}\to\mathbb{R}^{n}$.
By this property, on the one hand we have 
\begin{align*}
R_{1} & =\frac{1}{n}H(\mathbf{X})=\frac{1}{n}H(\mathbf{X}|\mathbf{Y})=\frac{1}{n}H(\mathbf{X}_{\mathcal{J}}|\mathbf{Y})\\
 & \le\frac{1}{n}H(\mathbf{X}_{\mathcal{J}}|\mathbf{Y}_{\mathcal{J}})\leq\frac{1}{n}\sum_{j\in\mathcal{J}}H(X_{j}|Y_{j})\\
 & =\frac{d}{n}H(X_{J}|Y_{J},J)\leq\frac{d}{n}H(X_{J}|Y_{J})=\frac{d}{n}H(\tilde{X}|\tilde{Y}),
\end{align*}
where $J\sim\mathrm{Unif}(\mathcal{J})$, $\tilde{X}:=X_{J},\tilde{Y}:=Y_{J}$,
with $J$ being independent of $(\mathbf{X},\mathbf{Y})$. Similarly,
we have 
\begin{align*}
R_{2} & =\frac{1}{n}H(\mathbf{Y})=\frac{1}{n}H(\mathbf{Y}|\mathbf{X})=\frac{1}{n}H(\mathbf{Y}_{\mathcal{J}^{c}}|\mathbf{X})\\
 & \le\frac{1}{n}H(\mathbf{Y}_{\mathcal{J}^{c}}|\mathbf{X}_{\mathcal{J}^{c}})\leq\frac{1}{n}\sum_{j\in\mathcal{J}^{c}}H(Y_{j}|X_{j})\\
 & =(1-\frac{d}{n})H(Y_{\hat{J}}|X_{\hat{J}},\hat{J})\\
 & \leq(1-\frac{d}{n})H(Y_{\hat{J}}|X_{\hat{J}})=(1-\frac{d}{n})H(\hat{Y}|\hat{X}),
\end{align*}
where $\hat{J}\sim\mathrm{Unif}(\mathcal{J}^{c})$, $\hat{X}:=X_{\hat{J}},\hat{Y}:=Y_{\hat{J}}$,
with $\hat{J}$ being independent of $(\mathbf{X},\mathbf{Y},J)$.
On the other hand, 
\begin{align*}
 & \frac{d}{n}P_{\tilde{X}\tilde{Y}}+(1-\frac{d}{n})P_{\hat{X}\hat{Y}}\\
 & =\frac{1}{n}\sum_{j\in\mathcal{J}}P_{X_{j}Y_{j}}+\frac{1}{n}\sum_{j\in\mathcal{J}^{c}}P_{X_{j}Y_{j}}\\
 & =\frac{1}{n}\sum_{j=1}^{n}P_{X_{j}Y_{j}}=\mathbb{E}_{(\mathbf{X},\mathbf{Y})}[T_{\mathbf{XY}}]=T_{XY},
\end{align*}
where $T_{\mathbf{XY}}$ denotes the joint type of a random pair $(\mathbf{X},\mathbf{Y})$
which is hence also random (but equals $T_{XY}$ pointwise). This
completes the proof of the outer bound.

\section{\label{sec:Proof-of-Lemma}Proof of Lemma \ref{lem:probexchangelemma} }

For a pair of orthogonal subspaces $(\calV,\calV^{\perp})$ with dimensions
respectively $n_{1},n-n_{1}$, let $\{\mathbf{u}_{j}:1\le j\le n_{1}\}$
be an orthogonal basis of $\calV$, and $\{\mathbf{u}_{j}:n_{1}+1\le j\le n\}$
be an orthogonal basis of $\calV^{\perp}$. Then $\{\mathbf{u}_{j}:1\le j\le n\}$
forms an orthogonal basis of $\mathbb{R}^{n}$. Denote by $\boldsymbol{\mathbf{U}}$
the $n\times n$ matrix with $j$-th row being $\mathbf{u}_{j}$.
Then $\boldsymbol{\mathbf{U}}$ is orthogonal. We now express $\mathbf{x}\in\calV$
and $\mathbf{y}\in\calV^{\perp}$, thought of as row vectors, in terms
of this orthogonal basis, i.e., 
\begin{equation}
\mathbf{x}=\hat{\mathbf{x}}\boldsymbol{\mathbf{U}},\qquad\mathbf{y}=\hat{\mathbf{y}}\boldsymbol{\mathbf{U}},\label{eq:-17-1}
\end{equation}
where $\hat{\mathbf{x}}:=\mathbf{x}\boldsymbol{\mathbf{U}}^{\top},\hat{\mathbf{y}}:=\mathbf{y}\boldsymbol{\mathbf{U}}^{\top}$,
and $\boldsymbol{\mathbf{U}}^{\top}$ is the transpose of $\boldsymbol{\mathbf{U}}$.
Since for any $\mathbf{x}\in\calV$ we have $\langle\mathbf{x},\mathbf{u}_{j}\rangle=0$
for all $n_{1}+1\le j\le n$, we obtain that $\hat{x}_{j}=0$ for
all $n_{1}+1\le j\le n$. Similarly, $\hat{y}_{j}=0$ for all $1\le j\le n_{1}$.
Hence we can rewrite $\hat{\mathbf{x}}=(\hat{\mathbf{x}}_{1},\mathbf{0}),\hat{\mathbf{y}}=(\mathbf{0},\hat{\mathbf{y}}_{2})$.
We write $\boldsymbol{\mathbf{U}}$ in a block form: $\boldsymbol{\mathbf{U}}=\begin{bmatrix}\boldsymbol{\mathbf{U}}_{1}\\
\boldsymbol{\mathbf{U}}_{2}
\end{bmatrix}$ where $\boldsymbol{\mathbf{U}}_{1},\boldsymbol{\mathbf{U}}_{2}$
are respectively of size $n_{1}\times n,(n-n_{1})\times n$. Then
\begin{equation}
\mathbf{x}=\hat{\mathbf{x}}_{1}\boldsymbol{\mathbf{U}}_{1},\qquad\mathbf{y}=\hat{\mathbf{y}}_{2}\boldsymbol{\mathbf{U}}_{2}.\label{eq:-17}
\end{equation}
We now need the following well-known exchange lemma. 
\begin{lem}[Exchange Lemma]
\label{lem:exchangek} \cite[Theorem 3.2]{greene1975some} Let $k\ge2$
be an integer. Let $\mathbf{B}$ be an $n\times n$ nonsingular matrix,
and $\{\mathcal{H}_{l},1\le l\le k\}$ be a partition of $[n]$. Then
there always exists another partition $\{\mathcal{L}_{l},1\le l\le k\}$
of $[n]$ with $|\mathcal{L}_{l}|=|\mathcal{H}_{l}|$ such that all
the sub-matrices $\mathbf{B}_{\mathcal{H}_{l},\mathcal{L}_{l}},1\le l\le k$
are nonsingular. 
\end{lem}
The proof of this lemma follows easily from repeated use of the Laplace
expansion for determinants. 
A short proof in the case $k=2$, which is the only case we use, goes
as follows. 
Let $\mathbf{B}=(b_{i,j})$ be an $n\times n$ matrix and $\mathcal{H}$
a subset of $[n]$. Then the determinant of $\mathbf{B}$ can be expanded
as follows:

\[
\det(\mathbf{B})=\sum_{\mathcal{L}\subseteq[n]:|\mathcal{L}|=|\mathcal{H}|}\varepsilon^{\mathcal{H},\mathcal{L}}\det(\mathbf{B}_{\mathcal{H},\mathcal{L}})\det(\mathbf{B}_{\mathcal{H}^{c},\mathcal{L}^{c}})
\]
where $\varepsilon^{\mathcal{H},\mathcal{L}}$ is the sign of the
permutation determined by $\mathcal{H}$ and $\mathcal{L}$, equal
to $(-1)^{(\sum_{h\in\mathcal{H}}h)+(\sum_{\ell\in\mathcal{L}}\ell)}$.
Since $\mathbf{B}$ is nonsingular, there must be at least one choice
of $|\mathcal{L}|$, with $|\mathcal{L}|=|\mathcal{H}|$, such that
both $\mathbf{B}_{\mathcal{H},\mathcal{L}}$ and $\mathbf{B}_{\mathcal{H}^{c},\mathcal{L}^{c}}$
are nonsingular, which is what is being claimed. 

Substituting $\mathbf{B}\leftarrow\boldsymbol{\mathbf{U}},\mathcal{H}_{1}\leftarrow[n_{1}],\mathcal{H}_{2}\leftarrow[n_{1}+1:n]$
in this lemma, we obtain that there exists a partition $\{\mathcal{J},\mathcal{J}^{c}\}$
of $[n]$ with $|\mathcal{J}|=n_{1}$ such that both the sub-matrices
$\mathbf{U}_{[n_{1}],\mathcal{J}},\mathbf{U}_{[n_{1}+1:n],\mathcal{J}^{c}}$
are nonsingular. Denote $\boldsymbol{\mathbf{U}}_{1,\mathcal{J}}$
as the submatrix of $\boldsymbol{\mathbf{U}}_{1}$ consisting of $\mathcal{J}$-
indexed columns of $\boldsymbol{\mathbf{U}}_{1}$, and define $\boldsymbol{\mathbf{U}}_{1,\mathcal{J}^{c}},\boldsymbol{\mathbf{U}}_{2,\mathcal{J}},\boldsymbol{\mathbf{U}}_{2,\mathcal{J}^{c}}$
similarly. Then, by definition, $\boldsymbol{\mathbf{U}}_{1,\mathcal{J}}=\mathbf{U}_{[n_{1}],\mathcal{J}},\boldsymbol{\mathbf{U}}_{2,\mathcal{J}^{c}}=\mathbf{U}_{[n_{1}+1:n],\mathcal{J}^{c}}$.
Therefore, from (\ref{eq:-17}), we have 
\[
\hat{\mathbf{x}}_{1}=\mathbf{x}_{\mathcal{J}}\boldsymbol{\mathbf{U}}_{1,\mathcal{J}}^{-1},\qquad\hat{\mathbf{y}}_{2}=\mathbf{y}_{\mathcal{J}^{c}}\boldsymbol{\mathbf{U}}_{2,\mathcal{J}^{c}}^{-1}.
\]
Substituting these back into (\ref{eq:-17}), we obtain that 
\begin{align*}
(\mathbf{x}_{\mathcal{J}},\mathbf{x}_{\mathcal{J}^{c}}) & =\mathbf{x}_{\mathcal{J}}\boldsymbol{\mathbf{U}}_{1,\mathcal{J}}^{-1}(\boldsymbol{\mathbf{U}}_{1,\mathcal{J}},\boldsymbol{\mathbf{U}}_{1,\mathcal{J}^{c}})\\
 & =(\mathbf{x}_{\mathcal{J}},\mathbf{x}_{\mathcal{J}}\boldsymbol{\mathbf{U}}_{1,\mathcal{J}}^{-1}\boldsymbol{\mathbf{U}}_{1,\mathcal{J}^{c}})
\end{align*}
and 
\[
(\mathbf{y}_{\mathcal{J}},\mathbf{y}_{\mathcal{J}^{c}})=(\mathbf{y}_{\mathcal{J}^{c}}\boldsymbol{\mathbf{U}}_{2,\mathcal{J}^{c}}^{-1}\boldsymbol{\mathbf{U}}_{2,\mathcal{J}},\mathbf{y}_{\mathcal{J}^{c}}).
\]
Hence the proof is completed.

\section{\label{sec:Proof-of-Proposition}Proof of Proposition \ref{prop:line}}

From Theorem \ref{thm:bicliquerateregion}, we know that $\mathcal{R}(T_{XY})=\mathcal{R}^{*}(T_{XY})$,
where $\mathcal{R}(T_{XY})$ is the asymptotic biclique rate region,
defined in \eqref{eq:asympbiclique} and $\mathcal{R}^{*}(T_{XY})$
is defined in Theorem \ref{thm:bicliquerateregion}. 
Furthermore, $\mathcal{R}^{*}(T_{XY})$ 
is a closed convex set (see Proposition \ref{prop:convex}). Hence
\[
\mathcal{R}(T_{XY})=\mathcal{R}_{\bigtriangleup}(T_{XY})
\]
if and only if 
\begin{equation}
\max_{\substack{0\le\alpha\le1,P_{XY},Q_{XY}:\\
\alpha P_{XY}+(1-\alpha)Q_{XY}=T_{XY}
}
}\varphi_{\alpha}(P_{XY},Q_{XY})\le1,\label{eq:-20}
\end{equation}
where $\varphi_{\alpha}(P_{XY},Q_{XY}):=\frac{\alpha}{\beta_{1}}H_{P}(X|Y)+\frac{1-\alpha}{\beta_{2}}H_{Q}(Y|X)$
with $\beta_{1}:=H_{T}(X|Y),\beta_{2}:=H_{T}(Y|X)$. Here the domain
of $\varphi_{\alpha}$ can be taken to be the set of pairs of probability
distributions $(P_{XY},Q_{XY})$ such that $\supp(P_{XY})=\supp(Q_{XY})\subseteq\supp(T_{XY})$.
Moreover, (\ref{eq:-20}) can be rewritten as that for any $0\le\alpha\le1,$
\begin{equation}
\max_{P_{XY},Q_{XY}:\alpha P_{XY}+(1-\alpha)Q_{XY}=T_{XY}}\varphi_{\alpha}(P_{XY},Q_{XY})\le1.\label{eq:-20-1}
\end{equation}
Observe that $\varphi_{\alpha}(T_{XY},T_{XY})=1$. Hence (\ref{eq:-20-1})
can be rewritten as that for any $0<\alpha<1,$ $P_{XY}=Q_{XY}=T_{XY}$
is 
an optimal solution to the LHS of (\ref{eq:-20-1}). Next we study
for what kind of $T_{XY}$ it holds for all $0<\alpha<1$ that $P_{XY}=Q_{XY}=T_{XY}$
is 
an optimal solution to the LHS of (\ref{eq:-20-1}). 

Given $0<\alpha<1$, observe that $\alpha P_{XY}+(1-\alpha)Q_{XY}$
is linear in $(P_{XY},Q_{XY})$, and $\varphi_{\alpha}(P_{XY},Q_{XY})$
is concave in $(P_{XY},Q_{XY})$ (which can be shown by the log sum
inequality \cite[Theorem 2.7.1]{Cover}). Hence the LHS of (\ref{eq:-20-1})
is a 
linearly-constrained convex optimization problem. 
This means that showing that the pair $(T_{XY},T_{XY})$ is an extremum
for this convex optimization problem iff $T_{XY}$ satisfies the conditions
given in Corollary \ref{prop:line}, is equivalent to establishing
that $(T_{XY},T_{XY})$ is an optimum for the convex optimization
problem (thus establishing \eqref{eq:-20-1} for $0<\alpha<1$) iff
$T_{XY}$ satisfies the conditions given in Corollary \ref{prop:line}.
Since the notion of extremality is local, to show this it suffices
to consider the modified version of this convex optimization problem
where the domain of $\varphi_{\alpha}$ is taken to be the set of
pairs of probability distributions $(P_{XY},Q_{XY})$ such that $\supp(P_{XY})=\supp(Q_{XY})=\supp(T_{XY})$.

We are thus led to consider the Lagrangian 
\begin{align*}
L & =\varphi_{\alpha}(P_{XY},Q_{XY})\\
 & \qquad+\sum_{(x,y)\in\supp(T_{XY})}\eta(x,y)\big(\alpha P(x,y)\\
 & \qquad\qquad+(1-\alpha)Q(x,y)-T(x,y)\big)\\
 & \qquad+\mu_{1}(\sum_{(x,y)\in\supp(T_{XY})}P(x,y)-1)\\
 & \qquad+\mu_{2}(\sum_{(x,y)\in\supp(T_{XY})}Q(x,y)-1).
\end{align*}
By checking the feasible solution $(P_{XY},Q_{XY})$ with $P_{XY}=Q_{XY}=T_{XY}$,
one can find that Slater's condition for the modified version of the
convex optimization problem in \eqref{eq:-20-1} (described above)
is satisfied, which implies that extrema of the modified version of
the optimization problem in \eqref{eq:-20-1} are given by the Karush--Kuhn--Tucker
(KKT) conditions: 
\begin{align}
\frac{\partial L}{\partial P(x,y)} & =-\frac{\alpha}{\beta_{1}}\log P(x|y)+\alpha\eta(x,y)+\mu_{1}\nonumber \\
 & =0,\forall(x,y)\in\supp(T_{XY}),\label{eq:-22}\\
\frac{\partial L}{\partial Q(x,y)} & =-\frac{1-\alpha}{\beta_{2}}\log Q(y|x)+(1-\alpha)\eta(x,y)+\mu_{2}\nonumber \\
 & =0,\forall(x,y)\in\supp(T_{XY}),\label{eq:-23}
\end{align}
\begin{align}
\alpha P(x,y)+(1-\alpha)Q(x,y) & =T(x,y),\nonumber \\
 & \forall(x,y)\in\supp(T_{XY}),\\
\sum_{(x,y)\in\supp(T_{XY})}P(x,y) & =1,\\
\sum_{(x,y)\in\supp(T_{XY})}Q(x,y) & =1,\\
P(x,y),Q(x,y) & >0,\forall(x,y)\in\supp(T_{XY}),\label{eq:-26}
\end{align}
for some reals $\eta(x,y),\mu_{1},\mu_{2}$ 
with $(x,y)\in\supp(T_{XY})$. Here the conditions in \eqref{eq:-26}
come from the restriction we have imposed on the domain of $\varphi_{\alpha}$.

We first prove ``if'' part. That is, for $T_{XY}$ satisfying the
conditions given in Corollary \ref{prop:line}, given any $0<\alpha<1$,
$P_{XY}=Q_{XY}=T_{XY}$ together with some reals $\eta(x,y),\mu_{1},\mu_{2}$
must satisfy (\ref{eq:-22})-(\ref{eq:-26}). To this end, we choose
$\eta(x,y)=\frac{1}{\beta_{1}}\log T(x|y)=\frac{1}{\beta_{2}}\log T(y|x)$,
$\mu_{1}=\mu_{2}=0$, which satisfy (\ref{eq:-22}) and (\ref{eq:-23}).



We next consider the ``only if'' part. Substituting $P=Q=T$ and
taking expectations with respect to the type $T_{XY}$ for the both
sides of (\ref{eq:-22}) and (\ref{eq:-23}), we obtain that 
\begin{equation}
\frac{\mu_{1}}{\alpha}=\frac{\mu_{2}}{1-\alpha}.\label{eq:-29}
\end{equation}
Substituting this back to (\ref{eq:-22}) and (\ref{eq:-23}) yields
that $T_{X|Y}(x|y)^{1/H_{T}(X|Y)}=T_{Y|X}(y|x)^{1/H_{T}(Y|X)}$ for
all $x,y$.

\section{\label{sec:Proof-of-Lemma-D}Proof of Lemma \ref{lem:InequalityD}}

The two equalities above can be verified easily. Here we only prove
the inequality above. Without loss of generality, we assume $0\le\beta\le\alpha\le\frac{1}{2}$.
Then, in the definition of $\DD(\alpha,\beta)$, we minimize $D_{\alpha,\beta}(p)$
over $0\le p\le\beta$. Furthermore, 
\begin{align*}
D_{\alpha,\beta}(p) & =-H(p,\alpha-p,\beta-p,1+p-\alpha-\beta)\\
 & \qquad-(1+2p-\alpha-\beta)\log(1+\rho)\\
 & \qquad-(\alpha+\beta-2p)\log(1-\rho)+\log4.
\end{align*}
Let $s:=\alpha+\beta-2p$. Then we have 
\begin{align*}
D_{\alpha,\beta}(p) & =-H(p,\alpha-p,\beta-p,1+p-\alpha-\beta)|_{p=\frac{\alpha+\beta-s}{2}}\\
 & \qquad-(1-s)\log(1+\rho)-s\log(1-\rho)+\log4.
\end{align*}
By definition, $\DD(\alpha,\beta)$ can be rewritten as the minimum
of $D_{\alpha,\beta}(p)$ over $\alpha-\beta\le s\le\alpha+\beta$.
Given $(\alpha,\beta)$, $D_{\alpha,\beta}(p)$ is convex in $s$
which follows by the convexity of the relative entropy. Moreover,
$H(p,\alpha-p,\beta-p,1+p-\alpha-\beta)$ is maximized at $p=\alpha\beta$
, i.e., at $s=\alpha+\beta-2\alpha\beta$. Hence, the derivative of
$D_{\alpha,\beta}(p)$ w.r.t. $s$ at $s=\alpha+\beta-2\alpha\beta$
is $\log\frac{1+\rho}{1-\rho}$, which is nonnegative. This implies
that the minimum of $D_{\alpha,\beta}(p)$ is attained at some point
$s$ such that $\alpha-\beta\le s\le\alpha+\beta-2\alpha\beta$ (or
equivalently, at some $p\in(\alpha\beta,\beta]$). In other words,
without changing the value of $\DD(\alpha,\beta)$, one can replace
$H(p,\alpha-p,\beta-p,1+p-\alpha-\beta)$ above with 
\begin{align*}
 & \tilde{H}(p,\alpha-p,\beta-p,1+p-\alpha-\beta)\\
 & :=\begin{cases}
H(p,\alpha-p,\beta-p,\\
\qquad1+p-\alpha-\beta), & p\in(\alpha\beta,\beta],\\
h(\alpha)+h(\beta), & p\in(-\infty,\alpha\beta].
\end{cases}
\end{align*}
That is, $\DD(\alpha,\beta)$ is equal to the minimum of 
\begin{align*}
 & -\tilde{H}(p,\alpha-p,\beta-p,1+p-\alpha-\beta)|_{p=\frac{\alpha+\beta-s}{2}}\\
 & \qquad-(1-s)\log(1+\rho)-s\log(1-\rho)+\log4
\end{align*}
over $s\ge\alpha-\beta$.

We next deal with $\DD(1-\alpha,\beta)$. In the definition of $\DD(1-\alpha,\beta)$,
we minimize $D_{1-\alpha,\beta}(p)$ over the same range $0\le p\le\beta$.
Furthermore, 
\begin{align*}
D_{1-\alpha,\beta}(p) & =-H(p,1-\alpha-p,\beta-p,\alpha+p-\beta)\\
 & \qquad-(\alpha+2p-\beta)\log(1+\rho)\\
 & \qquad-(1-\alpha+\beta-2p)\log(1-\rho)+\log4.
\end{align*}
Let $t:=1-\alpha+\beta-2p$. Then, similarly to the above, we have
\begin{align*}
D_{1-\alpha,\beta}(p) & =-H(p,\alpha-p,\beta-p,\\
 & \qquad1+p-\alpha-\beta)|_{p=\frac{\alpha+\beta-(1-t)}{2}}\\
 & \qquad-(1-t)\log(1+\rho)-t\log(1-\rho)+\log4,
\end{align*}
which can be seen by verifying that 
\begin{align*}
 & -H(p,\alpha-p,\beta-p,1+p-\alpha-\beta)|_{p=\frac{\alpha+\beta-(1-t)}{2}}\\
 & =-H(p,1-\alpha-p,\beta-p,\alpha+p-\beta)|_{p=\frac{1-\alpha+\beta-t}{2}}.
\end{align*}
Hence, $\DD(1-\alpha,\beta)$ is equal to the minimum of $D_{1-\alpha,\beta}(p)$
over $1-\alpha-\beta\le t\le1-\alpha+\beta$. Given $(\alpha,\beta)$,
$D_{1-\alpha,\beta}(p)$ is convex in $t$. Moreover, $H(p,\alpha-p,\beta-p,1+p-\alpha-\beta)$
is maximized at $p=\alpha\beta$, i.e., at $t=1-\alpha-\beta+2\alpha\beta$.
Hence the derivative of $D_{1-\alpha,\beta}(p)$ w.r.t. $t$ at $t=1-\alpha-\beta+2\alpha\beta$
is still $\log\frac{1+\rho}{1-\rho}$ which is nonnegative. Hence
$\DD(1-\alpha,\beta)$ is equal to the minimum of $D_{1-\alpha,\beta}(p)$
over $1-\alpha-\beta\le t\le1-\alpha-\beta+2\alpha\beta$.

To prove $\DD(1-\alpha,\beta)\ge\DD(\alpha,\beta)$, it suffices to
show that 
\begin{align}
 & -H(p,\alpha-p,\beta-p,1+p-\alpha-\beta)|_{p=\frac{\alpha+\beta-(1-s)}{2}}\nonumber \\
 & \geq-\tilde{H}(p,\alpha-p,\beta-p,1+p-\alpha-\beta)|_{p=\frac{\alpha+\beta-s}{2}}\label{eq:-12}
\end{align}
for all $1-\alpha-\beta\le s\le1-\alpha-\beta+2\alpha\beta$. By the
definition of $\tilde{H}$, we only need to check 
\begin{align}
g(s) & :=H(p,\alpha-p,\beta-p,1+p-\alpha-\beta)|_{p=\frac{\alpha+\beta-s}{2}}\nonumber \\
 & \qquad-H(p,\alpha-p,\beta-p,\nonumber \\
 & \qquad\qquad1+p-\alpha-\beta)|_{p=\frac{\alpha+\beta-(1-s)}{2}}\label{eq:-13}\\
 & \geq0\nonumber 
\end{align}
for $1-\alpha-\beta\le s\le\alpha+\beta-2\alpha\beta$. If $\alpha+\beta-2\alpha\beta<1-\alpha-\beta$
there is nothing to show. We may therefore assume that $1-\alpha-\beta\le\alpha+\beta-2\alpha\beta$.

Computing the derivative of $g$, we see that $g$ is nonincreasing
on $[1-\alpha-\beta,\alpha+\beta-2\alpha\beta]$. On the other hand,
observe that $g(\alpha+\beta-2\alpha\beta)\ge0$ since the maximum
of the first entropy in \eqref{eq:-13} is attained at $s=\alpha+\beta-2\alpha\beta$.
Hence, we have $g\ge0$ on $[1-\alpha-\beta,\alpha+\beta-2\alpha\beta]$.
This completes the proof. 
\begin{rem}
Although the proof above seems complicated, the intuition behind it
is simple. Observe that $\DD(\alpha,\beta)$ is equal to the asymptotic
exponent of $P_{XY}^{n}(\calA_{n}\times\calB_{n})$ where $\calA_{n},\calB_{n}$
are type classes with types asymptotically converging to $(\alpha,1-\alpha)$
and $(\beta,1-\beta)$ respectively. In other words, $\calA_{n},\calB_{n}$
are the concentric Hamming spheres with common center $(0,0,...,0)$
with radii $r_{n},s_{n}$ satisfying $r_{n}/n\to\alpha,s_{n}/n\to\beta$
as $n\to\infty$. 
Similarly, $\DD(1-\alpha,\beta)$ is equal to the asymptotic exponent
of $P_{XY}^{n}(\hat{\calA}_{n}\times\calB_{n})$ where $\hat{\calA}_{n}$
is the anti-concentric Hamming sphere of $\calA_{n}$. Hence, the
type of $\hat{\calA}_{n}$ converges to $(1-\alpha,\alpha)$ asymptotically.
On the other hand, we can write for $\mathbf{y}\in\calB_{n},$ 
\[
P_{X|Y}^{n}(\calA_{n}|\mathbf{y})=(\frac{1+\rho}{2})^{n}\sum_{\mathbf{x}\in\calA_{n}}(\frac{1-\rho}{1+\rho})^{d(\mathbf{x,}\mathbf{y})}.
\]
By permutation, one can observe that the expression above remains
the same for all $\mathbf{y}\in\calB_{n}$. Hence, 
\[
P_{XY}^{n}(\calA_{n}\times\calB_{n})=P_{Y}^{n}(\calB_{n})(\frac{1+\rho}{2})^{n}\sum_{\mathbf{x}\in\calA_{n}}(\frac{1-\rho}{1+\rho})^{d(\mathbf{x,}\mathbf{y})}.
\]
Denote $\eta:=\frac{1-\rho}{1+\rho}$ and denote $F$ as the CDF of
the distance $d(\mathbf{X},\mathbf{y})$ with $\mathbf{X}\sim\mathrm{Unif}(\calA_{n})$.
Then, we have 
\begin{align}
\frac{1}{|\calA_{n}|}\sum_{\mathbf{x}\in\calA_{n}}\eta^{d(\mathbf{x,}\mathbf{y})} & =\mathbb{E}_{\mathbf{X}\sim\mathrm{Unif}(\calA_{n})}\eta^{d(\mathbf{X,}\mathbf{y})}\nonumber \\
 & =\sum_{d=0}^{\infty}(F(d)-F(d-1))\eta^{d}\nonumber \\
 & =\sum_{d=0}^{\infty}F(d)(\eta^{d}-\eta^{d+1}).\label{eq:-14}
\end{align}
Similarly, 
\begin{align}
\frac{1}{|\hat{\calA}_{n}|}\sum_{\mathbf{x}\in\hat{\calA}_{n}}\eta^{d(\mathbf{x,}\mathbf{y})} & =\sum_{d=0}^{\infty}G(d)(\eta^{d}-\eta^{d+1}),\label{eq:-15}
\end{align}
where $G$ is the CDF of the distance $d(\mathbf{X},\mathbf{y})$
with $\mathbf{X}\sim\mathrm{Unif}(\hat{\calA}_{n})$. Since the sphere
$\calA_{n}$ is ``closer'' to $\mathbf{y}\in\calB_{n}$ than the
sphere $\hat{\calA}_{n}$, intuitively, $F(d)\ge G(d)$ for all $d\ge0$
which implies $P_{XY}^{n}(\calA_{n}\times\calB_{n})\ge P_{XY}^{n}(\hat{\calA}_{n}\times\calB_{n})$.
This further implies $\DD(1-\alpha,\beta)\ge\DD(\alpha,\beta)$. In
the proof above, we showed that the asymptotic exponent of \eqref{eq:-14}
is not larger than that of \eqref{eq:-15}, with $ns,nt$ denoting
the distances. This is a weaker version of $P_{XY}^{n}(\calA_{n}\times\calB_{n})\ge P_{XY}^{n}(\hat{\calA}_{n}\times\calB_{n})$,
but it still implies $\DD(1-\alpha,\beta)\ge\DD(\alpha,\beta)$. 
\end{rem}

\section{\label{sec:Proof-of-Theorem-1}Proof of Theorem \ref{thm:stronghypercontractivity-2}}

Our proof combines Theorem \ref{thm:sse} with ideas from \cite[Proof of Theorem 1.8]{kirshner2021moment}.
Observe that by the product construction, the optimal exponents 
\begin{align}
\underline{\Lambda}_{p,q}^{(n)}(\alpha,\beta) & :=-\frac{1}{n}\log\sup_{\substack{f,g:P_{X}^{n}(\mathrm{supp}(f))\leq2^{-n\alpha},\\
P_{Y}^{n}(\mathrm{supp}(g))\leq2^{-n\beta}
}
}\frac{\langle f,g\rangle}{\Vert f\Vert_{p}\Vert g\Vert_{q}},\label{eq:FI-3}\\
\overline{\Lambda}_{p,q}^{(n)}(\alpha,\beta) & :=-\frac{1}{n}\log\inf_{\substack{f,g:P_{X}^{n}(\mathrm{supp}(f))\leq2^{-n\alpha},\\
P_{Y}^{n}(\mathrm{supp}(g))\leq2^{-n\beta}
}
}\frac{\langle f,g\rangle}{\Vert f\Vert_{p}\Vert g\Vert_{q}},\label{eq:FI-1-1}
\end{align}
satisfy that $n\underline{\Lambda}_{p,q}^{(n)}(\alpha,\beta)$ is
subadditive and $n\overline{\Lambda}_{p,q}^{(n)}(\alpha,\beta)$ is
superadditive in $n$. So, by Fekete's lemma, $\inf_{n\ge1}\underline{\Lambda}_{p,q}^{(n)}(\alpha,\beta)=\lim_{n\to\infty}\underline{\Lambda}_{p,q}^{(n)}(\alpha,\beta)$
and $\sup_{n\ge1}\overline{\Lambda}_{p,q}^{(n)}(\alpha,\beta)=\lim_{n\to\infty}\overline{\Lambda}_{p,q}^{(n)}(\alpha,\beta)$,
which means that we only need to focus on the asymptotic case.

We may assume, by homogeneity, that $\Vert f\Vert_{1}=\Vert g\Vert_{1}=1$.
This means that $f\le1/P_{X,\min}^{n},g\le1/P_{Y,\min}^{n}$, and
moreover, $\frac{1}{n}\log\|f\|_{p}$ and $\frac{1}{n}\log\|g\|_{q}$
are uniformly bounded for all $n\ge1$. This is because given $\|f\|_{1}=1$,
for $p\ge1$, we have 
\begin{align}
1=\|f\|_{1}\le\|f\|_{p}\le\|f\|_{\infty}\le1/P_{X,\min}^{n},\label{eq:boundlp}
\end{align}
and for $0<p\le1$, we have 
\begin{align}
P_{X,\min}^{n(1-p)/p}\le\|f\|_{p}\le\|f\|_{1}=1.
\end{align}

For sufficiently large $a>0$, the points at which $f$ or $g<2^{-na}$
contribute little to $\|f\|_{p}$, $\|g\|_{q}$, and $\langle f,g\rangle$,
in the sense that if we set $f,g$ to be zero at these points (the
resulting functions denoted as $f_{a},g_{a}$), then $\frac{1}{n}\log\|f\|_{p}$,
$\frac{1}{n}\log\|g\|_{q}$, and $\frac{1}{n}\log\langle f,g\rangle$
only change by amounts of the order of $o_{n}(1)$, where $o_{n}(1)$
denotes a term vanishing as $n\to\infty$ uniformly over all $f$
and $g$ with $\|f\|_{1}=\|g\|_{1}=1$. This is because, 
\[
\|f_{a}\|_{p}^{p}\le\|f\|_{p}^{p}\le\|f_{a}\|_{p}^{p}+2^{-npa},
\]
and 
\[
\mathbb{E}[f_{a}g_{a}]\le\mathbb{E}[fg]\le\mathbb{E}[f_{a}g_{a}]+3\cdot2^{-an}.
\]

All the remaining points of $\mathcal{X}^{n}$ can be partitioned
into $r=r(a,b)$ level sets $\calA_{1},...,\calA_{r}$ such that $f$
varies by a factor of at most $2^{nb}$ in each level set, where $b>0$.
Similarly, all the remaining points of $\mathcal{Y}^{n}$ can be partitioned
into $s=s(a,b)$ level sets $\calB_{1},...,\calB_{s}$ such that $g$
varies by a factor of at most $2^{nb}$ in each level set. Let $\alpha_{i}:=-\frac{1}{n}\log P_{X}^{n}(\calA_{i}),\beta_{i}:=-\frac{1}{n}\log P_{Y}^{n}(\calB_{i})$,
and let $\mu_{i}=\frac{1}{n}\log(u_{i}),\nu_{i}=\frac{1}{n}\log(v_{i})$,
where $u_{i},v_{i}$ are respectively the median value of $f$ on
$\calA_{i}$ and the median value of $g$ on $\calB_{i}$. (If $\calA_{i}$
is empty then $u_{i}$ can be chosen to be any value within the level
set defining $\calA_{i}$, and similarly for $\calB_{i}$ and $v_{i}$.)
Note that $f({\bf x})\in[u_{i}2^{-nb},u_{i}2^{nb}]$ on the set $\calA_{i}$
and $g({\bf y})\in[v_{i}2^{-nb},v_{i}2^{nb}]$ on the set $\calB_{i}$.
Moreover, $\alpha_{i}\geq\alpha,\beta_{j}\geq\beta,\forall i,j$.
Then, 
\begin{align*}
\frac{1}{n}\log\Vert f\Vert_{p} & \ge\frac{1}{n}\log\Vert f_{a}\Vert_{p}\\
 & \ge\frac{1}{np}\log[\sum_{i=1}^{r}P_{X}^{n}(\calA_{i})u_{i}^{p}]-b\\
 & \ge N_{X}(p)-b,
\end{align*}
where $N_{X}(p):=\max_{1\le i\le r}\{-\frac{\alpha_{i}}{p}+\mu_{i}\}$.
Similarly, 
\begin{align*}
\frac{1}{n}\log\Vert g\Vert_{q}\ge N_{Y}(q)-b,
\end{align*}
where $N_{Y}(q):=\max_{1\le i\le s}\{-\frac{\beta_{i}}{q}+\nu_{i}\}$.

Utilizing these equations, we obtain 
\begin{align*}
 & \frac{1}{n}\log\langle f,g\rangle\\
 & \le\frac{1}{n}\log[\langle f_{a},g_{a}\rangle+3\cdot2^{-an}]\\
 & \le\frac{1}{n}\log[\sum_{i=1}^{r}\sum_{j=1}^{s}P_{XY}^{n}(\calA_{i}\times\calB_{j})u_{i}v_{j}\cdot2^{2nb}+3\cdot2^{-an}]\\
 & \le\frac{1}{n}\log[rs\cdot2^{n(\max_{1\le i\le r,1\le j\le s}\{-\underline{\Theta}^{*}(\alpha_{i},\beta_{j})+\mu_{i}+\nu_{j}\}+2b)}\\
 & \qquad+3\cdot2^{-an}]\\
 & \le\frac{1}{n}\log[rs\cdot2^{n\max_{1\le i\le r,1\le j\le s}\{-\underline{\Theta}^{*}(\alpha_{i},\beta_{j})+\frac{\alpha_{i}}{p}+\frac{\beta_{j}}{q}\}}\\
 & \qquad\cdot2^{n(N_{X}(p)+N_{Y}(q)+2b)}+3\cdot2^{-an}].
\end{align*}

Combining the inequalities above, we have 
\begin{align}
 & \frac{1}{n}\log\frac{\langle f,g\rangle}{\Vert f\Vert_{p}\Vert g\Vert_{q}}\nonumber \\
 & \le\frac{1}{n}\log[rs\cdot2^{n(-\underline{\Lambda}_{p,q}^{*}(\alpha,\beta)+N_{X}(p)+N_{Y}(q)+2b)}\nonumber \\
 & \qquad\cdot2^{-\log(\Vert f\Vert_{p}\Vert g\Vert_{q})}+\frac{3\cdot2^{-an}}{\Vert f\Vert_{p}\Vert g\Vert_{q}}]\nonumber \\
 & \le\frac{1}{n}\log[rs\cdot2^{n(-\underline{\Lambda}_{p,q}^{*}(\alpha,\beta)+4b)}+\frac{3\cdot2^{-an}}{\Vert f\Vert_{p}\Vert g\Vert_{q}}].\label{eq:ubexp}
\end{align}
From \eqref{eq:boundlp}, we know that if we choose $a$ sufficiently
large, the (negative) exponent of the second term in the last line
above can be arbitrarily large. On the other hand, if $a,b$ are fixed,
then $r,s$ are also fixed. Hence, \eqref{eq:ubexp} is upper bounded
by 
\begin{align*}
-\underline{\Lambda}_{p,q}^{*}(\alpha,\beta)+4b+o_{n}(1).
\end{align*}
Letting $n\to\infty$ and then $b\to0$, we obtain \eqref{eq:fh}.

We next prove (\ref{eq:rh}). First, observe that 
\begin{align*}
\frac{1}{n}\log\Vert f\Vert_{p} & \leq\frac{1}{n}\log\|f_{a}\|_{p}+\frac{1}{np}\log[1+\frac{2^{-npa}}{\|f_{a}\|_{p}^{p}}]\\
 & \leq\frac{1}{np}\log[\sum_{i=1}^{r}P_{X}^{n}(\calA_{i})u_{i}^{p}]+b\\
 & \qquad+\frac{1}{np}\log[1+\frac{2^{-npa}}{\|f_{a}\|_{p}^{p}}]\\
 & \leq N_{X}(p)+b+\epsilon_{n},
\end{align*}
where $\epsilon_{n}:=\frac{1}{np}\log r+\frac{1}{np}\log[1+\frac{2^{-npa}}{\|f_{a}\|_{p}^{p}}]$,
which tends to zero as $n\to\infty$ for large enough $a$ and any
fixed $b$

Similarly, we have 
\begin{align*}
\frac{1}{n}\log\Vert g\Vert_{q}\leq N_{Y}(q)+b+\hat{\epsilon}_{n},
\end{align*}
where $\hat{\epsilon}_{n}:=\frac{1}{nq}\log s+\frac{1}{nq}\log[1+\frac{2^{-nqa}}{\|g_{a}\|_{q}^{q}}]$,
which tends to zero as $n\to\infty$ for large enough $a$ and any
fixed $b$.

On the other hand, 
\begin{align}
 & \frac{1}{n}\log\langle f,g\rangle\nonumber \\
 & \geq\frac{1}{n}\log\langle f_{a},g_{a}\rangle\nonumber \\
 & \geq\frac{1}{n}\log\mathbb{E}[\sum_{i=1}^{r}\sum_{j=1}^{s}P_{XY}^{n}(\calA_{i}\times\calB_{j})u_{i}v_{j}\cdot2^{-2nb}]\nonumber \\
 & \geq\max_{1\le i\le r,1\le j\le s}\{-\overline{\Theta}^{*}(\alpha_{i},\beta_{j})+\mu_{i}+\nu_{j}\}-2b\label{eq:-4}\\
 & \geq-\overline{\Theta}^{*}(\alpha_{i^{*}},\beta_{j^{*}})+\mu_{i^{*}}+\nu_{j^{*}}-2b\label{eq:-30}\\
 & =-\overline{\Theta}^{*}(\alpha_{i^{*}},\beta_{j^{*}})+\frac{\alpha_{i^{*}}}{p}+\frac{\beta_{j^{*}}}{q}\nonumber \\
 & \qquad+N_{X}(p)+N_{Y}(q)-2b\nonumber \\
 & \geq\overline{\Lambda}_{p,q}^{*}(\alpha,\beta)+N_{X}(p)+N_{Y}(q)-2b,\nonumber 
\end{align}
where \eqref{eq:-4} follows from Theorem \ref{thm:sse}, with the
maximum being taken only over those pairs $(i,j)$ for which $P_{X}^{n}(\calA_{i})>0$
and $P_{Y}^{n}(\calB_{j})>0$, since $\overline{\Theta}^{*}(\alpha_{i},\beta_{j})$
is defined only for $\alpha_{i}\in[0,E_{1,\max}]$, $\beta_{j}\in[0,E_{2,\max}]$;
also, in \eqref{eq:-30}, $i^{*}$ is defined as the optimal $i$
attaining $N_{X}(p)$ and $j^{*}$ as the optimal $j$ attaining $N_{Y}(q)$.
Combining the inequalities above, we have 
\begin{align}
\frac{1}{n}\log\frac{\langle f,g\rangle}{\Vert f\Vert_{p}\Vert g\Vert_{q}} & \geq\overline{\Lambda}_{p,q}^{*}(\alpha,\beta)-4b-\epsilon_{n}-\hat{\epsilon}_{n}.\label{eq:ubexp-1}
\end{align}

We first choose $a$ sufficiently large, fix $a,b$, and let $n\to\infty$.
We have both $\epsilon_{n},\hat{\epsilon}_{n}\to0$. We then let $b\to0$,
and hence we obtain \eqref{eq:rh}.

 \bibliographystyle{unsrt}
\bibliography{ref}

\begin{thebibliography}{10}

\bibitem{yu2021type}
L.~Yu, V.~Anantharam, and J.~Chen.
\newblock Type graphs and small-set expansion.
\newblock In {\em 2021 IEEE International Symposium on Information Theory
  (ISIT)}, pages 993--998. IEEE, 2021.

\bibitem{nazari2009new}
A.~Nazari, S.~S. Pradhan, and A.~Anastasopoulos.
\newblock New bounds on the maximal error exponent for multiple-access
  channels.
\newblock In {\em 2009 IEEE International Symposium on Information Theory},
  pages 1704--1708. IEEE, 2009.

\bibitem{Csiszar}
I.~Csiszar and J.~K{\"o}rner.
\newblock {\em Information Theory: Coding Theorems for Discrete Memoryless
  Systems}.
\newblock Cambridge University Press, 2011.

\bibitem{macwilliams1977theory}
F.~J. MacWilliams and N.~J.~A. Sloane.
\newblock {\em The Theory of Error-Correcting Codes}, volume~16.
\newblock Elsevier, 1977.

\bibitem{yu2019on}
L.~Yu and V.~Y.~F. Tan.
\newblock On non-interactive simulation of binary random variables.
\newblock {\em IEEE Trans. Inf. Theory}, 67(4):2528--2538, 2021.

\bibitem{Ahls76}
R.~Ahlswede, P.~G\'{a}cs, and J.~K\"{o}rner.
\newblock {Bounds on conditional probabilities with applications in multi-user
  communication}.
\newblock {\em Z. Wahrscheinlichkeitstheorie verw. Gebiete}, 34(3):157--177,
  1976.

\bibitem{Csi97}
I.~Csisz\'{a}r and J.~{K\"{o}rner}.
\newblock {\em Information Theory: Coding Theorems for Discrete Memoryless
  Systems}.
\newblock Cambridge University Press, 2011.

\bibitem{han1985maximal}
T.~S. Han and K.~Kobayashi.
\newblock Maximal rectangular subsets contained in the set of partially jointly
  typical sequences for dependent random variables.
\newblock {\em Zeitschrift f{\"u}r Wahrscheinlichkeitstheorie und Verwandte
  Gebiete}, 70(1):15--32, 1985.

\bibitem{krithivasan2007large}
D.~Krithivasan and S.~S. Pradhan.
\newblock On large deviation analysis of sampling from typical sets.
\newblock In {\em Proc. Workshop on Information Theory and Applications (ITA)}.
  Citeseer, 2007.

\bibitem{nazari2010typicality}
A.~Nazari, D.~Krithivasan, S.~S. Pradhan, A.~Anastasopoulos, and
  R.~Venkataramanan.
\newblock Typicality graphs and their properties.
\newblock In {\em 2010 IEEE International Symposium on Information Theory},
  pages 520--524. IEEE, 2010.

\bibitem{gacs1973common}
P.~G{\'a}cs and J.~K{\"o}rner.
\newblock Common information is far less than mutual information.
\newblock {\em Problems of Control and Information Theory}, 2(2):149--162,
  1973.

\bibitem{witsenhausen1975sequences}
H.~S. Witsenhausen.
\newblock On sequences of pairs of dependent random variables.
\newblock {\em SIAM Journal on Applied Mathematics}, 28(1):100--113, 1975.

\bibitem{kahn1988influence}
J.~Kahn, G.~Kalai, and N.~Linial.
\newblock The influence of variables on {Boolean} functions.
\newblock In {\em 29th Annual Symposium on Foundations of Computer Science},
  pages 68--80. IEEE, 1988.

\bibitem{mossel2006non}
E.~Mossel, R.~O'Donnell, O.~Regev, J.~E. Steif, and B.~Sudakov.
\newblock Non-interactive correlation distillation, inhomogeneous {Markov}
  chains, and the reverse {Bonami-Beckner} inequality.
\newblock {\em Israel Journal of Mathematics}, 154(1):299--336, 2006.

\bibitem{O'Donnell14analysisof}
R.~O'Donnell.
\newblock {\em Analysis of {Boolean} Functions}.
\newblock Cambridge University Press, 2014.

\bibitem{kamath2016non}
S.~Kamath and V.~Anantharam.
\newblock On non-interactive simulation of joint distributions.
\newblock {\em IEEE Trans. Inf. Theory}, 62(6):3419--3435, 2016.

\bibitem{ordentlich2020note}
O.~Ordentlich, Y.~Polyanskiy, and O.~Shayevitz.
\newblock A note on the probability of rectangles for correlated binary
  strings.
\newblock {\em IEEE Trans. Inf. Theory}, 2020.

\bibitem{kirshner2021moment}
N.~Kirshner and A.~Samorodnitsky.
\newblock A moment ratio bound for polynomials and some extremal properties of
  {Krawchouk} polynomials and {Hamming} spheres.
\newblock {\em IEEE Trans. Inf. Theory}, 67(6):3509--3541, 2021.

\bibitem{borell1985geometric}
C.~Borell.
\newblock Geometric bounds on the {Ornstein--Uhlenbeck} velocity process.
\newblock {\em Probability Theory and Related Fields}, 70(1):1--13, 1985.

\bibitem{mossel2015robust}
E.~Mossel and J.~Neeman.
\newblock Robust optimality of {Gaussian} noise stability.
\newblock {\em Journal of the European Mathematical Society}, 17(2):433--482,
  2015.

\bibitem{mossel2005coin}
E.~Mossel and R.~O'Donnell.
\newblock Coin flipping from a cosmic source: {On} error correction of truly
  random bits.
\newblock {\em Random Structures \& Algorithms}, 26(4):418--436, 2005.

\bibitem{yu2021common}
L.~Yu and V.~Y.~F. Tan.
\newblock Common information, noise stability, and their extensions.
\newblock {\em Foundations and Trends{\textregistered} in Communications and
  Information Theory}, 19(2):107--389, 2022.

\bibitem{bonami1968ensembles}
A.~Bonami.
\newblock Ensembles {$\Lambda (p)$} dans le dual de {$D^{\infty}$}.
\newblock In {\em Annales de l'institut Fourier}, volume~18, pages 193--204,
  1968.

\bibitem{kiener1969uber}
K.~Kiener.
\newblock {\em Uber Produkte von quadratisch integrierbaren Funktionen
  endlicher Vielfalt}.
\newblock PhD thesis, PhD thesis, Dissertation, Universit{\"a}t Innsbruck,
  1969.

\bibitem{schreiber1969fermeture}
M.~Schreiber.
\newblock Fermeture en probabilit{\'e} de certains sous-espaces d'un espace
  {$L^2$}.
\newblock {\em Zeitschrift f{\"u}r Wahrscheinlichkeitstheorie und Verwandte
  Gebiete}, 14(1):36--48, 1969.

\bibitem{bonami1970etude}
A.~Bonami.
\newblock {\'E}tude des coefficients de {Fourier} des fonctions de {$ L^{p}(G)
  $}.
\newblock In {\em Annales de l'institut Fourier}, volume~20, pages 335--402,
  1970.

\bibitem{beckner1975inequalities}
W.~Beckner.
\newblock Inequalities in {F}ourier analysis.
\newblock {\em Annals of Mathematics}, pages 159--182, 1975.

\bibitem{gross1975logarithmic}
L.~Gross.
\newblock Logarithmic {Sobolev} inequalities.
\newblock {\em American Journal of Mathematics}, 97(4):1061--1083, 1975.

\bibitem{ahlswede1976spreading}
R.~Ahlswede and P.~G{\'a}cs.
\newblock Spreading of sets in product spaces and hypercontraction of the
  markov operator.
\newblock {\em The Annals of Probability}, pages 925--939, 1976.

\bibitem{borell1982positivity}
C.~Borell.
\newblock Positivity improving operators and hypercontractivity.
\newblock {\em Mathematische Zeitschrift}, 180(3):225--234, 1982.

\bibitem{bakry1994hypercontractivite}
D.~Bakry.
\newblock L'hypercontractivit{\'e} et son utilisation en th{\'e}orie des
  semigroupes.
\newblock In {\em Lectures on Probability Theory}, pages 1--114. Springer,
  1994.

\bibitem{mossel2013reverse}
E.~Mossel, K.~Oleszkiewicz, and A.~Sen.
\newblock On reverse hypercontractivity.
\newblock {\em Geometric and Functional Analysis}, 23(3):1062--1097, 2013.

\bibitem{carlen2009subadditivity}
E.~A. Carlen and D.~Cordero-Erausquin.
\newblock Subadditivity of the entropy and its relation to {Brascamp--Lieb}
  type inequalities.
\newblock {\em Geometric and Functional Analysis}, 19(2):373--405, 2009.

\bibitem{nair2014equivalent}
Chandra Nair.
\newblock Equivalent formulations of hypercontractivity using information
  measures.
\newblock In {\em International Zurich Seminar}, 2014.

\bibitem{kamath2015reverse}
S.~Kamath.
\newblock Reverse hypercontractivity using information measures.
\newblock In {\em 2015 53rd Annual Allerton Conference on Communication,
  Control, and Computing (Allerton)}, pages 627--633. IEEE, 2015.

\bibitem{beigi2016equivalent}
S.~Beigi and C.~Nair.
\newblock Equivalent characterization of reverse {Brascamp--Lieb-type}
  inequalities using information measures.
\newblock In {\em IEEE International Symposium on Information Theory (ISIT)},
  pages 1038--1042, 2016.

\bibitem{liu2016brascamp}
J.~Liu, T.~A. Courtade, P.~Cuff, and S.~Verd{\'u}.
\newblock {Brascamp-Lieb} inequality and its reverse: An information theoretic
  view.
\newblock In {\em 2016 IEEE International Symposium on Information Theory
  (ISIT)}, pages 1048--1052. IEEE, 2016.

\bibitem{liu2018forward}
J.~Liu, T.~A. Courtade, P.~W. Cuff, and S.~Verd{\'u}.
\newblock A forward-reverse brascamp-lieb inequality: Entropic duality and
  gaussian optimality.
\newblock {\em Entropy}, 20(6):418, 2018.

\bibitem{yu2021strong}
L.~Yu.
\newblock Strong {Brascamp-Lieb} inequalities.
\newblock {\em arXiv preprint arXiv:2102.06935}, Nov. 2021.

\bibitem{lovasz1979shannon}
L.~Lov{\'a}sz.
\newblock On the shannon capacity of a graph.
\newblock {\em IEEE Trans. Inf. Theory}, 25(1):1--7, 1979.

\bibitem{liu2022minoration}
J.~Liu.
\newblock Minoration via mixed volumes and {Cover's} problem for general
  channels.
\newblock {\em Probability Theory and Related Fields}, 183(1):315--357, 2022.

\bibitem{polyanskiy2019improved}
Y.~Polyanskiy and A.~Samorodnitsky.
\newblock Improved log-sobolev inequalities, hypercontractivity and uncertainty
  principle on the hypercube.
\newblock {\em Journal of Functional Analysis}, 277(11):108280, 2019.

\bibitem{Gamal}
A.~El Gamal and Y.-H. Kim.
\newblock {\em Network Information Theory}.
\newblock Cambridge University Press, 2011.

\bibitem{liu2019smoothing}
J.~Liu, T.~A. Courtade, P.~Cuff, and S.~Verd{\'u}.
\newblock Smoothing brascamp-lieb inequalities and strong converses of coding
  theorems.
\newblock {\em IEEE Trans. Inf. Theory}, 66(2):704--721, 2019.

\bibitem{yu2020exact}
L.~Yu and V.~Y.~F. Tan.
\newblock On exact and $\infty$-{R\'enyi} common informations.
\newblock {\em IEEE Trans. Inf. Theory}, 66(6):3366--3406, 2020.

\bibitem{ODonnell14analysisof}
R.~O'Donnell.
\newblock {\em Analysis of {Boolean} Functions}.
\newblock Cambridge University Press, 2014.

\bibitem{liu2018information}
J.~Liu.
\newblock {\em Information theory from a functional viewpoint}.
\newblock PhD thesis, Ph.D. dissertation, Dept. Electr. Eng., Princeton, NJ:
  Princeton University, 2018.

\bibitem{villani2008optimal}
C.~Villani.
\newblock {\em Optimal transport: old and new}, volume 338.
\newblock Springer Science \& Business Media, 2008.

\bibitem{yu2020asymptotics}
L.~Yu.
\newblock Asymptotics of {Strassen's} optimal transport problem.
\newblock {\em Annales de l'Institut Henri Poincar\'e, Probabilit\'es et
  Statistiques}, 59(4):1745 -- 1777, 2023.

\bibitem{gray1974source}
R.~M. Gray and A.~D. Wyner.
\newblock Source coding for a simple network.
\newblock {\em Bell System Technical Journal}, 53(9):1681--1721, 1974.

\bibitem{wyner1974recent}
A.~Wyner.
\newblock Recent results in the {Shannon} theory.
\newblock {\em IEEE Trans. Inf. Theory}, 20(1):2--10, 1974.

\bibitem{yu2022gray}
L.~Yu.
\newblock {Gray--Wyner} and mutual information regions for doubly symmetric
  binary sources and {Gaussian} sources.
\newblock {\em IEEE Trans. Inf. Theory}, 69(10):6251--6268, 2023.

\bibitem{leiyu2021june}
L.~Yu.
\newblock The convexity and concavity of envelopes of the
  minimum-relative-entropy region for the {DSBS}.
\newblock {\em arXiv preprint arXiv:2106.03654}, Jun. 2021.

\bibitem{lindstrom1969determination}
B.~Lindstr{\"o}m.
\newblock Determination of two vectors from the sum.
\newblock {\em Journal of Combinatorial Theory}, 6(4):402--407, 1969.

\bibitem{van1978upper}
H.~Van~Tilborg.
\newblock An upper bound for codes in a two-access binary erasure channel
  (corresp.).
\newblock {\em IEEE Trans. Inf. Theory}, 24(1):112--116, 1978.

\bibitem{kasami1978bounds}
T.~Kasami and S.~Lin.
\newblock Bounds on the achievable rates of block coding for a memoryless
  multiple-access channel.
\newblock {\em IEEE Trans. Inf. Theory}, 24(2):187--197, 1978.

\bibitem{weldon1978coding}
E.~J. Weldon~Jr.
\newblock Coding for a multiple-access channel.
\newblock {\em Information and Control}, 36(3):256--274, 1978.

\bibitem{kasami1983graph}
T.~Kasami, S.~Lin, V.~Wei, and S.~Yamamura.
\newblock Graph theoretic approaches to the code construction for the two-user
  multiple-access binary adder channel.
\newblock {\em IEEE Trans. Inf. Theory}, 29(1):114--130, 1983.

\bibitem{urbanke1998zero}
R.~Urbanke and Q.~Li.
\newblock The zero-error capacity region of the 2-user synchronous bac is
  strictly smaller than its shannon capacity region.
\newblock In {\em 1998 Information Theory Workshop (Cat. No. 98EX131)},
  page~61. IEEE, 1998.

\bibitem{ajjanagadde2015adder}
G.~Ajjanagadde and Y.~Polyanskiy.
\newblock Adder mac and estimates for r{\'e}nyi entropy.
\newblock In {\em 2015 53rd Annual Allerton Conference on Communication,
  Control, and Computing (Allerton)}, pages 434--441. IEEE, 2015.

\bibitem{ordentlich2016upper}
O.~Ordentlich and O.~Shayevitz.
\newblock An upper bound on the sizes of multiset-union-free families.
\newblock {\em SIAM Journal on Discrete Mathematics}, 30(2):1032--1045, 2016.

\bibitem{austrin2017sharper}
P.~Austrin, P.~Kaski, M.~Koivisto, and J.~Nederlof.
\newblock Sharper upper bounds for unbalanced uniquely decodable code pairs.
\newblock {\em IEEE Trans. Inf. Theory}, 64(2):1368--1373, 2017.

\bibitem{west2001introduction}
D.~B. West.
\newblock {\em Introduction to graph theory}, volume~2.
\newblock Prentice hall Upper Saddle River, 2001.

\bibitem{yu2018renyi}
L.~Yu and V.~Y.~F. Tan.
\newblock R{\'e}nyi resolvability and its applications to the wiretap channel.
\newblock {\em IEEE Trans. Inf. Theory}, 65(3):1862--1897, 2018.

\bibitem{greene1975some}
C.~Greene and T.~L. Magnanti.
\newblock Some abstract pivot algorithms.
\newblock {\em SIAM Journal on Applied Mathematics}, 29(3):530--539, 1975.

\bibitem{Cover}
T.~M. Cover and J.~A. Thomas.
\newblock {\em Elements of Information Theory}.
\newblock Wiley-Interscience, 2nd edition, 2006.

\end{thebibliography}

\begin{IEEEbiographynophoto}{Lei Yu} (Member, IEEE)  received the B.E. and Ph.D. degrees in electronic  
engineering from the University of Science and Technology of China (USTC)  
in 2010 and 2015, respectively. From 2015 to 2020, he worked as a 
Post-Doctoral Researcher at the USTC, National University of Singapore, and  
University of California at Berkeley. He is currently an Associate  
Professor at the School of Statistics and Data Science, LPMC, KLMDASR,  
and LEBPS, Nankai University, China. His research interests lie in the  
intersection of probability theory, information theory, and combinatorics.
\end{IEEEbiographynophoto}

\begin{IEEEbiographynophoto}{Venkat Anantharam} (Fellow, IEEE) received the B.Tech. degree in electronics
from IIT Madras in 1980, and the M.S. degree in electrical engineering,
the M.A. and C.Phil. degrees in mathematics, and the Ph.D. degree in electrical
engineering from UC Berkeley in 1982, 1983, 1984, and 1986, respectively.
From 1986 to 1994, he was on the Faculty of the School of EE, Cornell
University, before moving to the Department of Electrical Engineering and
Computer Sciences, UC Berkeley. He is currently on the Faculty with UC
Berkeley. His research interests include communication networking, game
theory, information theory, probability theory, and stochastic control. \end{IEEEbiographynophoto}

\begin{IEEEbiographynophoto}{Jun Chen}  (Senior Member, IEEE) received the B.E. degree in communication engineering from Shanghai Jiao Tong University, Shanghai, China, in 2001, and the M.S. and Ph.D. degrees in electrical and computer engineering from Cornell University, Ithaca, NY, USA, in 2004 and 2006, respectively.   

From September 2005 to July 2006, he was a Post-Doctoral Research Associate with the Coordinated Science Laboratory, University of Illinois at Urbana-Champaign, Urbana, IL, USA, and a Post-Doctoral Fellow with the IBM Thomas J. Watson Research Center, Yorktown Heights, NY, USA, from July 2006 to August 2007. Since September 2007, he has been with the Department of Electrical and Computer Engineering, McMaster University, Hamilton, ON, Canada, where he is currently a Professor. His research interests include information theory, machine learning, wireless communications, and signal processing.    

Dr. Chen was a recipient of the Josef Raviv Memorial Postdoctoral Fellowship in 2006, the Early Researcher Award from the Province of Ontario in 2010, the IBM Faculty Award in 2010, the ICC Best Paper Award in 2020, and the JSPS Invitational Fellowship in 2021. He held the title of the Barber-Gennum Chair of information technology from 2008 to 2013 and the title of the Joseph Ip Distinguished Engineering Fellow from 2016 to 2018. He was an Associate Editor of the IEEE Transactions on Information Theory (2014 - 2016, 2021 - 2024), an Editor of the IEEE Transactions on Green Communications and Networking (2020 - 2021), and a Guest Editor of the Special Issue on Modern Compression for the IEEE Journal on Selected Areas in Information Theory (2022). He is currently serving as an Associate Editor of the IEEE Transactions on Communications, a Lead Editor of the Special Issue Dedicated to the Memory of Toby Berger for the IEEE Journal on Selected Areas in Information Theory, and a Guest Editor of the Special Issue on Rethinking the Information Identification, Representation, and Transmission Pipeline for the IEEE Journal on Selected Areas in Communications.
\end{IEEEbiographynophoto}

\end{document}